\documentclass[aps,prl,twocolumn,showpacs,citeautoscript,longbibliography]{revtex4-1}

\usepackage{graphicx}  
\usepackage{dcolumn}   
\usepackage{bm}        
\usepackage{amssymb}   
\usepackage[none]{hyphenat}  
\usepackage{nicefrac}  
\usepackage{amsmath}   
\usepackage{footmisc}  
\usepackage{latexsym}  
\usepackage{romannum}  
\usepackage{hyperref}  
\usepackage{multirow}  

\renewcommand{\arraystretch}{1.5}

\newcommand{\ev}{\mathbf{\hat{e}}}
\newcommand{\rv}{\mathbf{r}}
\newcommand{\kv}{\mathbf{k}}
\newcommand{\Rv}{\mathbf{R}}
\newcommand{\Ev}{\mathbf{E}}
\newcommand{\Hv}{\mathbf{H}}
\newcommand{\Bv}{\mathbf{B}}
\newcommand{\Av}{\mathbf{A}}
\newcommand{\Jv}{\mathbf{J}}
\newcommand{\Gb}{\overline{\overline{\mathbf{G}}}}
\newcommand{\gb}{\overline{\overline{\mathbf{g}}}}
\newcommand{\kb}{\overline{\overline{\mathbf{k}}}}

\newcommand{\epsb}{\overline{\overline{\varepsilon}}}
\newcommand{\mub}{\overline{\overline{\mu}}}
\newcommand{\xib}{\overline{\overline{\xi}}}
\newcommand{\zetab}{\overline{\overline{\zeta}}}

\renewcommand{\Im}{\mathrm{Im}}
\renewcommand{\Re}{\mathrm{Re}}
\newcommand{\eps}{\varepsilon}

\begin{document}

\title{Thermal spin photonics in the near-field of nonreciprocal
  media}

\author{Chinmay Khandekar} \email{ckhandek@purdue.edu}
\affiliation{Birck Nanotechnology Center, School of Electrical and
  Computer Engineering, College of Engineering, Purdue University,
  West Lafayette, Indiana 47907, USA}

\author{Zubin Jacob} \email{zjacob@purdue.edu} \affiliation{Birck
  Nanotechnology Center, School of Electrical and Computer
  Engineering, College of Engineering, Purdue University, West
  Lafayette, Indiana 47907, USA}

\date{\today}

\begin{abstract} 
 The interplay of spin angular momentum and thermal radiation is a
 frontier area of interest to nanophotonics as well as topological
 physics. Here, we show that a thick planar slab of a nonreciprocal
 material, despite being at thermal equilibrium with its environment,
 can exhibit nonzero photon spin angular momentum and nonzero
 radiative heat flux in its vicinity. We identify them as the
 persistent thermal photon spin (PTPS) and the persistent planar heat
 current (PPHC) respectively. With a practical example system, we
 reveal that the fundamental origin of these phenomena is connected to
 the spin-momentum locking of thermally excited evanescent waves. We
 also discover spin magnetic moment of surface polaritons that further
 clarifies these features. We then propose an imaging experiment based
 on Brownian motion that allows one to witness these surprising
 features by directly looking at them using a lab microscope. We
 further demonstrate the universal behavior of these near-field
 thermal radiation phenomena through a comprehensive analysis of
 gyroelectric, gyromagnetic and magneto-electric nonreciprocal
 materials. Together, these results expose a surprisingly little
 explored research area of thermal spin photonics with prospects for
 new avenues related to non-Hermitian topological photonics and
 radiative heat transport.
\end{abstract}

 \maketitle
  
\section{Introduction}


Thermal spin photonics merges the fields of the thermal radiation and
the spin angular momentum of light. Thermal radiation plays an
important role in energy-conversion and renewable
technologies~\cite{fan2017thermal,tervo2018near} while the spin
angular momentum property of light is fundamentally relevant in the
context of spin-controlled
nanophotonics~\cite{le2015nanophotonic,mitsch2014quantum}, chiral
quantum optics~\cite{lodahl2015interfacing} and
spintronics~\cite{vzutic2004spintronics}. Despite extensive work in
the past few decades, there has been very little overlap between these
two areas. Important developments in this field include spin-polarized
(circularly polarized) far-field thermal radiation from chiral
absorbers~\cite{shitrit2013spin,wu2014spectrally,yin2013interpreting}
and the definition of the degree of polarization in the thermal
near-field~\cite{setala2002degree} of reciprocal media. In stark
contrast, the primary aim of this work is to explore thermal spin
photonics (spin-related thermal radiation phenomena) in the near-field
of non-reciprocal media.


Our work utilizes fluctuational electrodynamics and is fundamentally
beyond the regime of Kirchhoff's laws which is valid only for
far-field thermal emission from bodies at equilibrium. One striking
example where spin angular momentum of thermal radiation is not
captured by Kirchhoff's laws, is circularly polarized thermal emission
from coupled non-equilibrium antennas demonstrated in our recent
work~\cite{khandekar2019circular}. This approach of exploiting
interacting non-equilibrium bodies is fundamentally unrelated to
conventional approaches of achieving spin angular momentum of light
based on either polarization conversion or structural
chirality. Practically, this non-equilibrium mechanism enables
temperature-based reconfigurability of the spin state of emitted
thermal radiation. Our current work deals with bodies at thermal
equilibrium with their surroundings, and reveals surprising spin
angular momentum features in their near-field arising in presence of
nonreciprocity.

To show the universal nature of these non-reciprocal thermal spin
photonic effects, we develop a framework to analyze equilibrium
thermal-radiation properties of a planar slab of a generic
bianisotropic material described by arbitrary permittivity ($\epsb$),
permeability ($\mub$) and magneto-electric susceptibilities
($\xib,\zetab$). Such a material is reciprocal if the material
properties satisfy,
\begin{align*}
\epsb=\epsb^{T}, \hspace{20pt} \mub=\mub^{T}, \hspace{20pt}
\xib=-\zetab^{T}
\end{align*}
It is nonreciprocal if any one of these conditions is violated. With
fluctuational electrodynamic analysis, we show that a nonreciprocal
planar slab at thermal equilibrium with its environment can exhibit
nonzero spin angular momentum of thermal radiation in its
near-field. We identify it as the persistent thermal photon spin
(PTPS) because it exists without any temperature difference analogous
to well-known persistent electronic charge
current~\cite{bleszynski2009persistent,buttiker1983josephson} that
exists without any voltage difference. The PTPS is also accompanied by
locally nonzero radiative heat flux parallel to the surface which we
call as the persistent planar heat current (PPHC).

We reveal that the spin-momentum
locking~\cite{van2016universal,bliokh2015quantum,petersen2014chiral}
of thermally excited evanescent waves, plays a fundamental role in
facilitating both these phenomena with a practical example system. Our
work thus provides the first generalization of the spin-momentum
locking of light well-known in topological photonics and atomic
physics to thermally excited waves. We consider a doped Indium
Antimonide (InSb) slab at room temperature with an arbitrarily
directed magnetic field.  The thermally excited surface plasmon
polariton supported by InSb slab has transverse spin locked to its
momentum~\cite{van2016universal}. Our calculations reveal that the
spin-momentum locked polariton (electromagnetic wave) also carries
spin magnetic moment which leads to polaritonic energy/frequency shift
through Zeeman type interaction with the applied magnetic field. For
InSb sample with doping concentration $\sim 10^{17}$cm$^{-3}$, the
polaritonic spin magnetic moment is found to be around $10\mu_B$ where
$\mu_B$ is Bohr magneton. The polaritonic magnetic moment depends
asymmetrically on the momentum for forward and backward propagating
polaritons leading to asymmetric energy shifts. This clarifies the
fundamental origin of PTPS and PPHC, resulting from asymmetric
contributions of forward and backward propagating evanescent waves.

Detecting thermal radiation effects of non-reciprocal media is an open
challenge. We note however, that our discovered effects PTPS and PPHC
are significantly enhanced in the near-field due to a large density of
thermally excited evanescent states. In particular, we show the
striking result that at a distance $d \lesssim 0.5\mu$m from the slab
surface, the magnitude of PTPS exceeds the spin angular momentum
density contained in the laser light carrying typical power of $\sim
1$mW.  This immediately motivates experimental validation of our
predicted effects by probing optical forces and torques on small
absorptive particles in the thermal near-field of an InSb slab. We
predict that the Brownian movement of these particles will be
sufficiently influenced by the additional thermal spin photonic forces
and it can be directly viewed using a lab microscope.

We further demonstrate the universal behavior of both these near-field
thermal radiation phenomena with a comprehensive analysis of the key
classes of nonreciprocal media namely, gyroelectric ($\epsb\neq
\epsb^{T}$), gyromagnetic ($\mub \neq \mub^{T}$) and magneto-electric
($\xib \neq -\zetab^{T}$) materials. The general analysis describes
the origin and the nature of these features for any given material
type and further reveals that a nonreciprocal material is necessary
but not sufficient to observe PTPS and PPHC.


Our work advances science in multiple directions. It makes a new
fundamental connection between the spin-momentum locking of evanescent
waves~\cite{van2016universal,bliokh2015quantum,petersen2014chiral} and
the radiative heat transfer. The spin-momentum locking of thermally
excited waves opens a new degree of freedom for directional heat
transport at the nanoscale. The spin magnetic moment of gyrotropic
surface polaritons (which contain both s- and p-polarized waves)
invites related studies of spin-dependent quantum plasmonics and
spin-quantization. We also address the experimental detection of the
persistent thermal photonic phenomena which is not addressed by
previous works~\cite{zhu2016persistent,ott2018circular,
  silveirinha2017topological}. It is important for thermodynamic
revalidation of fundamental understanding of nonreciprocal systems and
also because there is no experiment till date probing the intriguing
effects of nonreciprocity on thermal radiation. While thermal
photonics is so far limited to isotropic, anisotropic and gyroelectric
materials~\cite{zhu2018theory,ekeroth2017thermal}, we provide a
universal description for all material types to motivate similar
studies of thermal-radiation phenomena with largely unexplored
material types such as topological insulators, multiferroic and
mangeto-electric materials. We note that the theoretical framework and
the tools employed here will be useful for studying not only
fluctuational~\cite{buhmann2012macroscopic,bermel2010design,
  zhu2014near,zhu2018theory,ben2016photon,ekeroth2017thermal} but also
quantum~\cite{clegg1995fluorescence,novotny2012principles,fuchs2017casimir,
  gangaraj2018optical,latella2017giant} electrodynamic effects by
using \emph{generic, bianisotropic materials}.


\section{Results} 

\begin{figure}[t!]
  \centering\includegraphics[width=0.8\linewidth]{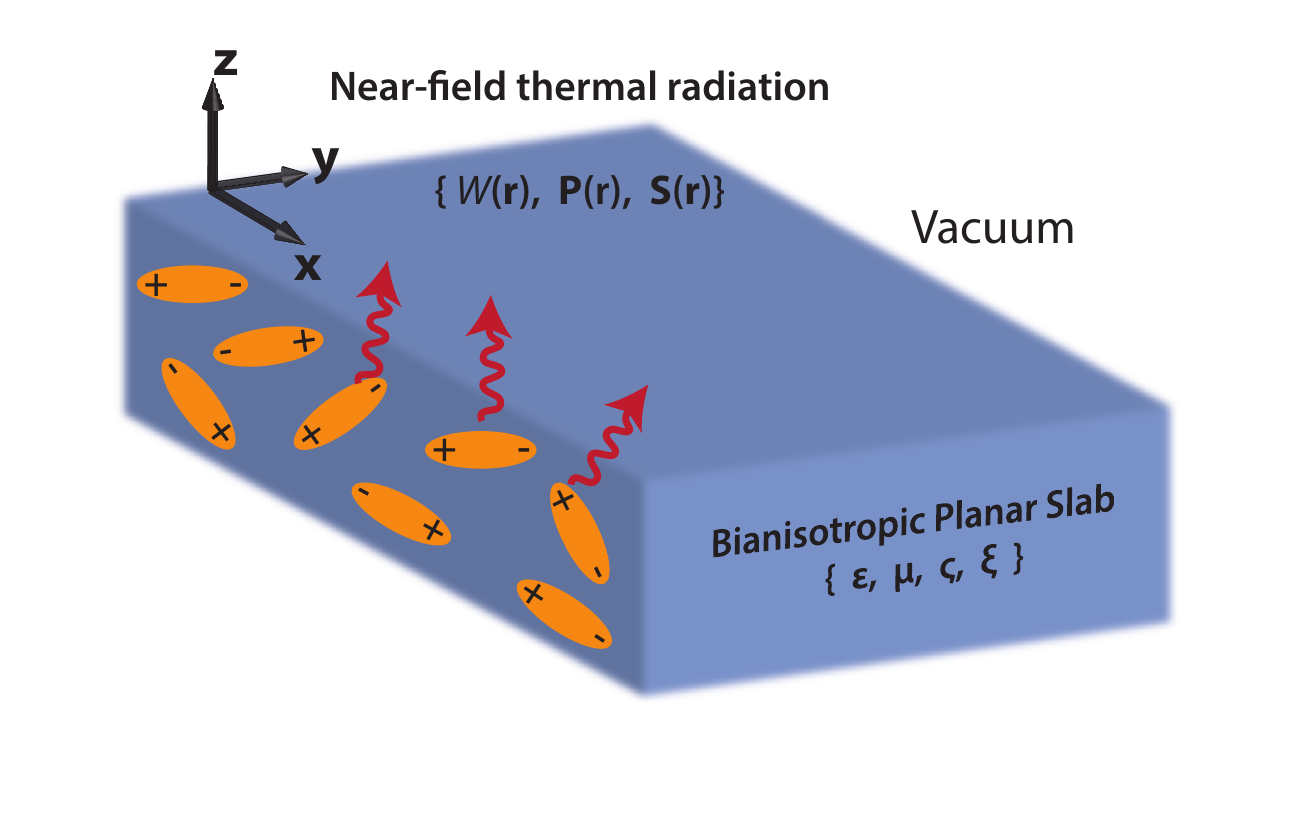}
  \caption{{\bf Geometry}. We analyze near-field thermal radiation
    properties namely energy density $W(\rv)$, Poynting flux
    $\mathbf{P}(\rv)$ and photon spin angular momentum density
    $\mathbf{S}(\rv)$ for a generic bianisotropic planar slab
    characterized by permittivity $\epsb$, permeability $\mub$ and
    magneto-electric coupling tensors $\xib,\zetab$. The yellow ovals
    indicate underlying fluctuating dipoles that emit thermal
    radiation. We show that thermal photon spin density
    $\mathbf{S}(\rv)$ and heat flux $\mathbf{P}(\rv)$ can be nonzero
    for a nonreciprocal material despite thermal equilibrium between
    vacuum and planar slab.}
  \label{scm1}
\end{figure}

{\bf Theory}. We consider a planar geometry shown in figure~\ref{scm1}
comprising of a semi-infinite half-space of generic homogeneous
material, interfacing with semi-infinite vacuum half-space at
$z=0$. We focus on the thermal radiation on the vacuum side of this
geometry where the physical quantities such as energy density $W$,
Poynting flux $\mathbf{P}$ and spin angular momentum density
$\mathbf{S}$ are
well-defined~\cite{joulain2003definition,barnett2016optical} and
measurable in suitable
experiments~\cite{kalhor2016universal,nieto2004near}:
\begin{align}
  \label{energydensity}
W(\rv) &= \frac{1}{2}\epsilon_0\langle\Ev^*(\rv)\cdot\Ev(\rv)\rangle +
\frac{1}{2}\mu_0\langle \Hv^*(\rv)\cdot\Hv(\rv) \rangle
\\ \label{poynting} \mathbf{P}(\rv) &= \Re\{\Ev^*(\rv)\times\Hv(\rv)\}
\\ \label{spindensity}\mathbf{S}(\rv) &=
\underbrace{\frac{\epsilon_0}{2\omega}\Im\langle \Ev^*(\rv)\times\Ev(\rv)
  \rangle}_{\mathbf{S}^{(\Ev)}} + \underbrace{\frac{\mu_0}{2\omega}\Im\langle
  \Hv^*(\rv)\times\Hv(\rv)\rangle}_{\mathbf{S}^{(\Hv)}}
\end{align}
where $\rv$ denotes the position vector and $\langle ... \rangle$
denotes the thermodynamic ensemble average. The spin angular momentum
density~\eqref{spindensity} has so far been studied primarily for
non-thermal light~\cite{joulain2003definition, barnett2016optical},
where it leads to proportionate optical torque on small, absorptive
particles~\cite{canaguier2013force,nieto2010optical}. We have
generalized it here and in our recent
work~\cite{khandekar2019circular} to thermally generated
electromagnetic fields in vacuum. We calculate both electric and
magnetic type thermal spin angular momentum density given by
$\mathbf{S}^{(\Ev)}$ and $\mathbf{S}^{(\Hv)}$ respectively. Throughout
the manuscript, all quantities are described in SI units and the
dependence on frequency $\omega$ (such as $\Ev(\omega),\Hv(\omega)$)
is suppressed assuming $e^{-i\omega t}$ time dependence in Maxwell's
equations. The above quantities $Q=\{W,\mathbf{P},\mathbf{S}\}$ are to
be integrated over frequency to obtain the total densities/flux rates
as $\hat{Q}=\int_{-\infty}^{\infty} \frac{d\omega}{2\pi}
Q(\omega)d\omega$. Keeping in mind the future explorations using
generic, bianisotropic materials, we prefer to use vector potential in
Landau gauge to obtain the electromagnetic fields ($\Ev=i\omega\Av$,
$\Hv=\nabla\times\Av$). The electromagnetic field correlations
required for calculation of densities and flux rates above are
obtained from the vector potential correlations. These correlations
evaluated at two spatial points $\rv_1,\rv_2$ are expressed in the
matrix form as $\langle \Av(\rv_1) \otimes \Av^*(\rv_2) \rangle =
\langle \Av(\rv_1) \Av(\rv_2)^{*^T}\rangle$. Here, the vector
quantities are written as column vectors such that $\Av=[A_x, A_y,
  A_z]^T$ where $[..]^T$ denotes the transpose and $[..]^*$ is complex
conjugation. We focus on the thermal equilibrium properties of
radiation where both vacuum and material half-spaces are at the same
thermodynamic temperature $T$. The vector potential correlations are
then obtained by making analogies with Kubo's formalism which
describes equilibrium correlations of fluctuating thermodynamic
quantities. The correlations are (see supplementary material for
derivation):
\begin{align}
\langle \Av(\rv_1)\otimes \Av^*(\rv_2) \rangle =
\frac{\Gb(\rv_1,\rv_2)-\Gb(\rv_2,\rv_1)^{*^T}}{2i}\frac{\mu_0}{\omega}\Theta(\omega,T)
\label{Afdt}
\end{align}
Here
$\Theta(\omega,T)=\hbar\omega/2+\hbar\omega/[\text{exp}(\hbar\omega/k_BT)-1]$
is the average thermal energy of the harmonic oscillator of frequency
$\omega$ at temperature $T$. The Green's tensor $\Gb(\rv_1,\rv_2)$
relates the vector potential $\Av(\rv_1)$ to all the source currents
$\Jv(\rv_2)$ such that
$\Av(\rv_1)=\int_{V_{\rv_2}}\Gb(\rv_1,\rv_2)\mu_0\Jv(\rv_2)
d^3\rv_2$. We derive the Green's function given in the methods section
for a planar slab of a generic, bianisotropic medium (see supplement
for derivation).

Finally, using the Green's function and vector potential correlations
above, we obtain the electromagnetic field correlations at two spatial
points $\rv_1$ and $\rv_2$ in vacuum:
\begin{align}
  \label{Ecor}
\langle \Ev(\rv_1)\otimes\Ev^*(\rv_2)\rangle&=\omega^2 \langle
\Av(\rv_1)\Av(\rv_2)^{*^T} \rangle \\ \label{EHcor}
\langle\Ev(\rv_1)\otimes\Hv^*(\rv_2)\rangle&=\frac{i\omega}{\mu_0}\langle
\Av(\rv_1) [\nabla_{r_2}\times\Av(\rv_2)]^{*^T} \rangle
\\ \label{Hcor} \langle
\Hv(\rv_1)\otimes\Hv^*(\rv_2)\rangle&=\frac{1}{\mu_0^2} \langle
\nabla_{r_1}\times\Av(\rv_1) [\nabla_{r_2}\times\Av(\rv_2)]^{*T}
\rangle
\end{align}
where $\nabla_{r_j}\times$ for $j=[1,2]$ is the differential curl
operator. The densities and flux rates are then calculated using
definitions \eqref{energydensity},\eqref{poynting} and
\eqref{spindensity} using above correlations with $\rv_1=\rv_2=\rv$
and making use of the fluctuation-dissipation relation given by
Eq.\eqref{Afdt}.

{\bf Possibility of observing nonzero spin angular momentum and heat
  flux despite thermal equilibrium}. We now show that a
\emph{nonreciprocal} medium can lead to nonzero spin angular momentum
density \eqref{spindensity} and nonzero Poynting flux \eqref{poynting}
in its thermal near-field. We make use of insightful expressions
derived here and time-reversal symmetry arguments, but we do not refer
to any specific material in this section.
  
The electromagnetic waves in this planar geometry are characterized by
their in-plane conserved propagation wavevector $\kv$. Poynting flux
and spin angular momentum of each wave are $\mathbf{P}(\kv)$ and
$\mathbf{S}(\kv)$ respectively. It follows from time-reversal symmetry
that heat and angular momentum associated with thermally excited
$\kv$-waves are negated by heat and angular momentum carried by
$-\kv$-waves ($\mathbf{P}(\kv)=-\mathbf{P}(-\kv)$,
$\mathbf{S}(\kv)=-\mathbf{S}(-\kv)$), resulting into zero flux rates at
thermal equilibrium. Because of violation of the time-reversal
symmetry for nonreciprocal media, this is no longer true and one can
expect to see nonzero flux rates in the absence of cancellation. We
identify the resulting nonzero spin angular momentum density as the
persistent thermal photon spin (PTPS) and nonzero heat flux as the
persistent planar heat current (PPHC). Although this analysis proves
that nonreciprocity is necessary to observe PTPS and PPHC, full
fluctuational electrodynamic calculations below, confirm and reveal
much more, including the result that nonreciprocity is not a
sufficient condition.

Before demonstrating the full calculations, we can make some general
comments regarding heat flux and thermal photon spin perpendicular to
the slab for which semi-analytic expressions are insightful (see
supplement for their derivation). The near-field Poynting flux in
$\ev_z$ direction,
\begin{align}
P_z &=\Theta(\omega,T)\int\int
\frac{k_{\parallel}dk_{\parallel}d\phi}{16
  \pi^2}\Re\{\big[2i\Im\{e^{-2ik_zd}(r_{ss}-r_{pp})\}\big]\} \nonumber
\\&= 0
\label{Pz}
\end{align}
for any material at thermal equilibrium with vacuum
half-space. Similarly, the electric and magnetic parts of spin angular
momentum density along $\ev_z$ direction are:
\begin{align}
S_{z}^{(\Ev)} &=\frac{\Theta(\omega,T)}{c^2}
\int\int\frac{k_{\parallel}dk_{\parallel}d\phi}{16 \pi^2
  k_0}\Im\{-(r_{ps}+r_{sp})e^{2ik_z d}\} \nonumber \\ &= -S_{z}^{(\Hv)}
\label{Sz}
\end{align}
It follows that the total perpendicular thermal spin
$S_z=S_z^{(\Ev)}+S_{z}^{(\Hv)}$, is zero for any material at thermal
equilibrium. Here $r_{ss},r_{pp},r_{sp},r_{pp}$ are the Fresnel
reflection coefficients for light incident on the planar slab having
perpendicular wavevector $k_z$, (conserved) parallel wavevector
$k_\parallel$, and making an azimuthal angle $\phi$ with x-axis of the
geometry. $k_0=\omega/c=\sqrt{k_z^2+k_\parallel^2}$ is the vacuum
wavevector (see methods and supplement). For reciprocal media,
$(r_{sp}+r_{ps})=0$~\cite{li2000symmetries} and the electric- and
magnetic-type persistent thermal photon spin (PTPS) are separately
zero. On the other hand, this condition is not necessarily true for
nonreciprocal media and interestingly, even though electric- and
magnetic-type PTPS can be separately nonzero, total PTPS perpendicular
to slab is always zero. We note that the semi-analytic expressions for
PTPS and PPHC parallel to the slab and given in the supplement are not
conducive for such general insights but full calculations of examples
below reveal their existence and nature.

\begin{figure*}[t!]
  \centering\includegraphics[width=\linewidth]{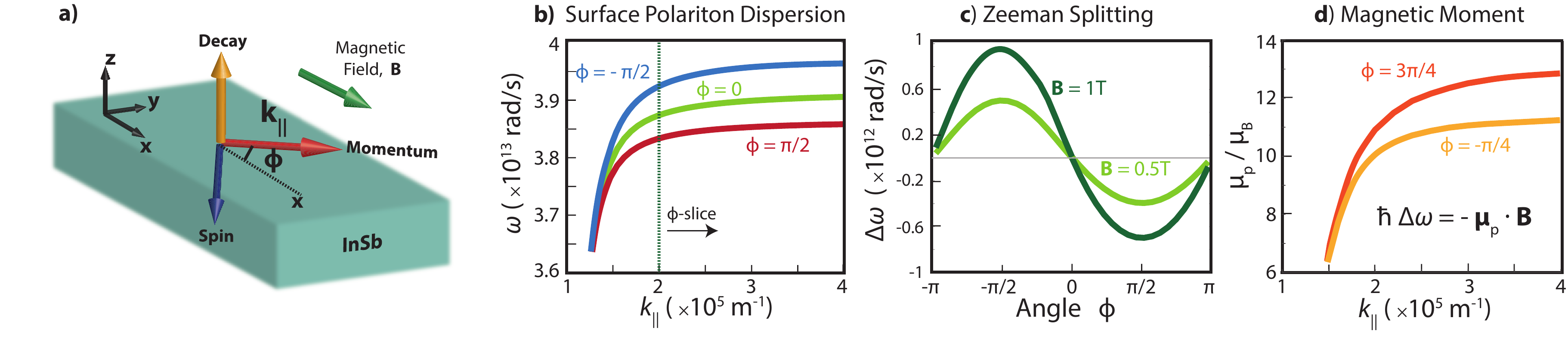}
  \caption{{\bf Polaritonic spin magnetic moment}. (a) We consider a
    practical example of a planar slab of doped InSb in presence of
    magnetic field (gyrotropy axis) parallel to its surface. As
    depicted, it supports surface plasmon polaritons characterized by
    conserved in-plane momentum $\kv_\parallel$ and carrying
    transverse spin locked to their momenta. A polariton of momentum
    $k_\parallel=|\kv_\parallel|$ makes an angle $\phi$ with the
    applied magnetic field. (b) The dispersion $\omega(k_\parallel)$
    for different angles $\phi$ is shown. Dispersion for $\phi=0$
    (green curve) is same for all angles in absence of $\Bv$ or with
    $\Bv$ (gyrotropy axis) perpendicular to surface. (c) The
    dependence of the frequency shift $\Delta\omega(\kv_\parallel)$ on
    the angle $\phi$ shows redshift for $\phi\geq 0$ (waves with spin
    component parallel to $\Bv$) and blueshift for $\phi < 0$ (waves
    with spin component anti-parallel to $\Bv$). For weak magnetic
    fields ($< 1$T), the frequency shift follows from the Zeeman
    interaction of the form
    $\hbar\Delta\omega=-\bm{\mu}_p(\kv_\parallel)\cdot\Bv$ where
    $\bm{\mu}_p$ is the polaritonic spin magnetic moment. (d)
    demonstrates the momentum dependence of $\mu_p=|\bm{\mu}_p|$ in
    units of Bohr magneton $\mu_B$. The asymmetry
    $\bm{\mu}_{p}(\kv_\parallel)\neq \bm{\mu}_{p}(-\kv_{\parallel})$
    for forward and backward waves demonstrated here with
    $\phi=-\frac{\pi}{4},-\frac{\pi}{4}+\pi$, leads to asymmetric
    polaritonic frequency shifts. This causes asymmetric contributions
    of spin angular momentum and Poynting flux carried by forward and
    backward waves resulting into PTPS and PPHC respectively.}
  \label{fig1}
\end{figure*}

It is important to point out here that the intuition confounding
presence of non-zero heat current at thermal equilibrium does not lead
to thermodynamic contradictions. In particular, because of the nonzero
heat flow parallel to the surface, it could be expected that one end
will be hotter than the other end. However, given the infinite
transverse extent of the system considered above, there is no end that
can be heated or cooled~\cite{ishimaru1962uni}. On the other hand,
since the two distinct half-spaces are separated by a well-defined
interface, no macroscopic flux rates can exist across the boundary by
definition of thermal equilibrium between the half-spaces. The
fluctuational electrodynamic theory produces a consistent result above
that the Poynting flux ($P_z$) and the total spin angular momentum
density ($S_z$) perpendicular to the surface are identically zero for
any material. For finite-size nonreciprocal systems such as finite
planar slabs or other geometries (cylinders, cubes) having
well-defined edges, it follows from the energy conservation under
global thermal equilibrium that, the energy exchange of any finite
subvolume $V$ of the system with the rest of the system is zero
i.e. $\int_{\partial V} \mathbf{P}.d\mathbf{A} = 0$ where $\partial V$
denotes the surface and $d\mathbf{A}$ is the differential area
vector. It then follows from the divergence theorem that the Poynting
flux $\mathbf{P}$ is divergence-free everywhere ($\nabla \cdot
\mathbf{P} = 0$). This means that there are no sources or sinks for
Poynting vector lines and they form closed loops. Therefore, the
persistent current in the near-field of a finite-size nonreciprocal
system will flow around the edges and form a closed loop, conserving
energy globally. While rigorous demonstration of fluctuational
electrodynamic confirmation in arbitrary geometries is challenging,
this is evident for finite spherical systems analyzed in
ref.~\cite{zhu2016persistent,ott2018circular}. We also note that our
analysis based on the assumption of infinite transverse extent is
suitable for describing a real (finite) nonreciprocal planar slab if
thermal equilibrium is achieved over transverse dimensions much larger
than the wavelengths associated with persistent features. In that
case, the \emph{local} persistent features can be analyzed within the
present theory by ignoring the edge effects. Since this situation is
quite realistic as we also describe in our experimental proposal
further below, our fluctuational electrodynamic analysis is
adequate. Also, we remark that the planar geometry of infinite
transverse extent is a reasonable theoretical approximation of planar
slabs. It is well-known and extensively used in the context of closely
related topics of Casimir force~\cite{capasso2007casimir} and
near-field radiative heat transport~\cite{song2015near} between planar
bodies. In the following, we consider a practical example system and
demonstrate these effects and further clarify their fundamental
connection with the spin-momentum locking of evanescent waves.


{\bf Practical Example of InSb slab}. We consider doped Indium
Antimonide (InSb) slab which has been most widely studied in context
of coupled magneto-plasmon surface
polaritons~\cite{kushwaha2001plasmons,hu2015surface} and whose
material permittivity dispersion has been well-characterized
experimentally~\cite{chochol2017experimental,
  hartstein1975investigation,palik1976coupled}. For
the sake of completeness of our study, we extend the known
permittivity model in
ref.~\cite{palik1976coupled,hartstein1975investigation}
to the case of an arbitrarily oriented magnetic field in our geometry
and obtain the semi-analytic form of permittivity given by
$\epsb=\varepsilon_{\infty}[1+(\omega_L^2-\omega_T^2)/(\omega_T^2-
  \omega^2-i\Gamma\omega)]\mathbb{I}_{3\times 3} +
\varepsilon_{\infty}\omega_p^2[\mathbb{L}_{3\times 3}(\omega)]^{-1}$
where:
\begin{align*}
\mathbb{L}_{3\times 3}(\omega) = \begin{bmatrix}
  -\omega^2-i\gamma\omega & -i\omega\omega_{cz} & i\omega\omega_{cy}
  \\ i\omega\omega_{cz} & -\omega^2-i\gamma\omega &
  -i\omega\omega_{cx} \\ -i\omega\omega_{cy} & i\omega\omega_{cx} &
  -\omega^2-i\gamma\omega \end{bmatrix}
\end{align*}
This is obtained from an extended Lorentz oscillator model in which
bound/free electrons (charge $q$, effective mass $m_f$, position
$\rv$) are described as mechanical oscillators, that further
experience additional Lorentz force ($q\dot{\rv}\times \Bv$) in
presence of applied magnetic field $\Bv$. Each $\omega_{cj}$ for
$j=[x,y,z]$ describes the cyclotron frequency in $\ev_j$ direction
given by $\omega_{cj}=q(\Bv\cdot\ev_j)/m_f$. In this model,
$\varepsilon_{\infty}$ is the high-frequency dielectric constant,
$\omega_L$ is the longitudinal optical phonon frequency, $\omega_T$ is
the transverse optical phonon frequency, $\omega_p=\sqrt{\frac{n
    q^2}{m_f \varepsilon_{\infty}\varepsilon_0}}$ is the plasma
frequency of free carriers of density $n$ and effective mass $m_f$
where $q$ is electron charge and $\varepsilon_0$ is vacuum
permittivity. $\Gamma$ denotes the optical phonon damping constant
while $\gamma$ is the free-carrier damping constant. All parameters
are obtained from Ref.~\cite{palik1976coupled} for
InSb sample of doping density
$n=10^{17}$cm$^{-3}$. $\varepsilon_{\infty}=15.7$,
$\omega_L=3.62\times 10^{13}$rad/s, $\omega_T=3.39\times
10^{13}$rad/s, $\omega_p=3.14\times 10^{13}$rad/s, $\Gamma=5.65\times
10^{11}$rad/s, $\gamma=3.39\times 10^{12}$rad/s, $m_f=0.022m_e$ where
$m_e=9.1094\times 10^{-31}$kg is electron mass. Because of the
anti-symmetric (gyroelectric-type) permittivity tensor ($\epsb =
-\epsb^T$) of InSb in presence of magnetic field, it is nonreciprocal
and can lead to persistent features in its thermal near-field. For the
above paramenters, InSb slab supports surface plasmon polaritons
(SPPs) at $\omega \sim 3.9\times 10^{13}$rad/s ($\sim 48\mu$m)
localized close to the interface with vacuum. Because of their
significant contribution to the near-field thermal radiation, we
analyze these polaritons and also clarify the connection of the
persistent features with the spin-momentum locking.

{\bf Spin magnetic moment of InSb surface polaritons}. As shown
schematically in fig.\ref{fig1}(a), we consider a magnetic field
applied along $\ev_x$ direction (parallel to the surface). The surface
plasmon polariton characterized by its conserved in-plane momentum
$\kv_{\parallel}$ makes an angle $\phi$ with $\ev_x$ (applied field
direction) such that $\phi\in [-\pi,\pi]$. Each such polariton also
carries a transverse spin locked to its momentum, depicted in the
schematic (spin momentum
locking~\cite{van2016universal}). Figure~\ref{fig1}(b) displays the
dispersion $\omega(\kv_{\parallel})$ of polaritons for different
angles $\phi$, obtained numerically as described in the methods
section. In the absence of magnetic field or with the magnetic field
perpendicular to the surface, the surface polaritons are p-polarized
and the dispersion is the same as the dispersion for $\phi=0$ (green
curve) for all angles. On the other hand, in presence of magnetic
field parallel to the surface, the surface polaritons contain both
s-polarized and p-polarized electromagnetic fields, requiring
numerical method for calculating the dispersion for arbitrary
propagation directions.  As shown in the figure~\ref{fig1}(b),
assuming magnetic field along $\ev_x$ direction, the polaritons
characterized by $\phi \geq 0$ (with positive spin component along
applied magnetic field) are redshifted while those characterized by
$\phi < 0$ (with negative spin component along applied magnetic field)
are blueshifted. This is further demonstrated in fig.~\ref{fig1}(c)
where $\omega(\kv_\parallel)$ is obtained as a function of angle
$\phi$ for a fixed $k_{\parallel}=|\kv_\parallel|$, for two different
values of magnetic field. We numerically find that the energy shift
for each $\kv_\parallel$ polariton increases \emph{linearly} with
magnetic field in the weak field regime ($\Bv \lesssim 1$T) while
the dependence is complicated for strong applied fields. All these
results strongly indicate that the polaritons have a spin magnetic
dipole moment ($\bm{\mu}_p$) parallel to the transverse spin that
interacts with the applied magnetic field. The energy of this Zeeman
interaction is described by the Hamiltonian,
\begin{align*}
  H_{\text{int}}=-\bm{\mu}_p \cdot\Bv
\end{align*}
It follows that the magnetic-field-induced frequency shift
($\Delta\omega$) for each polariton is:
\begin{align}
  \hbar\Delta\omega(\kv_\parallel)=-\bm{\mu}_p(\kv_\parallel)\cdot\Bv
\end{align}
We find that the magnetic moment $\mu_p=|\bm{\mu}_p|$ for each
$\kv_\parallel$ polariton depends not only on momentum
$k_\parallel=|\kv_\parallel|$ but also on the angle $\phi$. This is
evident upon closer inspection of fig~\ref{fig1}(c) where the energy
shift as a function of angle is not sinusoidal but exhibits slight
deviation (see maximum and minimum values). In fig.~\ref{fig1}(d), the
magnetic moment in units of $\mu_B$ (Bohr magneton) as a function of
$k_{\parallel}$ is displayed for two different angles
$\phi=\frac{-\pi}{4},\frac{3\pi}{4}$ showing that $\mu_p(\phi)\neq
\mu_p(\phi+\pi)$ or $\mu_p(\kv_\parallel)\neq
\mu_p(-\kv_\parallel)$. This asymmetry in polaritonic spin magnetic
moment and magnetic-field induced energy shift lies at the origin of
the asymmetry in the spin angular momentum and heat flux carried by
thermally excited $\pm\kv_\parallel$ polaritons, resulting into PTPS
and PPHC parallel to the surface.
\begin{figure*}[t!]
  \centering\includegraphics[width=\linewidth]{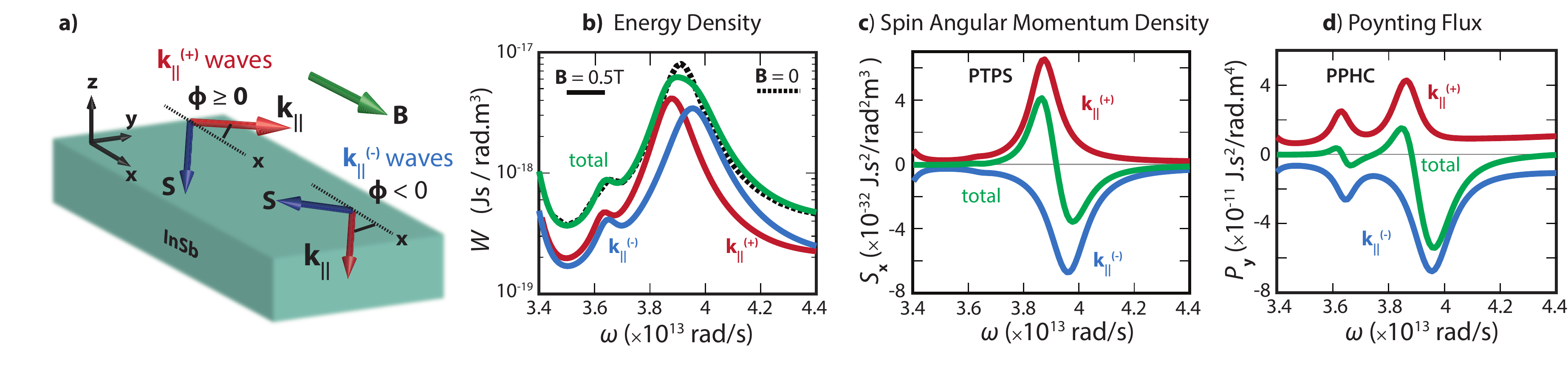}
  \caption{{\bf Fundamental connection of the persistent phenomena
      with spin-momentum locking}. We analyze near-field thermal
    radiation of InSb slab at thermal equilibrium with vacuum in
    presence of magnetic field $\Bv=0.5$T$\ev_x$. (a) Based on the
    analysis of spin-momentum locked polaritons, we divide the
    contribution of thermally excited waves into following two types:
    (1) $\kv_\parallel^{(+)}$ waves having $\phi \geq 0$ (spin
    component parallel to applied magnetic field) and (2)
    $\kv_\parallel^{(-)}$ waves having $\phi<0$ (spin component
    anti-parallel to applied magnetic field). (b,c,d) demonstrate the
    spectra of energy density, spin angular momentum density and
    Poynting flux, evaluated at a distance $d=1\mu$m from the surface
    assuming both vacuum and InSb to be at thermal equilibrium
    temperature of $T=300$K. The asymmetric overall contributions of
    $\kv_\parallel^{(+)}$ waves (red) and $\kv_\parallel^{(-)}$ waves
    (blue) evident from these figures result into nonzero spin angular
    momentum density and heat current despite thermal equilibrium
    i.e. PTPS and PPHC respectively, shown by green curves in (c,d).}
  \label{fig2}
\end{figure*}

We note that the Poynting vector and the spin of polaritonic waves
deviate from their usual directions ($\mathbf{P} \parallel
\kv_{\parallel}$, $S \perp \kv_{\parallel}$) for large magnetic
fields. However, these deviations are small for weak magnetic fields
($\Bv \lesssim 1$T) considered above and therefore, this analysis
suffices to qualitatively predict the existence of PTPS and PPHC. In
the following, we demonstrate them using full fluctuational
electrodynamic calculations.

{\bf PTPS and PPHC in thermal near-field of InSb slab}. We compute
the spin angular momentum density (Eq.\ref{spindensity}) and Poynting
flux (Eq.\ref{poynting}) in thermal near-field of InSb slab in
presence of magnetic field of strength $0.5$T along $\ev_x$
direction. Both vacuum and material are assumed to be at thermodynamic
equilibrium temperature of $T=300$K. Based on the discussion of
polaritons in the previous section, we calculate spin-resolved
quantities in the sense described schematically in
fig.\ref{fig2}(a). In particular, the contributions of electromagnetic
waves characterized by $\phi \geq 0$ ($\kv_\parallel^{(+)}$-waves) and
those characterized by $\phi < 0$ ($\kv_{\parallel}^{(-)}$-waves) are
calculated separately.

Figure \ref{fig2}(b,c,d) demonstrate the frequency spectra of energy
density $W(\omega)$, spin angular momentum density
$\mathbf{S}(\omega)$ and poynting flux $\mathbf{P}(\omega)$ at a
distance of $d=1\mu$m above the surface of InSb. All these figures
depict the separate contributions of $\kv_\parallel^{(+)}$ waves (red
curves) and $\kv_{\parallel}^{(-)}$ waves (blue curves) along with the
sum total (green curves). As evident from fig.\ref{fig2}(b), the
collective energy density of $\kv_{\parallel}^+$ waves is redshifted
and that of $\kv_{\parallel}^-$ waves is blueshifted similar to
polaritons, leading to broadening of total energy density spectrum
(green), compared to the spectrum in the absence of magnetic field
(black dashed line). The asymmetric overall contributions of
$\kv_\parallel^{(+)}$ and $\kv_{\parallel}^{(-)}$ waves result into
nonzero spin angular momentum density and nonzero Poynting flux at
thermal equilibrium i.e. PTPS and PPHC. Note that these persistent
quantities contain contributions from not only surface localized
polaritons but also other evanescent waves. For instance, another
small peak apparent in \ref{fig2}(b) and clearly visible in
\ref{fig2}(d) is not related to the polaritons studied in previous
section but instead arises from other nonreciprcoal surface waves
which make small contribution in comparison to surface plasmon
polaritons.

\begin{figure*}[t!]
  \centering\includegraphics[width=\linewidth]{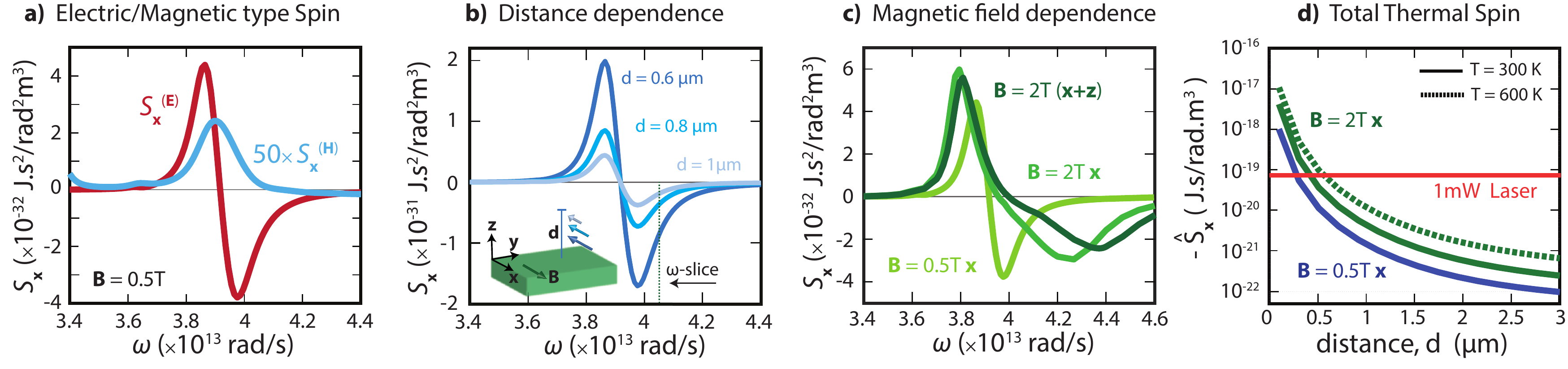}
  \caption{{\bf Persistent thermal photon spin}. We analyze the
    persistent thermal photon spin (PTPS) in detail at a distance $d$
    from InSb in presence of magnetic field $\Bv$ when both vacuum and
    InSb half-spaces are at temperature $T$. (a) Electric and magnetic
    contributions to PTPS indicate that the electric type spin
    dominates (b) The PTPS decays away from the surface at all
    frequencies while its direction remains the same. (c) Increasing
    magnetic field leads to PTPS over a broader range of
    frequencies. (d) Total frequency-integrated PTPS for various $\Bv$
    fields and temperatures $T$ is shown. An important comparison is
    made with total spin angular momentum density contained in a $1$mW
    laser light at the polaritonic frequency, focused to $1$mm$^2$
    spot size (orange). It shows that PTPS which is significantly
    enhanced due to large density of thermally excited states in the
    near-field, can surpass the spin angular momentum density of laser
    light at separations $\lesssim 0.5\mu$m.}
  \label{fig3}
\end{figure*}

Figure~\ref{fig3} describes PTPS spectrum [\ref{fig3}(a,b,c)] and also
demonstrates that the total frequency-integrated PTPS [\ref{fig3}(d)]
can compete with the angular momentum density contained in the laser
light. For brevity, we focus only on PTPS. As shown in
fig~\ref{fig3}(a), the electric-type PTPS is evidently much larger
than the magnetic-type PTPS. We describe later [fig~\ref{fig0}] that
this holds universally for any gyro-electric type nonreciprocal
material. Fig~\ref{fig3}(b) demonstrates the change in the spectrum as
a function of distance from the surface. At each frequency, the sign
(direction) of PTPS stays the same while the magnitude decays
exponentially as a function of distance from the surface. This also
indicates (although not shown separately) that PTPS and PPHC arise
from the waves that are evanescent on the vacuum side of the
geometry. Fig~\ref{fig3}(c) depicts the dependence of the spin angular
momentum density on the applied magnetic field. First, the PTPS
spectrum broadens as magnetic field is increased from $B=0.5T\ev_x$
(light green) to $B=2T\ev_x$ (green). Second, while perpendicular
magnetic field by itself does not lead to PTPS, it does affect the
spectrum observed in presence of parallel magnetic field. This is
evident from green and dark green ($B=2T(\ev_x+\ev_z)$) curves. When
magnetic field is at oblique angles to the surface such as this
example (dark green), the cyclotron motion in yz-plane due to $\Bv_x$
is intercoupled with that in xy-plane due to $\Bv_z$. Due to this
intercoupled cyclotron motion of underlying charges, the perpendicular
component of magnetic field (which does not lead to PTPS by itself)
affects the PTPS spectrum obtained with magnetic field parallel to the
surface.

Finally, we plot the total frequency integrated PTPS
($-\hat{S}_{\mathbf{x}}$) in fig~\ref{fig3}(d) as a function of
distance ($d$) for increasing values of magnetic field. The total PTPS
lies along $-\ev_x$ direction (anti-parallel to applied magnetic
field). In order to get a qualitative estimate of the overall strength
of PTPS, fig.\ref{fig3}(d) also displays the total spin angular
momentum density contained in monochromatic (polariton frequency),
circularly polarized laser light of power $1$mW focused to $1$mm$^2$
spot size. Evidently, PTPS which originates from the intrinsic
fluctuations in the medium can compete with the total spin angular
momentum density contained in the laser light. It can even surpass it
at separations $d \lesssim 0.5\mu$m. from the surface. This large
enhancement of PTPS arises from a large density of thermally excited
evanescent and surface waves in the near-field, otherwise inaccessible
in the far-field. The same figure also depicts the dependence on the
temperature with solid lines ($T=300$K) and dashed lines ($T=600$K)
which comes from the mean thermal energy given by
$\Theta(\omega,T)=\hbar\omega/2+\hbar\omega/[\text{exp}(\hbar\omega/k_BT)-1]$
. Since the mean thermal energy is approximately constant over the
frequency range of interest, PTPS increases/decreases proportionately
with the temperature.

We note that the total frequency-integrated Poynting flux
$\hat{P}_{\mathbf{y}}$ for this practical example (not shown) is along
$-\ev_{y}$ direction. The direction of integrated PPHC is related to
the underlying cyclotron motion of electrons induced by magnetic
field~\cite{silveirinha2017topological}. Inside the bulk of InSb, the
cyclotron motions of electrons cancel each other but at the surface,
this cancellation is incomplete. The direction of this incomplete
cyclotron motion co-incides with the direction of the PPHC.

We further note that the practical example considered here can also
lead to unidirectional energy transport because of
nonreciprocity~\cite{wang2009observation,rechtsman2013photonic,
  hu2015surface}. However, unidirectional transport is not a cause of
PTPS and PPHC and this is explained in the
following. Figure~\ref{fig1}(b) indicates that the unidirectional
transport due to polaritons can occur along directions for which
polaritonic momentum has a dominant $-\ev_y$ component (maximally
blue-shifted polaritons). Although PPHC spectrum in fig.\ref{fig2}(d)
has a predominant contribution along the same direction, both PPHC and
PTPS spectra shown in fig.~\ref{fig2}(c,d) show that smaller
frequencies with bidirectional polaritons also lead to nonzero
persistent quantities. While recent works~\cite{buddhiraju2018absence}
have started to explore the role of nonlocality in context of
nonreciprocity and unidirectional transport, we leave nonlocality
aside for future work.

\begin{figure}[t!]
  \centering\includegraphics[width=0.8\linewidth]{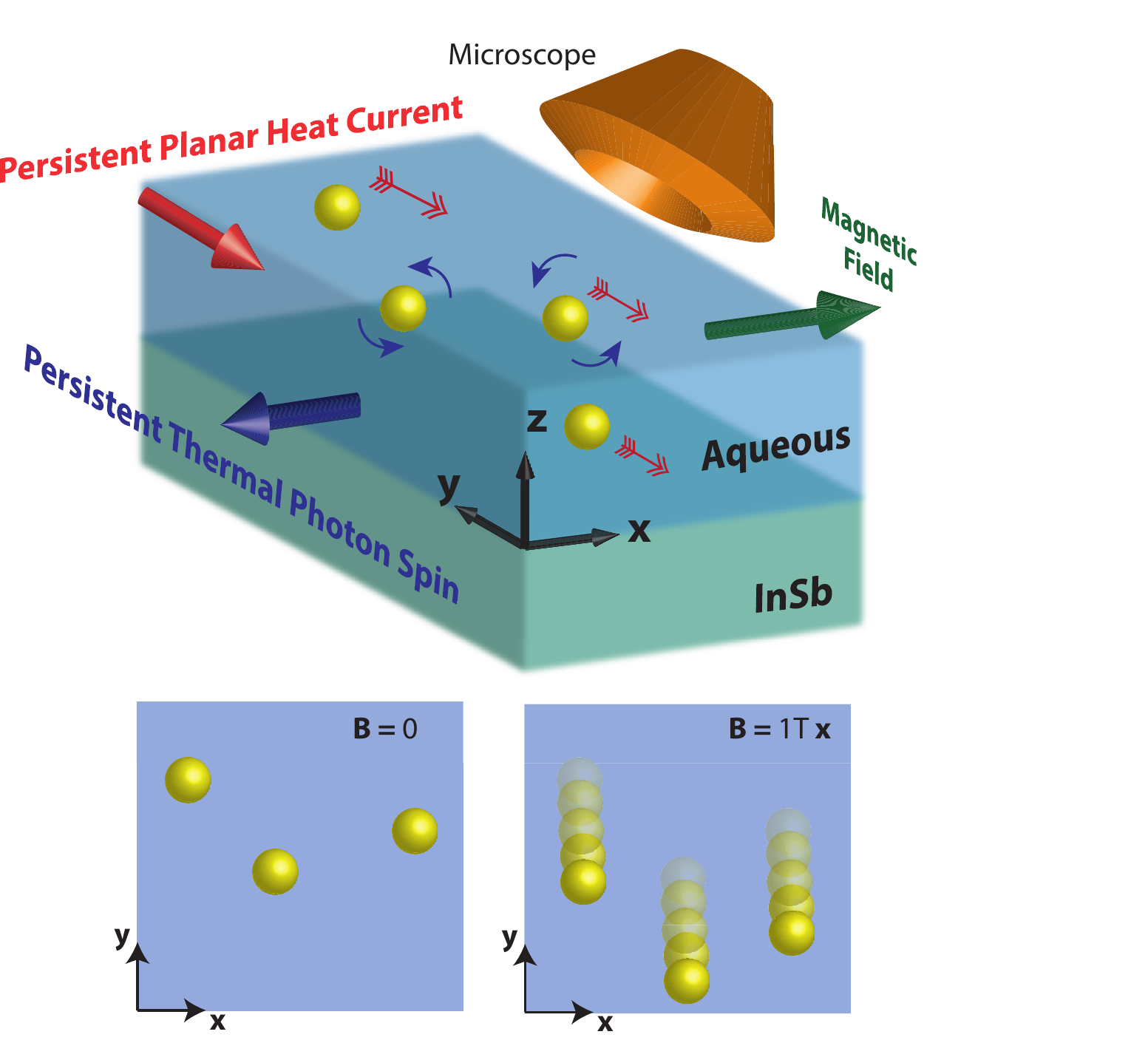}
  \caption{{\bf Experimental proposal}. A thick planar slab of Indium
    Antimonide (InSb) is covered with an aqueous medium containing
    suspended micrometer size nonmagnetic particles. The entire system
    is at room temperature and the particles perform Brownian movement
    about their mean positions at thermal equilibrium. When magnetic
    field is turned on, the particles experience additional
    translational and rotational diffusion due to PTPS and PPHC. The
    bottom two figures depict the expected view through the
    microscope, where magnetic field along $\ev_x$ direction leads to
    an observable overall vertical shift of the Brownian particles.}
  \label{scm2}
\end{figure}
\begin{figure*}[t!]
  \centering\includegraphics[width=\linewidth]{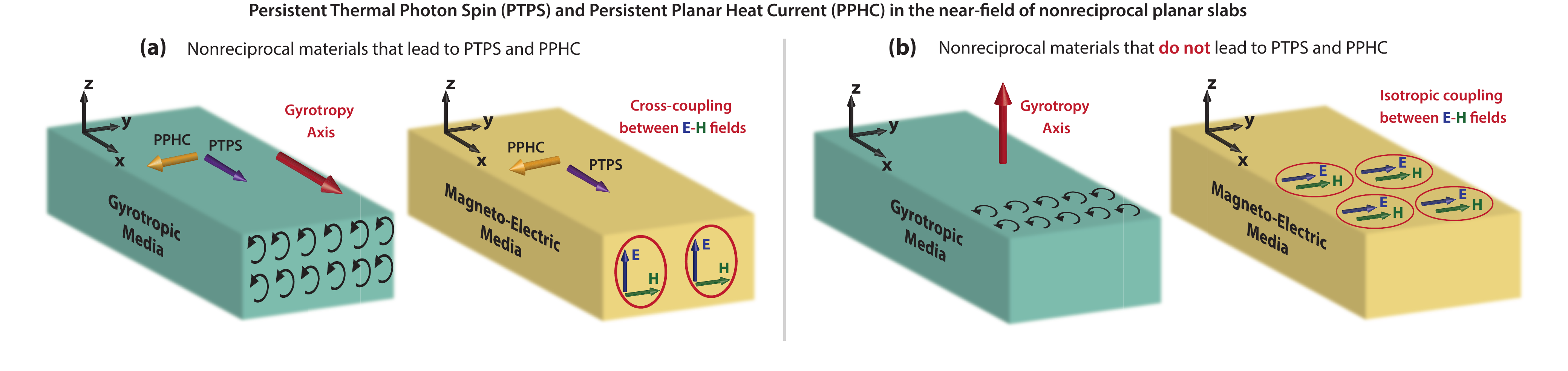}
  \caption{{\bf Universal behavior of the persistent phenomena}. (a)
    PTPS and PPHC are observed parallel to the surface for planar
    slabs of a gyroelectric material with gyrotropy axis parallel to
    the surface as well as a magneto-electric material with coupling
    between perpendicular components of $\Ev,\Hv$ fields. For
    gyroelectric material, PTPS is along the direction of gyrotropy
    axis (parallel or anti-parallel) and PPHC is perpendicular to
    PTPS. The electric/magnetic type contribution to PTPS dominates
    for gyro-electric/magnetic type media while both contributions are
    comparable for magneto-electric nonreciprocal media.(b) The
    persistent phenomena are not observed despite nonreciprocity, with
    planar slab of a gyroelectric material with gyrotropy axis
    perpendicular to surface or a magneto-electric material with
    isotropic coupling between $\Ev,\Hv$ fields.}
  \label{fig0}
\end{figure*}

{\bf Experimental Proposal}. Here we propose an experiment that can
provide a direct visual evidence of PPHC using the planar slab
considered above. It is well-known that light carrying momentum and
spin/orbital angular momentum can exert optical forces and torques on
small absorptive particles in its path. Many works have explored this
light-matter interaction in optical
tweezers~\cite{ukita2010optical,angelsky2012circular,
  canaguier2013force} by non-thermal means and in entirely passive
systems where forces and torques originate from intrinsic quantum and
thermal fluctuations~\cite{chan2001quantum,
  haslinger2018attractive,bao2018inhomogeneity}. We are interested in
the latter for non-intrusive (without disturbing thermal equilibrium)
detection of the persistent phenomena. We therefore consider a passive
system shown schematically in figure~\ref{scm2} where aqueous or
fluidic environment covers the gyrotropic InSb thick slab and contains
suspended small micrometer size absorptive and nonmagnetic
particles. The particles perform Brownian motion about their positions
at finite equilibrium temperature. Upon application of magnetic field,
the particles experience additional optical force and torque
associated with PPHC and PTPS. While the average motional energies of
particles remain constant at thermal equilibrium ($\sim \frac{1}{2}k_B
T$ by equipartition law), the additional forces and torques lead to
preferential changes in their mean positions and angular orientations
which can be detected using a
microscope~\cite{kirksey1988brownian,kawata1992movement}. In the
following, we estimate these changes with simplifying
approximations. A rigorous description of Brownian
dynamics~\cite{volpe2011microswimmers,mijalkov2013sorting} is beyond
the scope of this work and at this point unnecessary since the goal is
to merely detect the presence of the persistent thermal photonic
phenomena.

The particles are absorptive, nonmagnetic (not influenced by presence
or absence of magnetic field) and much smaller in size ($\lesssim
\mu$m) and hence dipolar in nature at surface-polariton wavelengths
($\gtrsim 48\mu$m) of InSb. The entire system is at thermal
equilibrium room temperature. Magnetic field of $1$T is applied along
$\ev_x$ direction resulting into persistent planar heat current
$\mathbf{P}=-P_y \ev_y$ and persistent thermal photon spin
$\mathbf{S}=-S_x \ev_x$. Our analysis above of planar geometry with
infinite transverse extent is valid when both magnetic field and
temperature are uniform over an area much larger than polaritonic
wavelengths which are of the order of $100\mu$m. Assuming that these
conditions are realized over an area of $1$cm$^2$ of InSb slab in an
actual experiment which is quite realistic, we extend our
fluctuational electrodynamic analysis to calculate the additional
forces acting on particles, originating from the \emph{local}
persistent features (not influenced by the edge effects).

The average stochastic optical force on a Brownian particle is written
as a sum of the following two
terms~\cite{henkel2002radiation,gordon1980motion}:
\begin{align*}
\mathbf{F}=\sum_{i={x,y,z}} \langle p_i^{(fl)} \nabla E_i^{(ind)}
\rangle + \langle p_i^{(ind)} \nabla E_i^{(fl)}\rangle
\end{align*}
Here $p$ is the dipole moment of the particle and $E$ is the total
electric field at the position of the particle. The first term
corresponds to the interaction of fluctuating dipole moment of the
particle with thermal fields induced by the particle itself. By
calculating induced thermal field using Green's function, it can be
shown that this does not lead to any lateral force on the particle
because of translational invariance~\cite{manjavacas2017lateral}. Note
that we are primarily interested in the lateral forces since the
perpendicular forces exist in the near-field of all
materials~\cite{henkel2002radiation} and cannot be used to detect
PPHC. The second term denotes the interaction of the thermal fields
with the induced dipole moment of the particle given by
$\mathbf{p}(\omega)=\epsilon_0\alpha(\omega)\Ev(\omega)$ where
$\alpha(\omega)=4\pi R^3
\frac{\epsilon(\omega)-\epsilon_b}{\epsilon(\omega)+\epsilon_b}$ is
the polarizability of the spherical nanoparticle of isotropic
permittivity $\epsb =\epsilon(\omega)$ and radius $R$, immersed in
water of permittivity $\epsilon_b \approx 1.77$. This interaction
leads to the following simplified expression~\cite{canaguier2013force}
for the lateral force which occurs only along $\ev_y$ direction for
magnetic field applied along $\ev_x$ direction:
\begin{align}
\mathbf{F}(z) = \int \frac{d\omega}{2\pi} \frac{\eps_b\omega}{2c_0^2}
\Im\{\alpha(\omega)\}[-P_y(\omega)- c^2\frac{\partial
    S_x^{(\Ev)}}{\partial z}(z,\omega)]\ev_y
\end{align}
Here, $z$ denotes the distance of the particle from InSb slab surface
and $c$ is speed of light. There are many readily available particles
such as chalk or milk particles and other commercially available
nanoparticles which can be used for the Brownian experiment. For
estimation purpose, we consider doped Silicon particles of diameter
$1\mu$m, mass density $\rho \sim 2329$kg$/$m$^3$ and Drude
permittivity dispersion
$\varepsilon(\omega)=11.7-\omega_p^2/(\omega^2+i\gamma\omega)$ where
$\omega_p=1.3\times 10^{14}$rad/s and $\gamma=\omega_p/100$. For these
parameters, at a distance of $z=2\mu$m from the slab surface, the
particles experience linear acceleration of $~4\mu$m/s$^2$ along
$-\ev_y$ i.e. direction of PPHC. While the particles also experience
torque due to PTPS, the resulting rotational changes are difficult to
observe with the proposed experiment. Nonetheless, the above
calculations indicate that there should be a noticeable displacement
of Brownian particles along the direction of PPHC which can be viewed
using a microscope as depicted in fig.~\ref{scm2}. Since the
additional lateral movement at thermal equilibrium is not possible in
absence of magnetic field or with other homogeneous reciprocal media,
its mere presence will be a clear indicator of the persistent thermal
photonic phenomena. It can be readily perceived upon seeing through a
simple lab microscope as shown in the insets of fig.~\ref{scm2} or by
methodically tracking the particle movements.

{\bf Universal behavior of PTPS and PPHC}. We now describe the
universal behavior of PTPS and PPHC with generic biansiotropic
material types. A bianisotropic medium is often considered in the
literature~\cite{kriegler2010bianisotropic,asadchy2018bianisotropic}
to represent a superset of all types of media, more commonly described
with following constitutive relations assuming local material response
(in the frequency domain):
\begin{align}
  \mathbf{D} &= \epsb\varepsilon_0 \Ev + \xib\frac{1}{c}\Hv \nonumber \\
  \mathbf{B} &= \zetab\frac{1}{c}\Ev + \mub\mu_0\Hv
\label{consti}
\end{align}
$\epsb,\mub$ are dimensionless permittivity and permeability tensors
and $\xib,\zetab$ are magneto-electric coupling tensors. Based on the
existing literature, we categorize bianisotropic materials into
following five well-studied material types:
\begin{itemize} 
\item Isotropic materials: Most naturally existing dielectric or
  metallic materials with scalar $\epsb,\mub$ and $\xib,\zetab=0$.
\item Uniaxial/biaxial anisotropic materials such as birefringent
  crystals that have diagonal $\epsb$ and $\mub$ with unequal diagonal
  entries and $\xib,\zetab=0$.
\item Gyroelectric (Magneto-optic) materials such as semiconductors in
  external magnetic fields~\cite{ishimaru2017electromagnetic} for
  which $\epsb$ has nonzero off-diagonal components, $\mub$ is a
  scalar and $\xib,\zetab=0$.
\item Gyromagnetic materials such as ferromagnets and
  ferrites~\cite{rodrigue1988generation} for which $\mub$ has nonzero
  off-diagonal components, $\epsb$ is a scalar and $\xib,\zetab=0$.
\item Magneto-electric materials such as chromite Cr$_2$O$_3$,
  multiferroics CuCrO$_2$~\cite{pyatakov2012magnetoelectric,
    albaalbaky2017magnetoelectric} and topological insulators such as
  Bi$_2$Se$_3$~\cite{laforge2010optical} for which $\epsb,\mub$ are
  diagonal and $\xib,\zetab$ are nonzero tensors.
\end{itemize}
  
\begin{table*}
  \centering \setlength{\tabcolsep}{0.1cm}
             {\renewcommand{\arraystretch}{1.4}
\begin{tabular}{ | c | m{3cm} | m{9cm} | m{1.3cm} | m{1.3cm} | m{1.2cm} | }
\hline \multicolumn{6}{ |c| }{PTPS and PPHC in thermal near-field of
  biansiotropic materials} \\ \hline No. & Material Type & Example &
$\frac{\mathbf{S}^{(\Ev)}c^3}{\omega\Theta(\omega,T)}$ &
$\frac{\mathbf{S}^{(\Hv)}c^3}{\omega\Theta(\omega,T)}$ &
$\frac{\mathbf{P}c^2}{\omega^2\Theta(\omega,T)}$ \\ \hline \hline $1$
& Uniaxial Anisotropic & $\epsb=(2+0.1i)\mathcal{I}_{3\times
  3}-4\ev_y\ev_y^T$, $\mub=\mathcal{I}_{3\times 3}$, $\xib,\zetab=0$ &
$0$ & $0$ & $0$ \\\hline $2$ & Gyroelectric (Magneto-optic) &
$\epsb=(-1+0.1i)\mathcal{I}_{3\times 3} +
0.1i(\ev_{y}\ev_{z}^T-\ev_z\ev_y^T)$, $\mub=\mathcal{I}_{3\times 3}$,
$\xib,\zetab=0$, Gyrotropy axis along $\ev_x$ (parallel to surface) &
$-65\ev_x$ & $0.2\ev_x$ & $-7.4\ev_y$ \\ \hline $3$ & Gyromagnetic &
$\epsb=4\mathcal{I}_{3\times 3}$, $\mub=(-1+0.1i)\mathcal{I}_{3\times
  3}+ 0.1i(\ev_{y}\ev_{z}^T-\ev_z\ev_y^T)$, $\xib,\zetab=0$, Gyrotropy
axis along $\ev_x$ (parallel to surface) & $0.2\ev_x$ & $-120\ev_x$ &
$-10.1\ev_y$ \\ \hline $4$ & Gyroelectric (Magneto-optic) &
$\epsb=(4+0.1i)\mathcal{I}_{3\times 3} +
0.1i(\ev_{x}\ev_{y}^T-\ev_y\ev_x^T)$, $\mub=\mathcal{I}_{3\times 3}$,
$\xib,\zetab=0$, Gyrotropy axis along $\ev_z$ (perpendicular to
surface) & $0.004\ev_z$ & $-0.004\ev_z$ & $0$ \\ \hline $5$ &
Magneto-Electric (Nonreciprocal) & $\epsb=4\mathcal{I}_{3\times 3}$,
$\mub=\mathcal{I}_{3\times 3}$, $\xib=\zetab=0.1\mathcal{I}_{3\times
  3}$, Isotropic magneto-electric coupling & $-0.01\ev_z$ &
$0.01\ev_z$ & $0$ \\ \hline $6$ & Magneto-Electric (Reciprocal) &
$\epsb=4\mathcal{I}_{3\times 3}$, $\mub=\mathcal{I}_{3\times 3}$,
$\xib=-\zetab=0.1i\mathcal{I}_{3\times 3}$ & $0$ & $0$ & $0$ \\ \hline
$7$ & Magneto-Electric (Nonreciprocal) & $\epsb=4\mathcal{I}_{3\times
  3}$, $\mub=\mathcal{I}_{3\times 3}$,
$\xib=\zetab=0.1(\ev_y\ev_z^T+\ev_z\ev_y^T)$, Magneto-electric cross
coupling between $\Ev,\Hv$ fields & $0.02\ev_y-0.003\ev_z$ &
$-0.05\ev_y+0.003\ev_z$ & $0.008\ev_x$ \\ \hline
\end{tabular}}
\caption{Electric and magnetic type persistent thermal photon spin
  (PTPS) and persistent planar heat current (PPHC) in the near-field
  of different types of bianisotropic media (including all forms of
  nonreciprocity) is analyzed. Using the material parameters (example
  column) at a given frequency $\omega$, the dimensionless PTPS and
  PPHC (last three columns) are calculated at a distance
  $d=0.1\frac{c}{\omega}$ from the surface. The examples are
  representative of the material types (second column) such that PTPS
  and PPHC are present or absent for that material type based on
  nonzero or zero values respectively in the last three columns.}
\label{table1}
\end{table*}

Apart from the naturally existing examples given above, there exists a
huge range of artificially designed metamaterials with/without bias
fields/currents that can effectively provide any combination of these
material
types~\cite{kriegler2010bianisotropic,asadchy2018bianisotropic}. It is
well-known that such a bianisotropic material is
reciprocal~\cite{caloz2018electromagnetic} when,
\begin{align}
\epsb = \epsb^{T}, \hspace{20pt} \mub = \mub^{T}, \hspace{20pt} \xib =
- \zetab^{T}
\end{align}
The material is nonreciprocal if atleast one of these conditions is
violated.  

Table~\ref{table1} summarizes PTPS and PPHC for generic bianisotropic
material classes with suitable representative examples that describe
the presence or absence of PTPS and PPHC. Both isotropic and
uniaxial/biaxial anisotropic materials being reciprocal in nature, do
not lead to any persistent spin or heat current. The first example
considers uniaxial anisotropic material with its anisotropy axis
parallel to the surface (breaking the rotational symmetry) and the
full calculations confirm the absence of the persistent phenomena. The
second and third examples in the table correspond to gyroelectric and
gyromagnetic materials having anti-symmetric (nonreciprocal)
permittivity $\epsb$ and permeability $\mub$ tensors respectively. For
both examples, the gyrotropy axis is assumed to be along $\ev_x$ which
leads to PTPS along $\ev_x$ and PPHC along $\ev_y$ direction. It is
also found that PTPS of gyroelectric material is mostly electric-type
while that of gyromagnetic material is mostly magnetic-type. While the
chosen parameters lead to plasmonic enhancement of the persistent
phenomena, other parameters (dielectric $\epsb$ and $\mub$) also show
the same (zero or nonzero) features. In the fourth example with
gyrotropy axis along $\ev_z$ (perpendicular to surface), no PPHC is
observed. The $\ev_z$ components of electric and magnetic type PTPS
are nonzero but cancel each other leading to zero total thermal
spin. This proves that nonreciprocity is not sufficient to observe
PTPS and PPHC.

The fifth and sixth examples consider isotropic, dielectric
permittivity and permeability ($\epsb,\mub$) and diagonal
magneto-electric susceptibilities ($\xib,\zetab$). For nonreciprocal
susceptibilities ($\xib \neq -\zetab^T$), it is found that both PPHC
and PTPS are zero although electric and magnetic type contributions to
PTPS are nonzero. This is qualitatively similar to gyrotropic media
with gyrotropy axis perpendicular to the surface. When the
off-diagonal components of $\xib,\zetab$ are nonzero as considered in
the seventh (last) example, PTPS and PPHC parallel to the surface are
observed. Interestingly, when $y$ and $z$-components of $\Ev,\Hv$
fields are coupled, PTPS is along $\ev_y$ direction while PPHC is along
$\ev_x$ direction. When $x$ and $y$-components of $\Ev,\Hv$ fields are
coupled, PTPS and PPHC are parallel to the surface but they are not
necessarily in a specific direction (not tabulated). For a
magneto-electric medium, both electric and magnetic contributions to
PTPS are comparable to each other as opposed to gyrotropic media where
one of them dominates. Figure~\ref{fig0} summarizes the important
findings based on this general analysis.

We note that the material parameters $\epsb,\mub,\zetab,\xib$ are not
entirely arbitrary but follow certain symmetry relations and are also
constrained by conditions of causality and
passivity~\cite{silveirinha2010comment,gustafsson2010sum}. The
causality constraint leads to Kramer-Kronig relations for frequency
dependent parameters and it requires separate examination for
different types of materials~\cite{silveirinha2011examining}. The
passivity requires that the material matrix,
\begin{align*}
  M=\begin{bmatrix} \epsb & \zetab \\ \xib & \mub \end{bmatrix},
\end{align*}
is such that $[(-iM)+(-iM)^{*^T}]/2$ is positive
definite~\cite{silveirinha2010comment}. We do not discuss the
frequency dependence of various parameters here since single frequency
calculations are sufficient to describe the nature (existence,
directions) of PTPS and PPHC for a given material type. However, we
make sure that all the parameters satisfy the constraint of passivity
and note that without such constraints, the persistent phenomena can
be incorrectly deduced for reciprocal systems that are non-passive
(nonequilibrium). The seven examples above and the analysis presented
here are sufficient to predict the presence or absence of the PTPS and
PPHC and their nature (directions, electric/magnetic type PTPS) for
any practical example of a bianisotropic material.

\section{Conclusion}

Modern thermal photonics utilizes fluctuational electrodynamic
paradigm to explore new phenomena (near-field radiative heat
transfer~\cite{song2015near}) and new regimes
(nonreciprocity~\cite{zhu2018theory},
nonlinearities~\cite{khandekar2017near} and
nonequilibrium~\cite{jin2016temperature}) which are inaccessible to
older paradigm of radiometry and Kirchhoff's laws. And yet, thermal
spin photonics is so far limited to inquiries based on Kirchhoff's
laws~\cite{shitrit2013spin,wu2014spectrally,yin2013interpreting}. This
work demonstrates intriguing spin angular momentum related thermal
radiation phenomena in the near-field of nonreciprocal materials
analyzed within fluctuational electrodynamic paradigm. It paves the
way for new fundamental and technological avenues in thermal spin
photonics. In particular, it will be useful in the near future for
shaping spin-angular-momentum related radiative heat transport
phenomena such as our recent work on circularly polarized thermal
light sources~\cite{khandekar2019circular}.

Our work revealed that the spin-momentum locking of thermally excited
evanescent waves plays a fundamental role in facilitating the
surprising thermal equilibrium features of PTPS and PPHC. The
connection between the spin-momentum locking and the radiative heat
transfer is important for exploring new ways of achieving directional
heat transport at the nanoscale. We found that the surface polaritons
of gyrotropic materials can carry spin magnetic moment which invites
separate related studies of spin-dependent quantum plasmonics and spin
quantization. We proposed an experiment based on Brownian motion that
can provide a visual confirmation of the persistent
phenomena. Currently, there are no experiments probing such intriguing
nonreciprocal thermal fluctuations and heat transport effects. Also,
the experimental detection of the predicted surprising effects is
important from the perspective of thermodynamic revalidation of
fundamental understanding of nonreciprocal systems.

We described the universal behavior of the thermal spin photonic
phenomena with a comprehensive analysis of key classes of
nonreciprocal materials namely, gyroelectric, gyromagnetic and
magneto-electric media. This general analysis motivates similar
studies of thermal radiation~\cite{li2018nanophotonic}, radiative heat
transfer~\cite{song2015near}, Casimir
forces/torques~\cite{capasso2007casimir} from generic bianisotropic
materials including largely unexplored material types in this context
such as topological insulators, multiferroic and magentoelectric
materials. The theoretical framework and the Green's function produced
here can also be used to study environment-assisted quantum
nanophotonic phenomena such as Forster resonance energy
transfer~\cite{clegg1995fluorescence}, atomic transition
shifts~\cite{novotny2012principles} with general, bianisotropic
materials. We leave all these promising directions of research aside
for future work.

\emph{Acknowledgments} This work was supported by the U.S. Department
of Energy, Office of Basic Energy Science under award number
DE-SC0017717, DARPA Nascent Light-Matter Interaction program and the
Lillian Gilbreth Postdoctoral Fellowship program at Purdue University
(C.K.).

\section{Methods}

{\bf Derivation of Green's function}. The Green's function relating
vector potential at $\rv_1=(\Rv_1,z_1)$ to source current at
$\rv_2=(\Rv_2,z_2)$ in vacuum where $\Rv=(x,y)$ denotes planar
co-ordinates, is:
\begin{align}
\Gb(\rv_1,\rv_2) = \int \frac{d^2\kv_{\parallel}}{(2\pi)^2}
e^{i\kv_{\parallel}\cdot (\Rv_1-\Rv_2)} \gb(\kv_{\parallel},z_1,z_2)
\end{align}
$\kv=(\kv_{\parallel},\pm k_z)$ is the total wavevector consisting of
conserved parallel component $\kv_{\parallel}$ and perpendicular
$z$-component in vacuum $\pm k_z$. The ($+$) and ($-$) signs denote
waves going away from and towards the interface respectively. It
follows from Maxwell's equations that they satisfy the dispersion
relation $k_{\parallel}^2+k_z^2=k_0^2=(\omega/c)^2$ where
$k_{\parallel}=|\kv_{\parallel}|$ is real and $k_z$ can be real
($k_\parallel < k_0$) or complex valued ($k_{\parallel} > k_0$). For
simplicity, we write
$\kv_{\parallel}=(k_{\parallel}\cos\phi,k_{\parallel}\sin\phi)$ where
$\phi$ is the angle subtended by $\kv_\parallel$ with $x$-axis.
Assuming $z_1 > z_2$, the integrand $\gb(\kv_{\parallel},z_1,z_2)$
tensor is written using the $s,p$-polarization vectors
($\ev_{s},\ev_{p}$):
\begin{widetext}
\begin{align}
  \gb(\kv_{\parallel},z_1,z_2) =
  \frac{i}{2k_z}\bigg[\overbrace{e^{ik_z(z_1-z_2)}[\ev_{s+}\ev_{s+}^T+
        \ev_{p+}\ev_{p+}^T]}^{\text{vacuum part $\gb_0$}} +
    \overbrace{e^{ik_z(z_1+z_2)}[\underbrace{(r_{ss}\ev_{s+}+r_{ps}\ev_{p+})
        \ev_{s-}^T}_{\text{reflection of $\ev_{s-}$ wave}} +
      \underbrace{(r_{sp}\ev_{s+}+
        r_{pp}\ev_{p+})\ev_{p-}^T}_{\text{reflection of $\ev_{p-}$
          wave}}]}^{\text{scattered/reflected part $\gb_{\text{ref}}$}}\bigg]
\label{gkz}  
\end{align}
\end{widetext}  
The polarization
vectors $\ev_{j\pm}$ for $j={s,p}$ with $\pm$ denoting waves going
along $\pm\ev_z$ directions are:
\begin{align}
\ev_{s\pm}=\begin{bmatrix}\sin\phi \\ -\cos\phi \\ 0 \end{bmatrix},
\ev_{p\pm}=\frac{-1}{k_0} \begin{bmatrix}\pm k_z\cos\phi \\ \pm
  k_z\sin\phi \\ -k_\parallel \end{bmatrix}
\label{spvectors}
\end{align}
The Fresnel reflection coefficient $r_{jk}$ for $j,k=[s,p]$ describes
the amplitude of $\ev_j$-polarized reflected light due to unit
amplitude $\ev_k$-polarized incident light.  The Green's function
above consists of two parts corresponding to the trajectories of
electromagnetic waves generated at the source position $\rv_2$ and
arriving at $\rv_1$ either directly ($\gb_0$) or upon reflection from
the interface ($\gb_{\text{ref}}$). The Green's function in
Eq.~\ref{gkz} is derived for $z_1 \geq z_2$. For $z_1 < z_2$, only the
vacuum part is modified to $\gb_0 =
e^{-ik_z(z_1-z_2)}[\ev_{s-}\ev_{s-}^T+ \ev_{p-}\ev_{p-}^T]$. The
Fresnel reflection coefficients can be obtained experimentally or
theoretically.

{\bf Fresnel reflection coefficients and polaritonic dispersion}. We
develop a tool to compute Fresnel reflection coefficients for a
generic, homogeneous medium that can be described using the following
constitutive relations assuming local response (in the frequency
domain):
\begin{align}
  \mathbf{D} &= \epsb\varepsilon_0 \Ev + \xib\frac{1}{c}\Hv \nonumber \\
  \mathbf{B} &= \zetab\frac{1}{c}\Ev + \mub\mu_0\Hv
\end{align}
$\epsb,\mub$ are dimensionless permittivity and permeability tensors
and $\xib,\zetab$ are magneto-electric coupling tensors. For isotropic
materials, $\xib,\zetab=0$ and $\epsb,\mub$ are scalars. For
gyro-electric (magneto-optic) and gyro-magnetic media, the tensors
$\epsb$ and $\mub$ respectively have off-diagonal components and
$\xib,\zetab=0$. The tensors $\xib,\zetab$ are nonzero for
magneto-electric media. By writing electromagnetic fields inside the
material as $[\Ev, \sqrt{\frac{\mu_0}{\varepsilon_0}}\Hv]^T
e^{i(\kv_{\parallel}\cdot\Rv+ k_z z -i\omega t)}$ and using above
constitutive relations in Maxwell's equations, we obtain the following
dimensionless dispersion equation for waves inside the material:
\begin{align}
  &\text{det}(M+M_k) = 0, \hspace{5pt} \text{for} \hspace{5pt}
  M=\begin{bmatrix} \epsb & \zetab \\ \xib & \mub \end{bmatrix},
  M_k=\begin{bmatrix} 0 & \kb/k_0 \\ -\kb/k_0 & 0 \end{bmatrix}
  \nonumber \\ &\kb=\begin{bmatrix} 0 & -k_z & k_\parallel\sin\phi
  \\ k_z & 0 & -k_\parallel\cos\phi \\ -k_\parallel\sin\phi &
  k_\parallel\cos\phi & 0 \end{bmatrix}
\end{align}
Here, $6\times 6$ material tensor $M$ describes the constitutive
relations and $M_k$ corresponds to the curl operator acting on plane
waves. Because of the generality of this problem, we obtain $k_z$
numerically by solving $\text{det}(M+M_k(k_z))=0$ for given
$(k_\parallel,\phi)$. Depending on the nature of the material, there
can be two (for isotropic media) or four (for anisotropic media)
solutions of $k_z$ corresponding to $\ev_z$-propagation of
electromagnetic waves. Overall, there are four eigensolutions spanning
the null-space of the matrix $M+M_k(k_z)$, out of which two solutions
correspond to waves propagating in $-\ev_z$ direction (transmitted
waves in our geometry).  The four Fresnel reflection coefficients are
then obtained by matching the tangential components at the interface
($E_x,E_y,H_x,H_y$) of incident and reflected fields with the
transmitted fields. Here, the transmitted fields are written in the
basis of former two null-space solutions while the incident and
reflected fields are written in the basis of
$\ev_s,\ev_p$-polarizations (Eq.~\ref{spvectors}). This procedure is
also extended in this work to compute the polaritonic dispersion
($\omega(\kv_{\parallel})$) of surface polaritons that decay on both
sides of the interface. While that calculation does not involve
Fresnel coefficients, the boundary conditions again lead to a
homogeneous, linear problem of the form $M_p(\omega,\kv_{\parallel})
X=0$ where $X$ contains the coefficients describing the decomposition
of polaritonic fields into four eigenstates ($s,p$-polarizations in
vacuum and two $-\ev_z$-propagating solutions inside the medium) at
the interface. By numerically solving
$\text{det}(M_p(\omega,\kv_{\parallel}))=0$, the polaritonic
dispersion $\omega(\kv_{\parallel})$ is obtained. The associated
null-space describes the polaritonic fields. Note that since
$\kv_{\parallel}$ is assumed to be real-valued and non-decaying,
$\omega$ is complex-valued with the imaginary part describing the
finite lifetime (quality factor) of the polaritons.

\bibliography{photon}

\begin{thebibliography}{74}%
\makeatletter
\providecommand \@ifxundefined [1]{%
 \@ifx{#1\undefined}
}%
\providecommand \@ifnum [1]{%
 \ifnum #1\expandafter \@firstoftwo
 \else \expandafter \@secondoftwo
 \fi
}%
\providecommand \@ifx [1]{%
 \ifx #1\expandafter \@firstoftwo
 \else \expandafter \@secondoftwo
 \fi
}%
\providecommand \natexlab [1]{#1}%
\providecommand \enquote  [1]{``#1''}%
\providecommand \bibnamefont  [1]{#1}%
\providecommand \bibfnamefont [1]{#1}%
\providecommand \citenamefont [1]{#1}%
\providecommand \href@noop [0]{\@secondoftwo}%
\providecommand \href [0]{\begingroup \@sanitize@url \@href}%
\providecommand \@href[1]{\@@startlink{#1}\@@href}%
\providecommand \@@href[1]{\endgroup#1\@@endlink}%
\providecommand \@sanitize@url [0]{\catcode `\\12\catcode `\$12\catcode
  `\&12\catcode `\#12\catcode `\^12\catcode `\_12\catcode `\%12\relax}%
\providecommand \@@startlink[1]{}%
\providecommand \@@endlink[0]{}%
\providecommand \url  [0]{\begingroup\@sanitize@url \@url }%
\providecommand \@url [1]{\endgroup\@href {#1}{\urlprefix }}%
\providecommand \urlprefix  [0]{URL }%
\providecommand \Eprint [0]{\href }%
\providecommand \doibase [0]{http://dx.doi.org/}%
\providecommand \selectlanguage [0]{\@gobble}%
\providecommand \bibinfo  [0]{\@secondoftwo}%
\providecommand \bibfield  [0]{\@secondoftwo}%
\providecommand \translation [1]{[#1]}%
\providecommand \BibitemOpen [0]{}%
\providecommand \bibitemStop [0]{}%
\providecommand \bibitemNoStop [0]{.\EOS\space}%
\providecommand \EOS [0]{\spacefactor3000\relax}%
\providecommand \BibitemShut  [1]{\csname bibitem#1\endcsname}%
\let\auto@bib@innerbib\@empty
\bibitem [{\citenamefont {Fan}(2017)}]{fan2017thermal}%
  \BibitemOpen
  \bibfield  {author} {\bibinfo {author} {\bibfnamefont {S.}~\bibnamefont
  {Fan}},\ }\bibfield  {title} {Thermal photonics and energy applications,\
  }\href {https://www.sciencedirect.com/science/article/pii/S2542435117300193}
  {\bibfield  {journal} {\bibinfo  {journal} {Joule}\ }\textbf {\bibinfo
  {volume} {1}},\ \bibinfo {pages} {264--273} (\bibinfo {year}
  {2017})}\BibitemShut {NoStop}%
\bibitem [{\citenamefont {Tervo}\ \emph {et~al.}(2018)\citenamefont {Tervo},
  \citenamefont {Bagherisereshki},\ and\ \citenamefont
  {Zhang}}]{tervo2018near}%
  \BibitemOpen
  \bibfield  {author} {\bibinfo {author} {\bibfnamefont {E.}~\bibnamefont
  {Tervo}}, \bibinfo {author} {\bibfnamefont {E.}~\bibnamefont
  {Bagherisereshki}}, \ and\ \bibinfo {author} {\bibfnamefont {Z.}~\bibnamefont
  {Zhang}},\ }\bibfield  {title} {Near-field radiative thermoelectric energy
  converters: a review,\ }\href
  {https://link.springer.com/article/10.1007/s11708-017-0517-z} {\bibfield
  {journal} {\bibinfo  {journal} {Front. Energy}\ }\textbf {\bibinfo {volume}
  {12}},\ \bibinfo {pages} {5--21} (\bibinfo {year} {2018})}\BibitemShut
  {NoStop}%
\bibitem [{\citenamefont {Le~Feber}\ \emph {et~al.}(2015)\citenamefont
  {Le~Feber}, \citenamefont {Rotenberg},\ and\ \citenamefont
  {Kuipers}}]{le2015nanophotonic}%
  \BibitemOpen
  \bibfield  {author} {\bibinfo {author} {\bibfnamefont {B.}~\bibnamefont
  {Le~Feber}}, \bibinfo {author} {\bibfnamefont {N.}~\bibnamefont {Rotenberg}},
  \ and\ \bibinfo {author} {\bibfnamefont {L.}~\bibnamefont {Kuipers}},\
  }\bibfield  {title} {Nanophotonic control of circular dipole emission,\
  }\href {https://www.nature.com/articles/ncomms7695} {\bibfield  {journal}
  {\bibinfo  {journal} {Nat. Commun.}\ }\textbf {\bibinfo {volume} {6}},\
  \bibinfo {pages} {6695} (\bibinfo {year} {2015})}\BibitemShut {NoStop}%
\bibitem [{\citenamefont {Mitsch}\ \emph {et~al.}(2014)\citenamefont {Mitsch},
  \citenamefont {Sayrin}, \citenamefont {Albrecht}, \citenamefont
  {Schneeweiss},\ and\ \citenamefont {\~Rauschenbeutel}}]{mitsch2014quantum}%
  \BibitemOpen
  \bibfield  {author} {\bibinfo {author} {\bibfnamefont {R.}~\bibnamefont
  {Mitsch}}, \bibinfo {author} {\bibfnamefont {C.}~\bibnamefont {Sayrin}},
  \bibinfo {author} {\bibfnamefont {B.}~\bibnamefont {Albrecht}}, \bibinfo
  {author} {\bibfnamefont {P.}~\bibnamefont {Schneeweiss}}, \ and\ \bibinfo
  {author} {\bibfnamefont {A.}~\bibnamefont {\~Rauschenbeutel}},\ }\bibfield
  {title} {Quantum state-controlled directional spontaneous emission of photons
  into a nanophotonic waveguide,\ }\href
  {https://www.nature.com/articles/ncomms6713} {\bibfield  {journal} {\bibinfo
  {journal} {Nat. Commun.}\ }\textbf {\bibinfo {volume} {5}},\ \bibinfo {pages}
  {5713} (\bibinfo {year} {2014})}\BibitemShut {NoStop}%
\bibitem [{\citenamefont {Lodahl}\ \emph {et~al.}(2015)\citenamefont {Lodahl},
  \citenamefont {Mahmoodian},\ and\ \citenamefont
  {Stobbe}}]{lodahl2015interfacing}%
  \BibitemOpen
  \bibfield  {author} {\bibinfo {author} {\bibfnamefont {P.}~\bibnamefont
  {Lodahl}}, \bibinfo {author} {\bibfnamefont {S.}~\bibnamefont {Mahmoodian}},
  \ and\ \bibinfo {author} {\bibfnamefont {S.}~\bibnamefont {Stobbe}},\
  }\bibfield  {title} {Interfacing single photons and single quantum dots with
  photonic nanostructures,\ }\href
  {https://journals.aps.org/rmp/abstract/10.1103/RevModPhys.87.347} {\bibfield
  {journal} {\bibinfo  {journal} {Rev. Mod. Phys.}\ }\textbf {\bibinfo {volume}
  {87}},\ \bibinfo {pages} {347} (\bibinfo {year} {2015})}\BibitemShut
  {NoStop}%
\bibitem [{\citenamefont {{\v{Z}}uti{\'c}}\ \emph {et~al.}(2004)\citenamefont
  {{\v{Z}}uti{\'c}}, \citenamefont {Fabian},\ and\ \citenamefont
  {Sarma}}]{vzutic2004spintronics}%
  \BibitemOpen
  \bibfield  {author} {\bibinfo {author} {\bibfnamefont {I.}~\bibnamefont
  {{\v{Z}}uti{\'c}}}, \bibinfo {author} {\bibfnamefont {J.}~\bibnamefont
  {Fabian}}, \ and\ \bibinfo {author} {\bibfnamefont {S.D.}\ \bibnamefont
  {Sarma}},\ }\bibfield  {title} {Spintronics: Fundamentals and applications,\
  }\href {https://journals.aps.org/rmp/abstract/10.1103/RevModPhys.76.323}
  {\bibfield  {journal} {\bibinfo  {journal} {Rev. Mod. Phys.}\ }\textbf
  {\bibinfo {volume} {76}},\ \bibinfo {pages} {323} (\bibinfo {year}
  {2004})}\BibitemShut {NoStop}%
\bibitem [{\citenamefont {Shitrit}\ \emph {et~al.}(2013)\citenamefont
  {Shitrit}, \citenamefont {Yulevich}, \citenamefont {Maguid}, \citenamefont
  {Ozeri}, \citenamefont {Veksler}, \citenamefont {Kleiner},\ and\
  \citenamefont {Hasman}}]{shitrit2013spin}%
  \BibitemOpen
  \bibfield  {author} {\bibinfo {author} {\bibfnamefont {N.}~\bibnamefont
  {Shitrit}}, \bibinfo {author} {\bibfnamefont {I.}~\bibnamefont {Yulevich}},
  \bibinfo {author} {\bibfnamefont {E.}~\bibnamefont {Maguid}}, \bibinfo
  {author} {\bibfnamefont {D.}~\bibnamefont {Ozeri}}, \bibinfo {author}
  {\bibfnamefont {D.}~\bibnamefont {Veksler}}, \bibinfo {author} {\bibfnamefont
  {V.}~\bibnamefont {Kleiner}}, \ and\ \bibinfo {author} {\bibfnamefont
  {E.}~\bibnamefont {Hasman}},\ }\bibfield  {title} {Spin-optical metamaterial
  route to spin-controlled photonics,\ }\href
  {https://science.sciencemag.org/content/340/6133/724} {\bibfield  {journal}
  {\bibinfo  {journal} {Science}\ }\textbf {\bibinfo {volume} {340}},\ \bibinfo
  {pages} {724--726} (\bibinfo {year} {2013})}\BibitemShut {NoStop}%
\bibitem [{\citenamefont {Wu}\ \emph {et~al.}(2014)\citenamefont {Wu},
  \citenamefont {Arju}, \citenamefont {Kelp}, \citenamefont {Fan},
  \citenamefont {Dominguez}, \citenamefont {Gonzales}, \citenamefont {Tutuc},
  \citenamefont {Brener},\ and\ \citenamefont {Shvets}}]{wu2014spectrally}%
  \BibitemOpen
  \bibfield  {author} {\bibinfo {author} {\bibfnamefont {C.}~\bibnamefont
  {Wu}}, \bibinfo {author} {\bibfnamefont {N.}~\bibnamefont {Arju}}, \bibinfo
  {author} {\bibfnamefont {G.}~\bibnamefont {Kelp}}, \bibinfo {author}
  {\bibfnamefont {J.~A.}\ \bibnamefont {Fan}}, \bibinfo {author} {\bibfnamefont
  {J.}~\bibnamefont {Dominguez}}, \bibinfo {author} {\bibfnamefont
  {E.}~\bibnamefont {Gonzales}}, \bibinfo {author} {\bibfnamefont
  {E.}~\bibnamefont {Tutuc}}, \bibinfo {author} {\bibfnamefont
  {I.}~\bibnamefont {Brener}}, \ and\ \bibinfo {author} {\bibfnamefont
  {G.}~\bibnamefont {Shvets}},\ }\bibfield  {title} {Spectrally selective
  chiral silicon metasurfaces based on infrared fano resonances,\ }\href
  {https://www.nature.com/articles/ncomms4892} {\bibfield  {journal} {\bibinfo
  {journal} {Nat. Commun.}\ }\textbf {\bibinfo {volume} {5}},\ \bibinfo {pages}
  {3892} (\bibinfo {year} {2014})}\BibitemShut {NoStop}%
\bibitem [{\citenamefont {Yin}\ \emph {et~al.}(2013)\citenamefont {Yin},
  \citenamefont {Schaferling}, \citenamefont {Metzger},\ and\ \citenamefont
  {Giessen}}]{yin2013interpreting}%
  \BibitemOpen
  \bibfield  {author} {\bibinfo {author} {\bibfnamefont {X.}~\bibnamefont
  {Yin}}, \bibinfo {author} {\bibfnamefont {M.}~\bibnamefont {Schaferling}},
  \bibinfo {author} {\bibfnamefont {B.}~\bibnamefont {Metzger}}, \ and\
  \bibinfo {author} {\bibfnamefont {H.}~\bibnamefont {Giessen}},\ }\bibfield
  {title} {Interpreting chiral nanophotonic spectra: the plasmonic born--kuhn
  model,\ }\href {https://pubs.acs.org/doi/10.1021/nl403705k} {\bibfield
  {journal} {\bibinfo  {journal} {Nano Lett.}\ }\textbf {\bibinfo {volume}
  {13}},\ \bibinfo {pages} {6238} (\bibinfo {year} {2013})}\BibitemShut
  {NoStop}%
\bibitem [{\citenamefont {Set{\"a}l{\"a}}\ \emph {et~al.}(2002)\citenamefont
  {Set{\"a}l{\"a}}, \citenamefont {Kaivola},\ and\ \citenamefont
  {Friberg}}]{setala2002degree}%
  \BibitemOpen
  \bibfield  {author} {\bibinfo {author} {\bibfnamefont {T.}~\bibnamefont
  {Set{\"a}l{\"a}}}, \bibinfo {author} {\bibfnamefont {M.}~\bibnamefont
  {Kaivola}}, \ and\ \bibinfo {author} {\bibfnamefont {A.T.}\ \bibnamefont
  {Friberg}},\ }\bibfield  {title} {Degree of polarization in near fields of
  thermal sources: effects of surface waves,\ }\href
  {https://journals.aps.org/prl/abstract/10.1103/PhysRevLett.88.123902}
  {\bibfield  {journal} {\bibinfo  {journal} {Phys. Rev. Lett.}\ }\textbf
  {\bibinfo {volume} {88}},\ \bibinfo {pages} {123902} (\bibinfo {year}
  {2002})}\BibitemShut {NoStop}%
\bibitem [{\citenamefont {Khandekar}\ and\ \citenamefont
  {Jacob}(2019)}]{khandekar2019circular}%
  \BibitemOpen
  \bibfield  {author} {\bibinfo {author} {\bibfnamefont {C.}~\bibnamefont
  {Khandekar}}\ and\ \bibinfo {author} {\bibfnamefont {Z.}~\bibnamefont
  {Jacob}},\ }\bibfield  {title} {Circularly polarized thermal radiation from
  nonequilibrium coupled antennas,\ }\href
  {https://journals.aps.org/prapplied/abstract/10.1103/PhysRevApplied.12.014053}
  {\bibfield  {journal} {\bibinfo  {journal} {Phys. Rev. Applied}\ }\textbf
  {\bibinfo {volume} {12}},\ \bibinfo {pages} {014053} (\bibinfo {year}
  {2019})}\BibitemShut {NoStop}%
\bibitem [{\citenamefont {Bleszynski-Jayich}\ \emph {et~al.}(2009)\citenamefont
  {Bleszynski-Jayich}, \citenamefont {Shanks}, \citenamefont {Peaudecerf},
  \citenamefont {Ginossar}, \citenamefont {Von~Oppen}, \citenamefont
  {Glazman},\ and\ \citenamefont {Harris}}]{bleszynski2009persistent}%
  \BibitemOpen
  \bibfield  {author} {\bibinfo {author} {\bibfnamefont {A.C.}\ \bibnamefont
  {Bleszynski-Jayich}}, \bibinfo {author} {\bibfnamefont {W.E.}\ \bibnamefont
  {Shanks}}, \bibinfo {author} {\bibfnamefont {B.}~\bibnamefont {Peaudecerf}},
  \bibinfo {author} {\bibfnamefont {E.}~\bibnamefont {Ginossar}}, \bibinfo
  {author} {\bibfnamefont {F.}~\bibnamefont {Von~Oppen}}, \bibinfo {author}
  {\bibfnamefont {L.}~\bibnamefont {Glazman}}, \ and\ \bibinfo {author}
  {\bibfnamefont {J.G.E.}\ \bibnamefont {Harris}},\ }\bibfield  {title}
  {Persistent currents in normal metal rings,\ }\href
  {https://science.sciencemag.org/content/326/5950/272} {\bibfield  {journal}
  {\bibinfo  {journal} {Science}\ }\textbf {\bibinfo {volume} {326}},\ \bibinfo
  {pages} {272--275} (\bibinfo {year} {2009})}\BibitemShut {NoStop}%
\bibitem [{\citenamefont {B{\"u}ttiker}\ \emph {et~al.}(1983)\citenamefont
  {B{\"u}ttiker}, \citenamefont {Imry},\ and\ \citenamefont
  {Landauer}}]{buttiker1983josephson}%
  \BibitemOpen
  \bibfield  {author} {\bibinfo {author} {\bibfnamefont {M.}~\bibnamefont
  {B{\"u}ttiker}}, \bibinfo {author} {\bibfnamefont {Y.}~\bibnamefont {Imry}},
  \ and\ \bibinfo {author} {\bibfnamefont {R.}~\bibnamefont {Landauer}},\
  }\bibfield  {title} {Josephson behavior in small normal one-dimensional
  rings,\ }\href
  {https://www.sciencedirect.com/science/article/pii/0375960183900117}
  {\bibfield  {journal} {\bibinfo  {journal} {Phys. Lett. A}\ }\textbf
  {\bibinfo {volume} {96}},\ \bibinfo {pages} {365--367} (\bibinfo {year}
  {1983})}\BibitemShut {NoStop}%
\bibitem [{\citenamefont {Van~Mechelen}\ and\ \citenamefont
  {Jacob}(2016)}]{van2016universal}%
  \BibitemOpen
  \bibfield  {author} {\bibinfo {author} {\bibfnamefont {T.}~\bibnamefont
  {Van~Mechelen}}\ and\ \bibinfo {author} {\bibfnamefont {Z.}~\bibnamefont
  {Jacob}},\ }\bibfield  {title} {Universal spin-momentum locking of evanescent
  waves,\ }\href@noop {} {\bibfield  {journal} {\bibinfo  {journal} {Optica}\
  }\textbf {\bibinfo {volume} {3}},\ \bibinfo {pages} {118--126} (\bibinfo
  {year} {2016})}\BibitemShut {NoStop}%
\bibitem [{\citenamefont {Bliokh}\ \emph {et~al.}(2015)\citenamefont {Bliokh},
  \citenamefont {Smirnova},\ and\ \citenamefont {Nori}}]{bliokh2015quantum}%
  \BibitemOpen
  \bibfield  {author} {\bibinfo {author} {\bibfnamefont {K.Y.}\ \bibnamefont
  {Bliokh}}, \bibinfo {author} {\bibfnamefont {D.}~\bibnamefont {Smirnova}}, \
  and\ \bibinfo {author} {\bibfnamefont {F.}~\bibnamefont {Nori}},\ }\bibfield
  {title} {Quantum spin hall effect of light,\ }\href
  {https://science.sciencemag.org/node/632049.full} {\bibfield  {journal}
  {\bibinfo  {journal} {Science}\ }\textbf {\bibinfo {volume} {348}},\ \bibinfo
  {pages} {1448--1451} (\bibinfo {year} {2015})}\BibitemShut {NoStop}%
\bibitem [{\citenamefont {Petersen}\ \emph {et~al.}(2014)\citenamefont
  {Petersen}, \citenamefont {Volz},\ and\ \citenamefont
  {Rauschenbeutel}}]{petersen2014chiral}%
  \BibitemOpen
  \bibfield  {author} {\bibinfo {author} {\bibfnamefont {J.}~\bibnamefont
  {Petersen}}, \bibinfo {author} {\bibfnamefont {J.}~\bibnamefont {Volz}}, \
  and\ \bibinfo {author} {\bibfnamefont {A.}~\bibnamefont {Rauschenbeutel}},\
  }\bibfield  {title} {Chiral nanophotonic waveguide interface based on
  spin-orbit interaction of light,\ }\href
  {https://science.sciencemag.org/content/346/6205/67} {\bibfield  {journal}
  {\bibinfo  {journal} {Science}\ }\textbf {\bibinfo {volume} {346}},\ \bibinfo
  {pages} {67--71} (\bibinfo {year} {2014})}\BibitemShut {NoStop}%
\bibitem [{\citenamefont {Zhu}\ and\ \citenamefont
  {Fan}(2016)}]{zhu2016persistent}%
  \BibitemOpen
  \bibfield  {author} {\bibinfo {author} {\bibfnamefont {L.}~\bibnamefont
  {Zhu}}\ and\ \bibinfo {author} {\bibfnamefont {S.}~\bibnamefont {Fan}},\
  }\bibfield  {title} {Persistent directional current at equilibrium in
  nonreciprocal many-body near field electromagnetic heat transfer,\ }\href
  {https://journals.aps.org/prl/abstract/10.1103/PhysRevLett.117.134303}
  {\bibfield  {journal} {\bibinfo  {journal} {Phys. Rev. Lett.}\ }\textbf
  {\bibinfo {volume} {117}},\ \bibinfo {pages} {134303} (\bibinfo {year}
  {2016})}\BibitemShut {NoStop}%
\bibitem [{\citenamefont {Ott}\ \emph {et~al.}(2018)\citenamefont {Ott},
  \citenamefont {Ben-Abdallah},\ and\ \citenamefont {Biehs}}]{ott2018circular}%
  \BibitemOpen
  \bibfield  {author} {\bibinfo {author} {\bibfnamefont {A.}~\bibnamefont
  {Ott}}, \bibinfo {author} {\bibfnamefont {P.}~\bibnamefont {Ben-Abdallah}}, \
  and\ \bibinfo {author} {\bibfnamefont {S-A.}\ \bibnamefont {Biehs}},\
  }\bibfield  {title} {Circular heat and momentum flux radiated by
  magneto-optical nanoparticles,\ }\href
  {https://journals.aps.org/prb/abstract/10.1103/PhysRevB.97.205414} {\bibfield
   {journal} {\bibinfo  {journal} {Phys. Rev. B}\ }\textbf {\bibinfo {volume}
  {97}},\ \bibinfo {pages} {205414} (\bibinfo {year} {2018})}\BibitemShut
  {NoStop}%
\bibitem [{\citenamefont {Silveirinha}(2017)}]{silveirinha2017topological}%
  \BibitemOpen
  \bibfield  {author} {\bibinfo {author} {\bibfnamefont {M.G.}\ \bibnamefont
  {Silveirinha}},\ }\bibfield  {title} {Topological angular momentum and
  radiative heat transport in closed orbits,\ }\href
  {https://journals.aps.org/prb/abstract/10.1103/PhysRevB.95.115103} {\bibfield
   {journal} {\bibinfo  {journal} {Phys. Rev. B}\ }\textbf {\bibinfo {volume}
  {95}},\ \bibinfo {pages} {115103} (\bibinfo {year} {2017})}\BibitemShut
  {NoStop}%
\bibitem [{\citenamefont {Zhu}\ \emph {et~al.}(2018)\citenamefont {Zhu},
  \citenamefont {Guo},\ and\ \citenamefont {Fan}}]{zhu2018theory}%
  \BibitemOpen
  \bibfield  {author} {\bibinfo {author} {\bibfnamefont {L.}~\bibnamefont
  {Zhu}}, \bibinfo {author} {\bibfnamefont {Y.}~\bibnamefont {Guo}}, \ and\
  \bibinfo {author} {\bibfnamefont {S.}~\bibnamefont {Fan}},\ }\bibfield
  {title} {Theory of many-body radiative heat transfer without the constraint
  of reciprocity,\ }\href
  {https://journals.aps.org/prb/abstract/10.1103/PhysRevB.97.094302} {\bibfield
   {journal} {\bibinfo  {journal} {Phys. Rev. B}\ }\textbf {\bibinfo {volume}
  {97}},\ \bibinfo {pages} {094302} (\bibinfo {year} {2018})}\BibitemShut
  {NoStop}%
\bibitem [{\citenamefont {Ekeroth}\ \emph {et~al.}(2017)\citenamefont
  {Ekeroth}, \citenamefont {Garc{\'\i}a-Mart{\'\i}n},\ and\ \citenamefont
  {Cuevas}}]{ekeroth2017thermal}%
  \BibitemOpen
  \bibfield  {author} {\bibinfo {author} {\bibfnamefont {R.M.A.}\ \bibnamefont
  {Ekeroth}}, \bibinfo {author} {\bibfnamefont {A.}~\bibnamefont
  {Garc{\'\i}a-Mart{\'\i}n}}, \ and\ \bibinfo {author} {\bibfnamefont {J.C.}\
  \bibnamefont {Cuevas}},\ }\bibfield  {title} {Thermal discrete dipole
  approximation for the description of thermal emission and radiative heat
  transfer of magneto-optical systems,\ }\href
  {https://www.sciencedirect.com/science/article/pii/S0022407313003543}
  {\bibfield  {journal} {\bibinfo  {journal} {Phys. Rev. B}\ }\textbf {\bibinfo
  {volume} {95}},\ \bibinfo {pages} {235428} (\bibinfo {year}
  {2017})}\BibitemShut {NoStop}%
\bibitem [{\citenamefont {Buhmann}\ \emph {et~al.}(2012)\citenamefont
  {Buhmann}, \citenamefont {Butcher},\ and\ \citenamefont
  {Scheel}}]{buhmann2012macroscopic}%
  \BibitemOpen
  \bibfield  {author} {\bibinfo {author} {\bibfnamefont {S.Y.}\ \bibnamefont
  {Buhmann}}, \bibinfo {author} {\bibfnamefont {D.T.}\ \bibnamefont {Butcher}},
  \ and\ \bibinfo {author} {\bibfnamefont {S.}~\bibnamefont {Scheel}},\
  }\bibfield  {title} {Macroscopic quantum electrodynamics in nonlocal and
  nonreciprocal media,\ }\href
  {https://iopscience.iop.org/article/10.1088/1367-2630/14/8/083034} {\bibfield
   {journal} {\bibinfo  {journal} {New J. Phys.}\ }\textbf {\bibinfo {volume}
  {14}},\ \bibinfo {pages} {083034} (\bibinfo {year} {2012})}\BibitemShut
  {NoStop}%
\bibitem [{\citenamefont {Bermel}\ \emph {et~al.}(2010)\citenamefont {Bermel},
  \citenamefont {Ghebrebrhan}, \citenamefont {Chan}, \citenamefont {Yeng},
  \citenamefont {Araghchini}, \citenamefont {Hamam}, \citenamefont {Marton},
  \citenamefont {Jensen}, \citenamefont {Solja{\v{c}}i{\'c}}, \citenamefont
  {Joannopoulos}, \citenamefont {Johnson},\ and\ \citenamefont
  {Celanovic}}]{bermel2010design}%
  \BibitemOpen
  \bibfield  {author} {\bibinfo {author} {\bibfnamefont {P.}~\bibnamefont
  {Bermel}}, \bibinfo {author} {\bibfnamefont {M.}~\bibnamefont {Ghebrebrhan}},
  \bibinfo {author} {\bibfnamefont {W.}~\bibnamefont {Chan}}, \bibinfo {author}
  {\bibfnamefont {Y.X.}\ \bibnamefont {Yeng}}, \bibinfo {author} {\bibfnamefont
  {M.}~\bibnamefont {Araghchini}}, \bibinfo {author} {\bibfnamefont
  {R.}~\bibnamefont {Hamam}}, \bibinfo {author} {\bibfnamefont {C.H.}\
  \bibnamefont {Marton}}, \bibinfo {author} {\bibfnamefont {K.F.}\ \bibnamefont
  {Jensen}}, \bibinfo {author} {\bibfnamefont {M.}~\bibnamefont
  {Solja{\v{c}}i{\'c}}}, \bibinfo {author} {\bibfnamefont {J.D.}\ \bibnamefont
  {Joannopoulos}}, \bibinfo {author} {\bibfnamefont {S.G.}\ \bibnamefont
  {Johnson}}, \ and\ \bibinfo {author} {\bibfnamefont {I.}~\bibnamefont
  {Celanovic}},\ }\bibfield  {title} {Design and global optimization of
  high-efficiency thermophotovoltaic systems,\ }\href
  {https://www.osapublishing.org/oe/abstract.cfm?uri=oe-18-s3-a314} {\bibfield
  {journal} {\bibinfo  {journal} {Opt. Exp.}\ }\textbf {\bibinfo {volume}
  {18}},\ \bibinfo {pages} {A314--A334} (\bibinfo {year} {2010})}\BibitemShut
  {NoStop}%
\bibitem [{\citenamefont {Zhu}\ and\ \citenamefont {Fan}(2014)}]{zhu2014near}%
  \BibitemOpen
  \bibfield  {author} {\bibinfo {author} {\bibfnamefont {L.}~\bibnamefont
  {Zhu}}\ and\ \bibinfo {author} {\bibfnamefont {S.}~\bibnamefont {Fan}},\
  }\bibfield  {title} {Near-complete violation of detailed balance in thermal
  radiation,\ }\href
  {https://journals.aps.org/prb/abstract/10.1103/PhysRevB.90.220301} {\bibfield
   {journal} {\bibinfo  {journal} {Phys. Rev. B}\ }\textbf {\bibinfo {volume}
  {90}},\ \bibinfo {pages} {220301} (\bibinfo {year} {2014})}\BibitemShut
  {NoStop}%
\bibitem [{\citenamefont {Ben-Abdallah}(2016)}]{ben2016photon}%
  \BibitemOpen
  \bibfield  {author} {\bibinfo {author} {\bibfnamefont {P.}~\bibnamefont
  {Ben-Abdallah}},\ }\bibfield  {title} {Photon thermal hall effect,\ }\href
  {https://journals.aps.org/prl/abstract/10.1103/PhysRevLett.116.084301}
  {\bibfield  {journal} {\bibinfo  {journal} {Phys. Rev. Lett.}\ }\textbf
  {\bibinfo {volume} {116}},\ \bibinfo {pages} {084301} (\bibinfo {year}
  {2016})}\BibitemShut {NoStop}%
\bibitem [{\citenamefont {Clegg}(1995)}]{clegg1995fluorescence}%
  \BibitemOpen
  \bibfield  {author} {\bibinfo {author} {\bibfnamefont {R.M.}\ \bibnamefont
  {Clegg}},\ }\bibfield  {title} {Fluorescence resonance energy transfer,\
  }\href {https://www.sciencedirect.com/science/article/pii/0958166995800166}
  {\bibfield  {journal} {\bibinfo  {journal} {Curr. Opin. Biotechnol.}\
  }\textbf {\bibinfo {volume} {6}},\ \bibinfo {pages} {103--110} (\bibinfo
  {year} {1995})}\BibitemShut {NoStop}%
\bibitem [{\citenamefont {Novotny}\ and\ \citenamefont
  {Hecht}(2012)}]{novotny2012principles}%
  \BibitemOpen
  \bibfield  {author} {\bibinfo {author} {\bibfnamefont {Lukas}\ \bibnamefont
  {Novotny}}\ and\ \bibinfo {author} {\bibfnamefont {Bert}\ \bibnamefont
  {Hecht}},\ }\href@noop {} {\emph {\bibinfo {title} {Principles of
  nano-optics}}}\ (\bibinfo  {publisher} {Cambridge university press},\
  \bibinfo {year} {2012})\BibitemShut {NoStop}%
\bibitem [{\citenamefont {Fuchs}\ \emph {et~al.}(2017)\citenamefont {Fuchs},
  \citenamefont {Crosse},\ and\ \citenamefont {Buhmann}}]{fuchs2017casimir}%
  \BibitemOpen
  \bibfield  {author} {\bibinfo {author} {\bibfnamefont {S.}~\bibnamefont
  {Fuchs}}, \bibinfo {author} {\bibfnamefont {J.A.}\ \bibnamefont {Crosse}}, \
  and\ \bibinfo {author} {\bibfnamefont {S.Y.}\ \bibnamefont {Buhmann}},\
  }\bibfield  {title} {Casimir-polder shift and decay rate in the presence of
  nonreciprocal media,\ }\href
  {https://journals.aps.org/pra/abstract/10.1103/PhysRevA.95.023805} {\bibfield
   {journal} {\bibinfo  {journal} {Phys. Rev. A}\ }\textbf {\bibinfo {volume}
  {95}},\ \bibinfo {pages} {023805} (\bibinfo {year} {2017})}\BibitemShut
  {NoStop}%
\bibitem [{\citenamefont {Gangaraj}\ \emph {et~al.}(2018)\citenamefont
  {Gangaraj}, \citenamefont {Silveirinha}, \citenamefont {Hanson},
  \citenamefont {Antezza},\ and\ \citenamefont
  {Monticone}}]{gangaraj2018optical}%
  \BibitemOpen
  \bibfield  {author} {\bibinfo {author} {\bibfnamefont {S.A.H.}\ \bibnamefont
  {Gangaraj}}, \bibinfo {author} {\bibfnamefont {M.G.}\ \bibnamefont
  {Silveirinha}}, \bibinfo {author} {\bibfnamefont {G.W.}\ \bibnamefont
  {Hanson}}, \bibinfo {author} {\bibfnamefont {M.}~\bibnamefont {Antezza}}, \
  and\ \bibinfo {author} {\bibfnamefont {F.}~\bibnamefont {Monticone}},\
  }\bibfield  {title} {Optical torque on a two-level system near a strongly
  nonreciprocal medium,\ }\href
  {https://journals.aps.org/prb/abstract/10.1103/PhysRevB.98.125146} {\bibfield
   {journal} {\bibinfo  {journal} {Phys. Rev. B}\ }\textbf {\bibinfo {volume}
  {98}},\ \bibinfo {pages} {125146} (\bibinfo {year} {2018})}\BibitemShut
  {NoStop}%
\bibitem [{\citenamefont {Latella}\ and\ \citenamefont
  {Ben-Abdallah}(2017)}]{latella2017giant}%
  \BibitemOpen
  \bibfield  {author} {\bibinfo {author} {\bibfnamefont {I.}~\bibnamefont
  {Latella}}\ and\ \bibinfo {author} {\bibfnamefont {P.}~\bibnamefont
  {Ben-Abdallah}},\ }\bibfield  {title} {Giant thermal magnetoresistance in
  plasmonic structures,\ }\href
  {https://journals.aps.org/prl/abstract/10.1103/PhysRevLett.118.173902}
  {\bibfield  {journal} {\bibinfo  {journal} {Phys. Rev. Lett.}\ }\textbf
  {\bibinfo {volume} {118}},\ \bibinfo {pages} {173902} (\bibinfo {year}
  {2017})}\BibitemShut {NoStop}%
\bibitem [{\citenamefont {Joulain}\ \emph {et~al.}(2003)\citenamefont
  {Joulain}, \citenamefont {Carminati}, \citenamefont {Mulet},\ and\
  \citenamefont {Greffet}}]{joulain2003definition}%
  \BibitemOpen
  \bibfield  {author} {\bibinfo {author} {\bibfnamefont {K.}~\bibnamefont
  {Joulain}}, \bibinfo {author} {\bibfnamefont {R.}~\bibnamefont {Carminati}},
  \bibinfo {author} {\bibfnamefont {J-P.}\ \bibnamefont {Mulet}}, \ and\
  \bibinfo {author} {\bibfnamefont {J-J.}\ \bibnamefont {Greffet}},\ }\bibfield
   {title} {Definition and measurement of the local density of electromagnetic
  states close to an interface,\ }\href
  {https://journals.aps.org/prb/abstract/10.1103/PhysRevB.68.245405} {\bibfield
   {journal} {\bibinfo  {journal} {Phys. Rev. B}\ }\textbf {\bibinfo {volume}
  {68}},\ \bibinfo {pages} {245405} (\bibinfo {year} {2003})}\BibitemShut
  {NoStop}%
\bibitem [{\citenamefont {Barnett}\ \emph {et~al.}(2016)\citenamefont
  {Barnett}, \citenamefont {Allen},\ and\ \citenamefont
  {Padgett}}]{barnett2016optical}%
  \BibitemOpen
  \bibfield  {author} {\bibinfo {author} {\bibfnamefont {S.M.}\ \bibnamefont
  {Barnett}}, \bibinfo {author} {\bibfnamefont {L.}~\bibnamefont {Allen}}, \
  and\ \bibinfo {author} {\bibfnamefont {M.J.}\ \bibnamefont {Padgett}},\
  }\href@noop {} {\emph {\bibinfo {title} {Optical angular momentum}}}\
  (\bibinfo  {publisher} {CRC Press},\ \bibinfo {year} {2016})\BibitemShut
  {NoStop}%
\bibitem [{\citenamefont {Kalhor}\ \emph {et~al.}(2016)\citenamefont {Kalhor},
  \citenamefont {Thundat},\ and\ \citenamefont {Jacob}}]{kalhor2016universal}%
  \BibitemOpen
  \bibfield  {author} {\bibinfo {author} {\bibfnamefont {F.}~\bibnamefont
  {Kalhor}}, \bibinfo {author} {\bibfnamefont {T.}~\bibnamefont {Thundat}}, \
  and\ \bibinfo {author} {\bibfnamefont {Z.}~\bibnamefont {Jacob}},\ }\bibfield
   {title} {Universal spin-momentum locked optical forces,\ }\href
  {https://www.osapublishing.org/optica/abstract.cfm?uri=optica-3-2-118}
  {\bibfield  {journal} {\bibinfo  {journal} {Appl. Phys. Lett.}\ }\textbf
  {\bibinfo {volume} {108}},\ \bibinfo {pages} {061102} (\bibinfo {year}
  {2016})}\BibitemShut {NoStop}%
\bibitem [{\citenamefont {Nieto-Vesperinas}\ \emph {et~al.}(2004)\citenamefont
  {Nieto-Vesperinas}, \citenamefont {Chaumet},\ and\ \citenamefont
  {Rahmani}}]{nieto2004near}%
  \BibitemOpen
  \bibfield  {author} {\bibinfo {author} {\bibfnamefont {M.}~\bibnamefont
  {Nieto-Vesperinas}}, \bibinfo {author} {\bibfnamefont {P.C.}\ \bibnamefont
  {Chaumet}}, \ and\ \bibinfo {author} {\bibfnamefont {A.}~\bibnamefont
  {Rahmani}},\ }\bibfield  {title} {Near-field photonic forces,\ }\href
  {https://royalsocietypublishing.org/doi/abs/10.1098/rsta.2003.1343}
  {\bibfield  {journal} {\bibinfo  {journal} {Philos. Trans. Royal Soc. A}\ ,\
  \bibinfo {pages} {719--738}} (\bibinfo {year} {2004})}\BibitemShut {NoStop}%
\bibitem [{\citenamefont {Canaguier-Durand}\ \emph {et~al.}(2013)\citenamefont
  {Canaguier-Durand}, \citenamefont {Cuche}, \citenamefont {Genet},\ and\
  \citenamefont {Ebbesen}}]{canaguier2013force}%
  \BibitemOpen
  \bibfield  {author} {\bibinfo {author} {\bibfnamefont {A.}~\bibnamefont
  {Canaguier-Durand}}, \bibinfo {author} {\bibfnamefont {A.}~\bibnamefont
  {Cuche}}, \bibinfo {author} {\bibfnamefont {C.}~\bibnamefont {Genet}}, \ and\
  \bibinfo {author} {\bibfnamefont {T.W.}\ \bibnamefont {Ebbesen}},\ }\bibfield
   {title} {Force and torque on an electric dipole by spinning light fields,\
  }\href {https://journals.aps.org/pra/abstract/10.1103/PhysRevA.88.033831}
  {\bibfield  {journal} {\bibinfo  {journal} {Phys. Rev. A}\ }\textbf {\bibinfo
  {volume} {88}},\ \bibinfo {pages} {033831} (\bibinfo {year}
  {2013})}\BibitemShut {NoStop}%
\bibitem [{\citenamefont {Nieto-Vesperinas}\ \emph {et~al.}(2010)\citenamefont
  {Nieto-Vesperinas}, \citenamefont {S{\'a}enz}, \citenamefont
  {G{\'o}mez-Medina},\ and\ \citenamefont {Chantada}}]{nieto2010optical}%
  \BibitemOpen
  \bibfield  {author} {\bibinfo {author} {\bibfnamefont {M.}~\bibnamefont
  {Nieto-Vesperinas}}, \bibinfo {author} {\bibfnamefont {J.J.}\ \bibnamefont
  {S{\'a}enz}}, \bibinfo {author} {\bibfnamefont {R.}~\bibnamefont
  {G{\'o}mez-Medina}}, \ and\ \bibinfo {author} {\bibfnamefont
  {L.}~\bibnamefont {Chantada}},\ }\bibfield  {title} {Optical forces on small
  magnetodielectric particles,\ }\href
  {https://www.osapublishing.org/oe/abstract.cfm?uri=oe-18-11-11428} {\bibfield
   {journal} {\bibinfo  {journal} {Opt. Exp.}\ }\textbf {\bibinfo {volume}
  {18}},\ \bibinfo {pages} {11428--11443} (\bibinfo {year} {2010})}\BibitemShut
  {NoStop}%
\bibitem [{\citenamefont {Li}(2000)}]{li2000symmetries}%
  \BibitemOpen
  \bibfield  {author} {\bibinfo {author} {\bibfnamefont {L.}~\bibnamefont
  {Li}},\ }\bibfield  {title} {Symmetries of cross-polarization diffraction
  coefficients of gratings,\ }\href
  {https://www.osapublishing.org/josaa/abstract.cfm?uri=josaa-17-5-881}
  {\bibfield  {journal} {\bibinfo  {journal} {J. Opt. Spc. Am A}\ }\textbf
  {\bibinfo {volume} {17}},\ \bibinfo {pages} {881--887} (\bibinfo {year}
  {2000})}\BibitemShut {NoStop}%
\bibitem [{\citenamefont {Ishimaru}(1962)}]{ishimaru1962uni}%
  \BibitemOpen
  \bibfield  {author} {\bibinfo {author} {\bibfnamefont {A.}~\bibnamefont
  {Ishimaru}},\ }\bibfield  {title} {Unidirectional waves in anisotropic media
  and the resolution of the thermodynamic paradox,\ }\href
  {https://apps.dtic.mil/dtic/tr/fulltext/u2/297019.pdf} {\bibfield  {journal}
  {\bibinfo  {journal} {Tech. Rep.}\ } (\bibinfo {year} {1962})}\BibitemShut
  {NoStop}%
\bibitem [{\citenamefont {Capasso}\ \emph {et~al.}(2007)\citenamefont
  {Capasso}, \citenamefont {Munday}, \citenamefont {Iannuzzi},\ and\
  \citenamefont {Chan}}]{capasso2007casimir}%
  \BibitemOpen
  \bibfield  {author} {\bibinfo {author} {\bibfnamefont {F.}~\bibnamefont
  {Capasso}}, \bibinfo {author} {\bibfnamefont {J.N.}\ \bibnamefont {Munday}},
  \bibinfo {author} {\bibfnamefont {D.}~\bibnamefont {Iannuzzi}}, \ and\
  \bibinfo {author} {\bibfnamefont {H.B.}\ \bibnamefont {Chan}},\ }\bibfield
  {title} {Casimir forces and quantum electrodynamical torques: Physics and
  nanomechanics,\ }\href {https://ieeexplore.ieee.org/document/4159963}
  {\bibfield  {journal} {\bibinfo  {journal} {IEEE J. Sel. Top. Quantum
  Electron.}\ }\textbf {\bibinfo {volume} {13}},\ \bibinfo {pages} {400--414}
  (\bibinfo {year} {2007})}\BibitemShut {NoStop}%
\bibitem [{\citenamefont {Song}\ \emph {et~al.}(2015)\citenamefont {Song},
  \citenamefont {Fiorino}, \citenamefont {Meyhofer},\ and\ \citenamefont
  {Reddy}}]{song2015near}%
  \BibitemOpen
  \bibfield  {author} {\bibinfo {author} {\bibfnamefont {B.}~\bibnamefont
  {Song}}, \bibinfo {author} {\bibfnamefont {A.}~\bibnamefont {Fiorino}},
  \bibinfo {author} {\bibfnamefont {E.}~\bibnamefont {Meyhofer}}, \ and\
  \bibinfo {author} {\bibfnamefont {P.}~\bibnamefont {Reddy}},\ }\bibfield
  {title} {Near-field radiative thermal transport: From theory to experiment,\
  }\href {https://aip.scitation.org/doi/10.1063/1.4919048} {\bibfield
  {journal} {\bibinfo  {journal} {AIP Adv.}\ }\textbf {\bibinfo {volume} {5}},\
  \bibinfo {pages} {053503} (\bibinfo {year} {2015})}\BibitemShut {NoStop}%
\bibitem [{\citenamefont {Kushwaha}(2001)}]{kushwaha2001plasmons}%
  \BibitemOpen
  \bibfield  {author} {\bibinfo {author} {\bibfnamefont {M.S.}\ \bibnamefont
  {Kushwaha}},\ }\bibfield  {title} {Plasmons and magnetoplasmons in
  semiconductor heterostructures,\ }\href
  {https://www.sciencedirect.com/science/article/pii/S0167572900000078}
  {\bibfield  {journal} {\bibinfo  {journal} {Surf. Sci. Rep.}\ }\textbf
  {\bibinfo {volume} {41}},\ \bibinfo {pages} {1--416} (\bibinfo {year}
  {2001})}\BibitemShut {NoStop}%
\bibitem [{\citenamefont {Hu}\ \emph {et~al.}(2015)\citenamefont {Hu},
  \citenamefont {Zhang},\ and\ \citenamefont {Wang}}]{hu2015surface}%
  \BibitemOpen
  \bibfield  {author} {\bibinfo {author} {\bibfnamefont {B.}~\bibnamefont
  {Hu}}, \bibinfo {author} {\bibfnamefont {Y.}~\bibnamefont {Zhang}}, \ and\
  \bibinfo {author} {\bibfnamefont {Q.J.}\ \bibnamefont {Wang}},\ }\bibfield
  {title} {Surface magneto plasmons and their applications in the infrared
  frequencies,\ }\href
  {https://www.degruyter.com/view/j/nanoph.2015.4.issue-4/nanoph-2014-0026/nanoph-2014-0026.xml}
  {\bibfield  {journal} {\bibinfo  {journal} {Nanophotonics}\ }\textbf
  {\bibinfo {volume} {4}},\ \bibinfo {pages} {383--396} (\bibinfo {year}
  {2015})}\BibitemShut {NoStop}%
\bibitem [{\citenamefont {Chochol}\ \emph {et~al.}(2017)\citenamefont
  {Chochol}, \citenamefont {Postava}, \citenamefont {{\v{C}}ada},\ and\
  \citenamefont {Pi{\v{s}}tora}}]{chochol2017experimental}%
  \BibitemOpen
  \bibfield  {author} {\bibinfo {author} {\bibfnamefont {J.}~\bibnamefont
  {Chochol}}, \bibinfo {author} {\bibfnamefont {K.}~\bibnamefont {Postava}},
  \bibinfo {author} {\bibfnamefont {M.}~\bibnamefont {{\v{C}}ada}}, \ and\
  \bibinfo {author} {\bibfnamefont {J.}~\bibnamefont {Pi{\v{s}}tora}},\
  }\bibfield  {title} {Experimental demonstration of magnetoplasmon polariton
  at insb (inas)/dielectric interface for terahertz sensor application,\ }\href
  {https://www.nature.com/articles/s41598-017-13394-0} {\bibfield  {journal}
  {\bibinfo  {journal} {Sci. Rep.}\ }\textbf {\bibinfo {volume} {7}},\ \bibinfo
  {pages} {13117} (\bibinfo {year} {2017})}\BibitemShut {NoStop}%
\bibitem [{\citenamefont {Hartstein}\ \emph {et~al.}(1975)\citenamefont
  {Hartstein}, \citenamefont {Burstein}, \citenamefont {Palik}, \citenamefont
  {Gammon},\ and\ \citenamefont {Henvis}}]{hartstein1975investigation}%
  \BibitemOpen
  \bibfield  {author} {\bibinfo {author} {\bibfnamefont {A.}~\bibnamefont
  {Hartstein}}, \bibinfo {author} {\bibfnamefont {E.}~\bibnamefont {Burstein}},
  \bibinfo {author} {\bibfnamefont {E.D.}\ \bibnamefont {Palik}}, \bibinfo
  {author} {\bibfnamefont {R.W.}\ \bibnamefont {Gammon}}, \ and\ \bibinfo
  {author} {\bibfnamefont {B.W.}\ \bibnamefont {Henvis}},\ }\bibfield  {title}
  {Investigation of optic-phonon-magnetoplasmon-type surface polaritons on
  n-insb,\ }\href
  {https://journals.aps.org/prb/abstract/10.1103/PhysRevB.12.3186} {\bibfield
  {journal} {\bibinfo  {journal} {Phys. Rev. B}\ }\textbf {\bibinfo {volume}
  {12}},\ \bibinfo {pages} {3186} (\bibinfo {year} {1975})}\BibitemShut
  {NoStop}%
\bibitem [{\citenamefont {Palik}\ \emph {et~al.}(1976)\citenamefont {Palik},
  \citenamefont {Kaplan}, \citenamefont {Gammon}, \citenamefont {Kaplan},
  \citenamefont {Wallis},\ and\ \citenamefont {Quinn}}]{palik1976coupled}%
  \BibitemOpen
  \bibfield  {author} {\bibinfo {author} {\bibfnamefont {E.D.}\ \bibnamefont
  {Palik}}, \bibinfo {author} {\bibfnamefont {R.}~\bibnamefont {Kaplan}},
  \bibinfo {author} {\bibfnamefont {R.W.}\ \bibnamefont {Gammon}}, \bibinfo
  {author} {\bibfnamefont {H.}~\bibnamefont {Kaplan}}, \bibinfo {author}
  {\bibfnamefont {R.F.}\ \bibnamefont {Wallis}}, \ and\ \bibinfo {author}
  {\bibfnamefont {J.J.}\ \bibnamefont {Quinn}},\ }\bibfield  {title} {Coupled
  surface magnetoplasmon-optic-phonon polariton modes on insb,\ }\href
  {https://journals.aps.org/prb/pdf/10.1103/PhysRevB.13.2497} {\bibfield
  {journal} {\bibinfo  {journal} {Phys. Rev. B}\ }\textbf {\bibinfo {volume}
  {13}},\ \bibinfo {pages} {2497} (\bibinfo {year} {1976})}\BibitemShut
  {NoStop}%
\bibitem [{\citenamefont {Wang}\ \emph {et~al.}(2009)\citenamefont {Wang},
  \citenamefont {Chong}, \citenamefont {Joannopoulos},\ and\ \citenamefont
  {Solja{\v{c}}i{\'c}}}]{wang2009observation}%
  \BibitemOpen
  \bibfield  {author} {\bibinfo {author} {\bibfnamefont {Z.}~\bibnamefont
  {Wang}}, \bibinfo {author} {\bibfnamefont {Y.}~\bibnamefont {Chong}},
  \bibinfo {author} {\bibfnamefont {J.D.}\ \bibnamefont {Joannopoulos}}, \ and\
  \bibinfo {author} {\bibfnamefont {M.}~\bibnamefont {Solja{\v{c}}i{\'c}}},\
  }\bibfield  {title} {Observation of unidirectional backscattering-immune
  topological electromagnetic states,\ }\href
  {https://www.nature.com/articles/nature08293} {\bibfield  {journal} {\bibinfo
   {journal} {Nature}\ }\textbf {\bibinfo {volume} {461}},\ \bibinfo {pages}
  {772} (\bibinfo {year} {2009})}\BibitemShut {NoStop}%
\bibitem [{\citenamefont {Rechtsman}\ \emph {et~al.}(2013)\citenamefont
  {Rechtsman}, \citenamefont {Zeuner}, \citenamefont {Plotnik}, \citenamefont
  {Lumer}, \citenamefont {Podolsky}, \citenamefont {Dreisow}, \citenamefont
  {Nolte}, \citenamefont {Segev},\ and\ \citenamefont
  {Szameit}}]{rechtsman2013photonic}%
  \BibitemOpen
  \bibfield  {author} {\bibinfo {author} {\bibfnamefont {M.C.}\ \bibnamefont
  {Rechtsman}}, \bibinfo {author} {\bibfnamefont {J.M.}\ \bibnamefont
  {Zeuner}}, \bibinfo {author} {\bibfnamefont {Y.}~\bibnamefont {Plotnik}},
  \bibinfo {author} {\bibfnamefont {Y.}~\bibnamefont {Lumer}}, \bibinfo
  {author} {\bibfnamefont {D.}~\bibnamefont {Podolsky}}, \bibinfo {author}
  {\bibfnamefont {F.}~\bibnamefont {Dreisow}}, \bibinfo {author} {\bibfnamefont
  {S.}~\bibnamefont {Nolte}}, \bibinfo {author} {\bibfnamefont
  {M.}~\bibnamefont {Segev}}, \ and\ \bibinfo {author} {\bibfnamefont
  {A.}~\bibnamefont {Szameit}},\ }\bibfield  {title} {Photonic floquet
  topological insulators,\ }\href {https://www.nature.com/articles/nature12066}
  {\bibfield  {journal} {\bibinfo  {journal} {Nature}\ }\textbf {\bibinfo
  {volume} {496}},\ \bibinfo {pages} {196} (\bibinfo {year}
  {2013})}\BibitemShut {NoStop}%
\bibitem [{\citenamefont {Buddhiraju}\ \emph {et~al.}(2018)\citenamefont
  {Buddhiraju}, \citenamefont {Shi}, \citenamefont {Song}, \citenamefont
  {Wojcik}, \citenamefont {Minkov}, \citenamefont {Williamson}, \citenamefont
  {Dutt},\ and\ \citenamefont {Fan}}]{buddhiraju2018absence}%
  \BibitemOpen
  \bibfield  {author} {\bibinfo {author} {\bibfnamefont {S.}~\bibnamefont
  {Buddhiraju}}, \bibinfo {author} {\bibfnamefont {Y.}~\bibnamefont {Shi}},
  \bibinfo {author} {\bibfnamefont {A.}~\bibnamefont {Song}}, \bibinfo {author}
  {\bibfnamefont {C.}~\bibnamefont {Wojcik}}, \bibinfo {author} {\bibfnamefont
  {M.}~\bibnamefont {Minkov}}, \bibinfo {author} {\bibfnamefont {I.A.D.}\
  \bibnamefont {Williamson}}, \bibinfo {author} {\bibfnamefont
  {A.}~\bibnamefont {Dutt}}, \ and\ \bibinfo {author} {\bibfnamefont
  {S.}~\bibnamefont {Fan}},\ }\bibfield  {title} {Absence of unidirectionally
  propagating surface plasmon-polaritons in nonreciprocal plasmonics,\ }\href
  {https://arxiv.org/abs/1809.05100} {\bibfield  {journal} {\bibinfo  {journal}
  {arXiv preprint arXiv:1809.05100}\ } (\bibinfo {year} {2018})}\BibitemShut
  {NoStop}%
\bibitem [{\citenamefont {Ukita}\ and\ \citenamefont
  {Kawashima}(2010)}]{ukita2010optical}%
  \BibitemOpen
  \bibfield  {author} {\bibinfo {author} {\bibfnamefont {H.}~\bibnamefont
  {Ukita}}\ and\ \bibinfo {author} {\bibfnamefont {H.}~\bibnamefont
  {Kawashima}},\ }\bibfield  {title} {Optical rotor capable of controlling
  clockwise and counterclockwise rotation in optical tweezers by displacing the
  trapping position,\ }\href
  {https://www.osapublishing.org/ao/abstract.cfm?uri=ao-49-10-1991} {\bibfield
  {journal} {\bibinfo  {journal} {Appl. Opt.}\ }\textbf {\bibinfo {volume}
  {49}},\ \bibinfo {pages} {1991--1996} (\bibinfo {year} {2010})}\BibitemShut
  {NoStop}%
\bibitem [{\citenamefont {Angelsky}\ \emph {et~al.}(2012)\citenamefont
  {Angelsky}, \citenamefont {Bekshaev}, \citenamefont {Maksimyak},
  \citenamefont {Maksimyak}, \citenamefont {Mokhun}, \citenamefont {Hanson},
  \citenamefont {Zenkova},\ and\ \citenamefont
  {Tyurin}}]{angelsky2012circular}%
  \BibitemOpen
  \bibfield  {author} {\bibinfo {author} {\bibfnamefont {O.V.}\ \bibnamefont
  {Angelsky}}, \bibinfo {author} {\bibfnamefont {A.Y.}\ \bibnamefont
  {Bekshaev}}, \bibinfo {author} {\bibfnamefont {P.P.}\ \bibnamefont
  {Maksimyak}}, \bibinfo {author} {\bibfnamefont {A.P.}\ \bibnamefont
  {Maksimyak}}, \bibinfo {author} {\bibfnamefont {I.I.}\ \bibnamefont
  {Mokhun}}, \bibinfo {author} {\bibfnamefont {S.G.}\ \bibnamefont {Hanson}},
  \bibinfo {author} {\bibfnamefont {C.Y.}\ \bibnamefont {Zenkova}}, \ and\
  \bibinfo {author} {\bibfnamefont {A.V.}\ \bibnamefont {Tyurin}},\ }\bibfield
  {title} {Circular motion of particles suspended in a gaussian beam with
  circular polarization validates the spin part of the internal energy flow,\
  }\href {https://www.osapublishing.org/oe/abstract.cfm?uri=oe-20-10-11351}
  {\bibfield  {journal} {\bibinfo  {journal} {Opt. Exp.}\ }\textbf {\bibinfo
  {volume} {20}},\ \bibinfo {pages} {11351--11356} (\bibinfo {year}
  {2012})}\BibitemShut {NoStop}%
\bibitem [{\citenamefont {Chan}\ \emph {et~al.}(2001)\citenamefont {Chan},
  \citenamefont {Aksyuk}, \citenamefont {Kleiman}, \citenamefont {Bishop},\
  and\ \citenamefont {Capasso}}]{chan2001quantum}%
  \BibitemOpen
  \bibfield  {author} {\bibinfo {author} {\bibfnamefont {H.B.}\ \bibnamefont
  {Chan}}, \bibinfo {author} {\bibfnamefont {V.A.}\ \bibnamefont {Aksyuk}},
  \bibinfo {author} {\bibfnamefont {R.N.}\ \bibnamefont {Kleiman}}, \bibinfo
  {author} {\bibfnamefont {D.J.}\ \bibnamefont {Bishop}}, \ and\ \bibinfo
  {author} {\bibfnamefont {F.}~\bibnamefont {Capasso}},\ }\bibfield  {title}
  {Quantum mechanical actuation of microelectromechanical systems by the
  casimir force,\ }\href
  {https://science.sciencemag.org/content/291/5510/1941.abstract} {\bibfield
  {journal} {\bibinfo  {journal} {Science}\ }\textbf {\bibinfo {volume}
  {291}},\ \bibinfo {pages} {1941--1944} (\bibinfo {year} {2001})}\BibitemShut
  {NoStop}%
\bibitem [{\citenamefont {Haslinger}\ \emph {et~al.}(2018)\citenamefont
  {Haslinger}, \citenamefont {Jaffe}, \citenamefont {Xu}, \citenamefont
  {Schwartz}, \citenamefont {Sonnleitner}, \citenamefont {Ritsch-Marte},
  \citenamefont {Ritsch},\ and\ \citenamefont
  {M{\"u}ller}}]{haslinger2018attractive}%
  \BibitemOpen
  \bibfield  {author} {\bibinfo {author} {\bibfnamefont {P.}~\bibnamefont
  {Haslinger}}, \bibinfo {author} {\bibfnamefont {M.}~\bibnamefont {Jaffe}},
  \bibinfo {author} {\bibfnamefont {V.}~\bibnamefont {Xu}}, \bibinfo {author}
  {\bibfnamefont {O.}~\bibnamefont {Schwartz}}, \bibinfo {author}
  {\bibfnamefont {M.}~\bibnamefont {Sonnleitner}}, \bibinfo {author}
  {\bibfnamefont {M.}~\bibnamefont {Ritsch-Marte}}, \bibinfo {author}
  {\bibfnamefont {H.}~\bibnamefont {Ritsch}}, \ and\ \bibinfo {author}
  {\bibfnamefont {H.}~\bibnamefont {M{\"u}ller}},\ }\bibfield  {title}
  {Attractive force on atoms due to blackbody radiation,\ }\href@noop {}
  {\bibfield  {journal} {\bibinfo  {journal} {Nat. Phys.}\ }\textbf {\bibinfo
  {volume} {14}},\ \bibinfo {pages} {257} (\bibinfo {year} {2018})}\BibitemShut
  {NoStop}%
\bibitem [{\citenamefont {Bao}\ \emph {et~al.}(2018)\citenamefont {Bao},
  \citenamefont {Shi}, \citenamefont {Cao}, \citenamefont {Evans},\ and\
  \citenamefont {He}}]{bao2018inhomogeneity}%
  \BibitemOpen
  \bibfield  {author} {\bibinfo {author} {\bibfnamefont {F.}~\bibnamefont
  {Bao}}, \bibinfo {author} {\bibfnamefont {K.}~\bibnamefont {Shi}}, \bibinfo
  {author} {\bibfnamefont {G.}~\bibnamefont {Cao}}, \bibinfo {author}
  {\bibfnamefont {J.S.}\ \bibnamefont {Evans}}, \ and\ \bibinfo {author}
  {\bibfnamefont {S.}~\bibnamefont {He}},\ }\bibfield  {title}
  {Inhomogeneity-induced casimir transport of nanoparticles,\ }\href
  {https://journals.aps.org/prl/abstract/10.1103/PhysRevLett.121.130401}
  {\bibfield  {journal} {\bibinfo  {journal} {Phys. Rev. Lett.}\ }\textbf
  {\bibinfo {volume} {121}},\ \bibinfo {pages} {130401} (\bibinfo {year}
  {2018})}\BibitemShut {NoStop}%
\bibitem [{\citenamefont {Kirksey}(1988)}]{kirksey1988brownian}%
  \BibitemOpen
  \bibfield  {author} {\bibinfo {author} {\bibfnamefont {H.G.}\ \bibnamefont
  {Kirksey}},\ }\bibfield  {title} {Brownian motion: A classroom demonstration
  and student experiment,\ }\href
  {https://pubs.acs.org/doi/abs/10.1021/ed065p1091} {\bibfield  {journal}
  {\bibinfo  {journal} {J. Chem. Educ.}\ }\textbf {\bibinfo {volume} {65}},\
  \bibinfo {pages} {1091} (\bibinfo {year} {1988})}\BibitemShut {NoStop}%
\bibitem [{\citenamefont {Kawata}\ and\ \citenamefont
  {Sugiura}(1992)}]{kawata1992movement}%
  \BibitemOpen
  \bibfield  {author} {\bibinfo {author} {\bibfnamefont {S.}~\bibnamefont
  {Kawata}}\ and\ \bibinfo {author} {\bibfnamefont {T.}~\bibnamefont
  {Sugiura}},\ }\bibfield  {title} {Movement of micrometer-sized particles in
  the evanescent field of a laser beam,\ }\href
  {https://www.osapublishing.org/ol/abstract.cfm?uri=ol-17-11-772} {\bibfield
  {journal} {\bibinfo  {journal} {Opt. Lett.}\ }\textbf {\bibinfo {volume}
  {17}},\ \bibinfo {pages} {772--774} (\bibinfo {year} {1992})}\BibitemShut
  {NoStop}%
\bibitem [{\citenamefont {Volpe}\ \emph {et~al.}(2011)\citenamefont {Volpe},
  \citenamefont {Buttinoni}, \citenamefont {Vogt}, \citenamefont
  {K{\"u}mmerer},\ and\ \citenamefont {Bechinger}}]{volpe2011microswimmers}%
  \BibitemOpen
  \bibfield  {author} {\bibinfo {author} {\bibfnamefont {G.}~\bibnamefont
  {Volpe}}, \bibinfo {author} {\bibfnamefont {I.}~\bibnamefont {Buttinoni}},
  \bibinfo {author} {\bibfnamefont {D.}~\bibnamefont {Vogt}}, \bibinfo {author}
  {\bibfnamefont {H-J.}\ \bibnamefont {K{\"u}mmerer}}, \ and\ \bibinfo {author}
  {\bibfnamefont {C.}~\bibnamefont {Bechinger}},\ }\bibfield  {title}
  {Microswimmers in patterned environments,\ }\href
  {https://pubs.rsc.org/en/content/articlelanding/2011/sm/c1sm05960b#!divAbstract}
  {\bibfield  {journal} {\bibinfo  {journal} {Soft Matt.}\ }\textbf {\bibinfo
  {volume} {7}},\ \bibinfo {pages} {8810--8815} (\bibinfo {year}
  {2011})}\BibitemShut {NoStop}%
\bibitem [{\citenamefont {Mijalkov}\ and\ \citenamefont
  {Volpe}(2013)}]{mijalkov2013sorting}%
  \BibitemOpen
  \bibfield  {author} {\bibinfo {author} {\bibfnamefont {M.}~\bibnamefont
  {Mijalkov}}\ and\ \bibinfo {author} {\bibfnamefont {G.}~\bibnamefont
  {Volpe}},\ }\bibfield  {title} {Sorting of chiral microswimmers,\ }\href
  {https://pubs.rsc.org/en/content/articlelanding/2013/sm/c3sm27923e#!divAbstract}
  {\bibfield  {journal} {\bibinfo  {journal} {Soft Matt.}\ }\textbf {\bibinfo
  {volume} {9}},\ \bibinfo {pages} {6376--6381} (\bibinfo {year}
  {2013})}\BibitemShut {NoStop}%
\bibitem [{\citenamefont {Henkel}\ \emph {et~al.}(2002)\citenamefont {Henkel},
  \citenamefont {Joulain}, \citenamefont {Mulet},\ and\ \citenamefont
  {Greffet}}]{henkel2002radiation}%
  \BibitemOpen
  \bibfield  {author} {\bibinfo {author} {\bibfnamefont {C.}~\bibnamefont
  {Henkel}}, \bibinfo {author} {\bibfnamefont {K.}~\bibnamefont {Joulain}},
  \bibinfo {author} {\bibfnamefont {J-P.}\ \bibnamefont {Mulet}}, \ and\
  \bibinfo {author} {\bibfnamefont {J-J.}\ \bibnamefont {Greffet}},\ }\bibfield
   {title} {Radiation forces on small particles in thermal near fields,\ }\href
  {https://iopscience.iop.org/article/10.1088/1464-4258/4/5/356/meta}
  {\bibfield  {journal} {\bibinfo  {journal} {J Opt A-Pure Appl Opt.}\ }\textbf
  {\bibinfo {volume} {4}},\ \bibinfo {pages} {S109} (\bibinfo {year}
  {2002})}\BibitemShut {NoStop}%
\bibitem [{\citenamefont {Gordon}\ and\ \citenamefont
  {Ashkin}(1980)}]{gordon1980motion}%
  \BibitemOpen
  \bibfield  {author} {\bibinfo {author} {\bibfnamefont {J.P.}\ \bibnamefont
  {Gordon}}\ and\ \bibinfo {author} {\bibfnamefont {A.}~\bibnamefont
  {Ashkin}},\ }\bibfield  {title} {Motion of atoms in a radiation trap,\ }\href
  {https://journals.aps.org/pra/pdf/10.1103/PhysRevA.21.1606} {\bibfield
  {journal} {\bibinfo  {journal} {Phys. Rev. A}\ }\textbf {\bibinfo {volume}
  {21}},\ \bibinfo {pages} {1606} (\bibinfo {year} {1980})}\BibitemShut
  {NoStop}%
\bibitem [{\citenamefont {Manjavacas}\ \emph {et~al.}(2017)\citenamefont
  {Manjavacas}, \citenamefont {Zayats},\ and\ \citenamefont
  {de~Abajo}}]{manjavacas2017lateral}%
  \BibitemOpen
  \bibfield  {author} {\bibinfo {author} {\bibfnamefont {F.J.}\ \bibnamefont
  {Manjavacas}, \bibfnamefont {A.and Rodr{\'\i}guez-Fortu{\~n}o}}, \bibinfo
  {author} {\bibfnamefont {A.V.}\ \bibnamefont {Zayats}}, \ and\ \bibinfo
  {author} {\bibfnamefont {F.J.G.}\ \bibnamefont {de~Abajo}},\ }\bibfield
  {title} {Lateral casimir force on a rotating particle near a planar surface,\
  }\href {https://journals.aps.org/prl/abstract/10.1103/PhysRevLett.118.133605}
  {\bibfield  {journal} {\bibinfo  {journal} {Phys. Rev. Lett.}\ }\textbf
  {\bibinfo {volume} {118}},\ \bibinfo {pages} {133605} (\bibinfo {year}
  {2017})}\BibitemShut {NoStop}%
\bibitem [{\citenamefont {Kriegler}\ \emph {et~al.}(2010)\citenamefont
  {Kriegler}, \citenamefont {Rill}, \citenamefont {Linden},\ and\ \citenamefont
  {Wegener}}]{kriegler2010bianisotropic}%
  \BibitemOpen
  \bibfield  {author} {\bibinfo {author} {\bibfnamefont {C.E.}\ \bibnamefont
  {Kriegler}}, \bibinfo {author} {\bibfnamefont {M.S.}\ \bibnamefont {Rill}},
  \bibinfo {author} {\bibfnamefont {S.}~\bibnamefont {Linden}}, \ and\ \bibinfo
  {author} {\bibfnamefont {M.}~\bibnamefont {Wegener}},\ }\bibfield  {title}
  {Bianisotropic photonic metamaterials,\ }\href
  {https://ieeexplore.ieee.org/abstract/document/5075657} {\bibfield  {journal}
  {\bibinfo  {journal} {IEEE J. Sel. Top. Quantum Electron.}\ }\textbf
  {\bibinfo {volume} {16}},\ \bibinfo {pages} {367--375} (\bibinfo {year}
  {2010})}\BibitemShut {NoStop}%
\bibitem [{\citenamefont {Asadchy}\ \emph {et~al.}(2018)\citenamefont
  {Asadchy}, \citenamefont {D{\'\i}az-Rubio},\ and\ \citenamefont
  {Tretyakov}}]{asadchy2018bianisotropic}%
  \BibitemOpen
  \bibfield  {author} {\bibinfo {author} {\bibfnamefont {V.S.}\ \bibnamefont
  {Asadchy}}, \bibinfo {author} {\bibfnamefont {A.}~\bibnamefont
  {D{\'\i}az-Rubio}}, \ and\ \bibinfo {author} {\bibfnamefont {S.A.}\
  \bibnamefont {Tretyakov}},\ }\bibfield  {title} {Bianisotropic metasurfaces:
  physics and applications,\ }\href
  {https://www.degruyter.com/view/j/nanoph.2018.7.issue-6/nanoph-2017-0132/nanoph-2017-0132.xml}
  {\bibfield  {journal} {\bibinfo  {journal} {Nanophotonics}\ }\textbf
  {\bibinfo {volume} {7}},\ \bibinfo {pages} {1069--1094} (\bibinfo {year}
  {2018})}\BibitemShut {NoStop}%
\bibitem [{\citenamefont {Ishimaru}(2017)}]{ishimaru2017electromagnetic}%
  \BibitemOpen
  \bibfield  {author} {\bibinfo {author} {\bibfnamefont {A.}~\bibnamefont
  {Ishimaru}},\ }\href@noop {} {\emph {\bibinfo {title} {Electromagnetic wave
  propagation, radiation, and scattering: from fundamentals to applications}}}\
  (\bibinfo  {publisher} {John Wiley \& Sons},\ \bibinfo {year}
  {2017})\BibitemShut {NoStop}%
\bibitem [{\citenamefont {Rodrigue}(1988)}]{rodrigue1988generation}%
  \BibitemOpen
  \bibfield  {author} {\bibinfo {author} {\bibfnamefont {G.P.}\ \bibnamefont
  {Rodrigue}},\ }\bibfield  {title} {A generation of microwave ferrite
  devices,\ }\href {https://ieeexplore.ieee.org/document/4389} {\bibfield
  {journal} {\bibinfo  {journal} {Proc. IEEE}\ }\textbf {\bibinfo {volume}
  {76}},\ \bibinfo {pages} {121--137} (\bibinfo {year} {1988})}\BibitemShut
  {NoStop}%
\bibitem [{\citenamefont {Pyatakov}\ and\ \citenamefont
  {Zvezdin}(2012)}]{pyatakov2012magnetoelectric}%
  \BibitemOpen
  \bibfield  {author} {\bibinfo {author} {\bibfnamefont {A.P.}\ \bibnamefont
  {Pyatakov}}\ and\ \bibinfo {author} {\bibfnamefont {A.K.}\ \bibnamefont
  {Zvezdin}},\ }\bibfield  {title} {Magnetoelectric and multiferroic media,\
  }\href {https://iopscience.iop.org/article/10.3367/UFNe.0182.201206b.0593}
  {\bibfield  {journal} {\bibinfo  {journal} {Phys.-Uspekhi}\ }\textbf
  {\bibinfo {volume} {55}},\ \bibinfo {pages} {557--581} (\bibinfo {year}
  {2012})}\BibitemShut {NoStop}%
\bibitem [{\citenamefont {Albaalbaky}\ \emph {et~al.}(2017)\citenamefont
  {Albaalbaky}, \citenamefont {Kvashnin}, \citenamefont {Ledue}, \citenamefont
  {Patte},\ and\ \citenamefont {Fr{\'e}sard}}]{albaalbaky2017magnetoelectric}%
  \BibitemOpen
  \bibfield  {author} {\bibinfo {author} {\bibfnamefont {A.}~\bibnamefont
  {Albaalbaky}}, \bibinfo {author} {\bibfnamefont {Y.}~\bibnamefont
  {Kvashnin}}, \bibinfo {author} {\bibfnamefont {D.}~\bibnamefont {Ledue}},
  \bibinfo {author} {\bibfnamefont {R.}~\bibnamefont {Patte}}, \ and\ \bibinfo
  {author} {\bibfnamefont {R.}~\bibnamefont {Fr{\'e}sard}},\ }\bibfield
  {title} {Magnetoelectric properties of multiferroic cucro2 studied by means
  of ab initio calculations and monte carlo simulations,\ }\href
  {https://journals.aps.org/prb/abstract/10.1103/PhysRevB.96.064431} {\bibfield
   {journal} {\bibinfo  {journal} {Phys. Rev. B}\ }\textbf {\bibinfo {volume}
  {96}},\ \bibinfo {pages} {064431} (\bibinfo {year} {2017})}\BibitemShut
  {NoStop}%
\bibitem [{\citenamefont {LaForge}\ \emph {et~al.}(2010)\citenamefont
  {LaForge}, \citenamefont {Frenzel}, \citenamefont {Pursley}, \citenamefont
  {Lin}, \citenamefont {Liu}, \citenamefont {Shi},\ and\ \citenamefont
  {Basov}}]{laforge2010optical}%
  \BibitemOpen
  \bibfield  {author} {\bibinfo {author} {\bibfnamefont {A.D.}\ \bibnamefont
  {LaForge}}, \bibinfo {author} {\bibfnamefont {A.}~\bibnamefont {Frenzel}},
  \bibinfo {author} {\bibfnamefont {B.C.}\ \bibnamefont {Pursley}}, \bibinfo
  {author} {\bibfnamefont {T.}~\bibnamefont {Lin}}, \bibinfo {author}
  {\bibfnamefont {X.}~\bibnamefont {Liu}}, \bibinfo {author} {\bibfnamefont
  {J.}~\bibnamefont {Shi}}, \ and\ \bibinfo {author} {\bibfnamefont {D.N.}\
  \bibnamefont {Basov}},\ }\bibfield  {title} {Optical characterization of
  bi2se3 in a magnetic field: Infrared evidence for magnetoelectric coupling in
  a topological insulator material,\ }\href
  {https://journals.aps.org/prb/abstract/10.1103/PhysRevB.81.125120} {\bibfield
   {journal} {\bibinfo  {journal} {Phys. Rev. B}\ }\textbf {\bibinfo {volume}
  {81}},\ \bibinfo {pages} {125120} (\bibinfo {year} {2010})}\BibitemShut
  {NoStop}%
\bibitem [{\citenamefont {Caloz}\ \emph {et~al.}(2018)\citenamefont {Caloz},
  \citenamefont {Al{\`u}}, \citenamefont {Tretyakov}, \citenamefont {Sounas},
  \citenamefont {Achouri},\ and\ \citenamefont
  {Deck-L{\'e}ger}}]{caloz2018electromagnetic}%
  \BibitemOpen
  \bibfield  {author} {\bibinfo {author} {\bibfnamefont {C.}~\bibnamefont
  {Caloz}}, \bibinfo {author} {\bibfnamefont {A.}~\bibnamefont {Al{\`u}}},
  \bibinfo {author} {\bibfnamefont {S.}~\bibnamefont {Tretyakov}}, \bibinfo
  {author} {\bibfnamefont {D.}~\bibnamefont {Sounas}}, \bibinfo {author}
  {\bibfnamefont {K.}~\bibnamefont {Achouri}}, \ and\ \bibinfo {author}
  {\bibfnamefont {Z-L.}\ \bibnamefont {Deck-L{\'e}ger}},\ }\bibfield  {title}
  {Electromagnetic nonreciprocity,\ }\href
  {https://journals.aps.org/prapplied/abstract/10.1103/PhysRevApplied.10.047001}
  {\bibfield  {journal} {\bibinfo  {journal} {Phys. Rev. Appl.}\ }\textbf
  {\bibinfo {volume} {10}},\ \bibinfo {pages} {047001} (\bibinfo {year}
  {2018})}\BibitemShut {NoStop}%
\bibitem [{\citenamefont {Silveirinha}\ and\ \citenamefont
  {Maslovski}(2010)}]{silveirinha2010comment}%
  \BibitemOpen
  \bibfield  {author} {\bibinfo {author} {\bibfnamefont {M.G.}\ \bibnamefont
  {Silveirinha}}\ and\ \bibinfo {author} {\bibfnamefont {S.I.}\ \bibnamefont
  {Maslovski}},\ }\bibfield  {title} {Comment on “repulsive casimir force in
  chiral metamaterials”,\ }\href
  {https://journals.aps.org/prl/pdf/10.1103/PhysRevLett.105.189301} {\bibfield
  {journal} {\bibinfo  {journal} {Phys. Rev. Lett.}\ }\textbf {\bibinfo
  {volume} {105}},\ \bibinfo {pages} {189301} (\bibinfo {year}
  {2010})}\BibitemShut {NoStop}%
\bibitem [{\citenamefont {Gustafsson}\ and\ \citenamefont
  {Sj{\"o}berg}(2010)}]{gustafsson2010sum}%
  \BibitemOpen
  \bibfield  {author} {\bibinfo {author} {\bibfnamefont {M.}~\bibnamefont
  {Gustafsson}}\ and\ \bibinfo {author} {\bibfnamefont {D.}~\bibnamefont
  {Sj{\"o}berg}},\ }\bibfield  {title} {Sum rules and physical bounds on
  passive metamaterials,\ }\href
  {https://iopscience.iop.org/article/10.1088/1367-2630/12/4/043046} {\bibfield
   {journal} {\bibinfo  {journal} {New J. Phys.}\ }\textbf {\bibinfo {volume}
  {12}},\ \bibinfo {pages} {043046} (\bibinfo {year} {2010})}\BibitemShut
  {NoStop}%
\bibitem [{\citenamefont {Silveirinha}(2011)}]{silveirinha2011examining}%
  \BibitemOpen
  \bibfield  {author} {\bibinfo {author} {\bibfnamefont {M.G.}\ \bibnamefont
  {Silveirinha}},\ }\bibfield  {title} {Examining the validity of
  kramers-kronig relations for the magnetic permeability,\ }\href
  {https://journals.aps.org/prb/abstract/10.1103/PhysRevB.83.165119} {\bibfield
   {journal} {\bibinfo  {journal} {Phys. Rev. B}\ }\textbf {\bibinfo {volume}
  {83}},\ \bibinfo {pages} {165119} (\bibinfo {year} {2011})}\BibitemShut
  {NoStop}%
\bibitem [{\citenamefont {Khandekar}\ and\ \citenamefont
  {Rodriguez}(2017)}]{khandekar2017near}%
  \BibitemOpen
  \bibfield  {author} {\bibinfo {author} {\bibfnamefont {C.}~\bibnamefont
  {Khandekar}}\ and\ \bibinfo {author} {\bibfnamefont {A.~W.}\ \bibnamefont
  {Rodriguez}},\ }\bibfield  {title} {Near-field thermal upconversion and
  energy transfer through a kerr medium,\ }\href
  {https://www.osapublishing.org/oe/abstract.cfm?uri=oe-25-19-23164} {\bibfield
   {journal} {\bibinfo  {journal} {Opt. Exp.}\ }\textbf {\bibinfo {volume}
  {25}},\ \bibinfo {pages} {23164--23180} (\bibinfo {year} {2017})}\BibitemShut
  {NoStop}%
\bibitem [{\citenamefont {Jin}\ \emph {et~al.}(2016)\citenamefont {Jin},
  \citenamefont {Polimeridis},\ and\ \citenamefont
  {Rodriguez}}]{jin2016temperature}%
  \BibitemOpen
  \bibfield  {author} {\bibinfo {author} {\bibfnamefont {W.}~\bibnamefont
  {Jin}}, \bibinfo {author} {\bibfnamefont {A.~G.}\ \bibnamefont
  {Polimeridis}}, \ and\ \bibinfo {author} {\bibfnamefont {A.~W.}\ \bibnamefont
  {Rodriguez}},\ }\bibfield  {title} {Temperature control of thermal radiation
  from composite bodies,\ }\href
  {https://journals.aps.org/prb/abstract/10.1103/PhysRevB.93.121403} {\bibfield
   {journal} {\bibinfo  {journal} {Phys. Rev. B.}\ }\textbf {\bibinfo {volume}
  {93}},\ \bibinfo {pages} {121403} (\bibinfo {year} {2016})}\BibitemShut
  {NoStop}%
\bibitem [{\citenamefont {Li}\ and\ \citenamefont
  {Fan}(2018)}]{li2018nanophotonic}%
  \BibitemOpen
  \bibfield  {author} {\bibinfo {author} {\bibfnamefont {W.}~\bibnamefont
  {Li}}\ and\ \bibinfo {author} {\bibfnamefont {S.}~\bibnamefont {Fan}},\
  }\bibfield  {title} {Nanophotonic control of thermal radiation for energy
  applications,\ }\href
  {https://www.osapublishing.org/oe/abstract.cfm?uri=oe-26-12-15995} {\bibfield
   {journal} {\bibinfo  {journal} {Opt. Exp.}\ }\textbf {\bibinfo {volume}
  {26}},\ \bibinfo {pages} {15995--16021} (\bibinfo {year} {2018})}\BibitemShut
  {NoStop}%
\end{thebibliography}%


\begin{thebibliography}{1}%
\makeatletter
\providecommand \@ifxundefined [1]{%
 \@ifx{#1\undefined}
}%
\providecommand \@ifnum [1]{%
 \ifnum #1\expandafter \@firstoftwo
 \else \expandafter \@secondoftwo
 \fi
}%
\providecommand \@ifx [1]{%
 \ifx #1\expandafter \@firstoftwo
 \else \expandafter \@secondoftwo
 \fi
}%
\providecommand \natexlab [1]{#1}%
\providecommand \enquote  [1]{``#1''}%
\providecommand \bibnamefont  [1]{#1}%
\providecommand \bibfnamefont [1]{#1}%
\providecommand \citenamefont [1]{#1}%
\providecommand \href@noop [0]{\@secondoftwo}%
\providecommand \href [0]{\begingroup \@sanitize@url \@href}%
\providecommand \@href[1]{\@@startlink{#1}\@@href}%
\providecommand \@@href[1]{\endgroup#1\@@endlink}%
\providecommand \@sanitize@url [0]{\catcode `\\12\catcode `\$12\catcode
  `\&12\catcode `\#12\catcode `\^12\catcode `\_12\catcode `\%12\relax}%
\providecommand \@@startlink[1]{}%
\providecommand \@@endlink[0]{}%
\providecommand \url  [0]{\begingroup\@sanitize@url \@url }%
\providecommand \@url [1]{\endgroup\@href {#1}{\urlprefix }}%
\providecommand \urlprefix  [0]{URL }%
\providecommand \Eprint [0]{\href }%
\providecommand \doibase [0]{http://dx.doi.org/}%
\providecommand \selectlanguage [0]{\@gobble}%
\providecommand \bibinfo  [0]{\@secondoftwo}%
\providecommand \bibfield  [0]{\@secondoftwo}%
\providecommand \translation [1]{[#1]}%
\providecommand \BibitemOpen [0]{}%
\providecommand \bibitemStop [0]{}%
\providecommand \bibitemNoStop [0]{.\EOS\space}%
\providecommand \EOS [0]{\spacefactor3000\relax}%
\providecommand \BibitemShut  [1]{\csname bibitem#1\endcsname}%
\let\auto@bib@innerbib\@empty
\bibitem [{\citenamefont {Pitaevskii}\ \emph {et~al.}(2014)\citenamefont
  {Pitaevskii}, \citenamefont {Landau},\ and\ \citenamefont {E.}}]{landau}%
  \BibitemOpen
  \bibfield  {author} {\bibinfo {author} {\bibfnamefont {L}~\bibnamefont
  {Pitaevskii}}, \bibinfo {author} {\bibfnamefont {L.}~\bibnamefont {Landau}},
  \ and\ \bibinfo {author} {\bibfnamefont {Lifshitz}\ \bibnamefont {E.}},\
  }\href@noop {} {\emph {\bibinfo {title} {Statistical Physics - Course of
  Theoretical Physics Vol.9}}}\ (\bibinfo  {publisher} {Elsevier},\ \bibinfo
  {year} {2014})\BibitemShut {NoStop}%
\end{thebibliography}%

\end{document}


\title{Thermal spin photonics in the near-field of nonreciprocal media: Supplementary Materials}

\author{Chinmay Khandekar} \email{ckhandek@purdue.edu}
\affiliation{Birck Nanotechnology Center, School of Electrical and
  Computer Engineering, College of Engineering, Purdue University,
  West Lafayette, Indiana 47907, USA}

\author{Zubin Jacob}
\email{zjacob@purdue.edu}
\affiliation{Birck Nanotechnology Center, School of Electrical and Computer Engineering, College of Engineering, Purdue University, West Lafayette, Indiana 47907, USA}

\date{\today}

\begin{abstract}
  We consider a semi-infinite half-space of a generic bianisotropic
  medium at thermal equilibrium with vacuum. To analyze the thermal
  radiation on the vacuum side of the geometry, we derive the Green's
  function, equilibrium correlations of vector potential (fluctuation
  dissipation relation) and equilibrium correlations of
  electromagentic fields. Finally, we provide semi-analytic
  expressions for spin angular momentum density and Poynting flux
  perpendicular and parallel to the surface.
\end{abstract}

\pacs{}
\maketitle

\onecolumngrid

\section{Derivation of Green's function}
\label{sec1}

The vector potential $\Av(\rv_1)$ at $\rv_1$ produced by source
current density $\mathbf{J}(\rv_2)$ located at $\rv_2$ can be
calculated using Green's function $\Gb(\rv_1,\rv_2)$ using the
relation:
\begin{align}
  \Av(\rv_1)=\int_{V_{\rv_2}}\Gb(\rv_1,\rv_2)\mu_0\Jv(\rv_2) d^3\rv_2
\end{align}
Since we use Landau gauge $\Ev=i\omega\Av$, this is same as the
electric-type Green's function that is commonly employed in the
literature in the form of following equation:
\begin{align}
  \Ev(\rv_1)= i\omega\mu_0 \int_{V_{\rv_2}}\Gb(\rv_1,\rv_2) \Jv(\rv_2)
  d^3\rv_2 = \omega^2\mu_0 \int_{V_{\rv_2}}\Gb(\rv_1,\rv_2)
  \mathbf{p}(\rv_2) d^3\rv_2
\label{Gee}  
\end{align}
where $\mathbf{p}$ is polarization (dipole moment) density. Since the
derivation of this Green's function is quite well-known for isotropic
media, we do not reproduce that derivation here but focus mainly on
its extension to the case of bianisotropic half-space considered in
the manuscript. We use Weyl's angular spectrum representation of
Green's function. We write the position vectors as $\rv_j=(\Rv_j,z_j)$
with transverse co-ordinates $\Rv_j=(x_j,y_j)$ for $j=[1,2]$ and
wave-vectors $\kv=(\kv_{\parallel},k_z)$ with transverse wavevector
$\kv_{\parallel}=k_{\parallel}(\cos\phi\ev_x + \sin\phi\ev_y)$ where
we introduce $\phi$ for simplicity of expressions below. In vacuum,
the dispersion relation $k_{\parallel}^2+k_z^2=k_0^2=(\omega/c)^2$
follows from Maxwell's equation, where $k_{\parallel}$ is real-valued
and $k_z$ can be real (for $k_{\parallel} \leq k_0$) or complex-valued
(for $k_{\parallel} > k_0$). In vacuum, the electric field at $\rv_1$
produced by the dipole moment $\mathbf{p}=\mathbf{d}\delta(\rv-\rv_2)$
located at $\rv_2$ is written in the angular spectrum representation
for $z_1 \geq z_2$ as:
\begin{align}
\Ev_0(\Rv_1,z_1)=\omega^2\mu_0 \int
\frac{d^2\kv_{\parallel}}{(2\pi)^2} e^{i\kv_{\parallel}\cdot
  (\Rv_1-\Rv_2)}
\frac{i}{2k_z}e^{ik_z(z_1-z_2)}[\ev_{s+}(\ev_{s+}\cdot
  \mathbf{d})+\ev_{p+}(\ev_{p+}\cdot \mathbf{d})]
\end{align}
The polarization vectors $\ev_{j\pm}$ for $j={s,p}$ with $\pm$
denoting waves going along $\pm\ev_z$ directions are:
\begin{align}
\ev_{s\pm}=\begin{bmatrix}\sin\phi \\ -\cos\phi \\ 0 \end{bmatrix},
\ev_{p\pm}=\frac{-1}{k_0} \begin{bmatrix}\pm k_z\cos\phi \\ \pm
  k_z\sin\phi \\ -k_\parallel \end{bmatrix}
\label{spvectors}
\end{align}
For $z_1 < z_2$, the integrand is modified and contains the term
$e^{-ik_z(z_1-z_2)}[\ev_{s-}(\ev_{s-}\cdot
  \mathbf{d})+\ev_{p-}(\ev_{p-}\cdot \mathbf{d})]$. For consistency,
we will stick with $z_1\geq z_2$ in the following discussion. The
scattered/reflected field is calculated by considering the reflection
of the incident field at the interface ($z=0$). Since the waves
propagate in $-\ev_z$ direction to reach the interface, the incident
field will be of the form $e^{ik_z z_2}[\ev_{s-}(\ev_{s-}\cdot
  \mathbf{d})+\ev_{p-}(\ev_{p-}\cdot \mathbf{d})]$. It undergoes
reflection at the interface where polarization vectors change to
$\ev_{s-} \rightarrow r_{ss}\ev_{s+} + r_{ps}\ev_{p+}$ and $\ev_{p-}
\rightarrow r_{sp}\ev_{s+}+r_{pp}\ev_{p+}$. The Fresnel reflection
coefficient $r_{jk}$ for $j,k=[s,p]$ describes the amplitude of
$\ev_j$-polarized reflected light due to unit amplitude
$\ev_k$-polarized incident light. For isotropic media,
cross-polarization Fresnel coefficients $r_{sp},r_{ps}$ are zero which
simplifies the calculation. But they are not necessarily zero for
general bianisotropic media. The reflected field then acquires an
additional phase of $e^{ik_z z_1}$ upon reaching the position $\rv_1$
along with the overall transverse phase accrual same as
$e^{ik_{\parallel}\cdot(\Rv_1-\Rv_2)}$. This results in the
scattered/reflected field at $\rv_1$ given below:
\begin{align}
\Ev_{\text{ref}}(\Rv_1,z_1)=\omega^2\mu_0 \int
\frac{d^2\kv_{\parallel}}{(2\pi)^2} e^{i\kv_{\parallel}\cdot
  (\Rv_1-\Rv_2)}
\frac{i}{2k_z}e^{ik_z(z_1+z_2)}[(r_{ss}\ev_{s+}+r_{ps}\ev_{p+})
  (\ev_{s-}\cdot
  \mathbf{d})+(r_{sp}\ev_{s+}+r_{pp}\ev_{p+})(\ev_{p-}\cdot
  \mathbf{d})]
\end{align}
By writing the total field
$\Ev(\rv_1)=\Ev_0(\rv_1)+\Ev_{\text{ref}}(\rv_2)$ and using
Eq~\ref{Gee}, the Green's function $\Gb(\rv_1,\rv_2)$ is derived for
the geometry considered in the manuscript. 

\section{Derivation of Fluctuation Dissipation Theorem (FDT)}

We follow Landau's discussion in Ref.~\cite{landau} (Statistical
Physics Part 2, Chapter 8) to obtain the vector potential correlations
in the vacuum half-space when both the vacuum and the material
half-spaces are at the same thermodynamic temperature $T$ (FDT of
first kind). Let's consider the linear response theory developed by
Kubo. In this theory, we consider a discrete set of quantities denoted
by $x_a$ for $(a=1,2,...)$ which describe the behavior of the system
under certain external interactions. These interactions are described
by external forces $f_a$ such that interaction energy has the form:
\begin{align}
V(\mathbf{r}) = - \sum_{a} f_a(\mathbf{r}) x_a(\mathbf{r}) 
\end{align}
The quantities $x_a$ are further related to to the forces $f_a$
through linear generalized susceptibilities
$\alpha_{ab}(\mathbf{r},\mathbf{r}')$ (linear response). In the
Fourier domain, they can be written as:
\begin{align}
x_a(\omega,\mathbf{r}) = \int \sum_b
\alpha_{ab}(\omega;\mathbf{r},\mathbf{r}') f_b(\omega,\mathbf{r}')
d^3\mathbf{r}'
\end{align}
The spectral distribution of the fluctuating quantities
$x_a(\omega,r)$ is related to the generalized susceptibilities by
Kubo's fluctuation dissipation relation given by: 
\begin{align}
\langle x_a(\mathbf{r},\omega) x_b^*(\mathbf{r}',\omega') \rangle =
\frac{\alpha_{ab}(\omega;\mathbf{r},\mathbf{r}')-
  \alpha_{ba}^*(\omega;\mathbf{r}',\mathbf{r})}{2i}\frac{1}{\omega}
\underbrace{\bigg(\frac{\hbar\omega}{2}+\frac{\hbar\omega}{
    \text{exp}[\hbar\omega/k_BT]-1}\bigg)
  \delta(\omega-\omega')}_{\Theta(\omega,T)}
\label{kubofdt}
\end{align}
For electromagnetic fields, $x_a(\omega,\mathbf{r}) \rightarrow
A_j(\omega,\mathbf{r})$ ($j=x,y,z$ component of vector potential). The
interaction with the externally induced current is given by $V =
-\mathbf{j}\cdot \mathbf{A}$ where $\mathbf{j}(\omega,\mathbf{r})$ is
the generalized force. Since vector potential and current density are
related by the Green's function:
\begin{align}
A_j(\mathbf{r},\omega) = \int
G_{jk}(\omega;\mathbf{r},\mathbf{r}')\mu_0 J_{k}(\omega,\mathbf{r}')
d^3\mathbf{r}' 
\end{align}
the generalized susceptibility becomes
$\alpha_{ab}(\omega;\mathbf{r},\mathbf{r}')=\mu_0
G_{jk}(\omega;\mathbf{r}, \mathbf{r}')$. Making these substitutions in
Kubo's linear FDT given by Eq.~\ref{kubofdt}, one retrieves FDT for
vector potential components written in the matrix form as:
\begin{align}
  \langle \Av(\rv_1)\otimes \Av^*(\rv_2) \rangle =
  \frac{\Gb(\rv_1,\rv_2)-\Gb(\rv_2,\rv_1)^{*^T}}{2i}
    \frac{\mu_0}{\omega}\Theta(\omega,T)
\end{align}
$[..]^T$ denotes the matrix transpose and $[..]^*$ denotes complex
conjugation. The vector quantities are written as column vectors such
that $\Av=[A_x, A_y, A_z]^T$. Substituting the Green's function
obtained in section~\ref{sec1}, the vector potential correlations are:
\begin{align}
  \langle \Av(\rv_1)\otimes \Av^*(\rv_2) \rangle =
  \frac{\mu_0\Theta(\omega,T)}{\omega}\int
  \frac{k_{\parallel}dk_{\parallel}d\phi}{2i(2\pi)^2}
  &e^{i\kv_{\parallel}\cdot(\Rv_1-\Rv_2)}\bigg[
    \frac{i}{2k_z}e^{ik_z(z_1-z_2)}[\ev_{s+}\ev_{s+}^T+
      \ev_{p+}\ev_{p+}^T]\nonumber \\ &+
    \frac{i}{2k_z}e^{ik_z(z_1+z_2)}[(r_{ss}\ev_{s+}+
      r_{ps}\ev_{p+})\ev_{s-}^T +
      (r_{sp}\ev_{s+}+r_{pp}\ev_{p+})\ev_{p-}^T] \nonumber \\ &+
    \frac{i}{2k_z^*}e^{-ik_z^*(z_1-z_2)}[\ev_{s-}\ev_{s-}^T+
      \ev_{p-}\ev_{p-}^T]^{*^T} \nonumber \\ &+
    \frac{i}{2k_z^*}e^{-ik_z^*(z_1+z_2)}[(r_{ss}\ev_{s+}+
      r_{ps}\ev_{p+})\ev_{s-}^T +
      (r_{sp}\ev_{s+}+r_{pp}\ev_{p+})\ev_{p-}^T]^{*^T} \bigg]
\end{align}
Note that even though we eventually compute the correlations for
$\rv_1=\rv_2$, we still need to expand all the terms since
$\Gb(\rv_2,\rv_1)^{*^T} \neq \Gb(\rv_1,\rv_2)^{*^T}$. Furthermore,
following calculations involve curl operators that act differently on
different terms due to different phase factors making it necessary to
calculate each term carefully.

\section{Derivation of Electromagnetic field correlations}

The calculation of field correlations is straightforward in Landau
gauge since $\mathbf{E}=i\omega \mathbf{A},
\mathbf{B}=\nabla\times\mathbf{A}$:
\begin{align}
\label{Ecor}
&\langle \mathbf{E}(\mathbf{r}_1)\otimes \mathbf{E}^*(\mathbf{r}_2)
\rangle = \langle i\omega\mathbf{A}(\mathbf{r}_1))
(i\omega\mathbf{A}(\mathbf{r}_2))^{*^T} \rangle = \omega^2 \langle
\Av(\rv_1)\otimes \Av^*(\rv_2)\rangle \\ \label{EHcor} &\langle
\mathbf{E}(\mathbf{r}_1,\omega)\otimes
\mathbf{H}^*(\mathbf{r}_2,\omega) \rangle = \frac{1}{\mu_0}\langle
(i\omega\mathbf{A}(\mathbf{r}_1,\omega))
(\nabla_{r_2}\times\mathbf{A}(\mathbf{r}_2,\omega))^{*^T} \rangle =
\frac{i\omega}{\mu_0} \langle \Av(\rv_1)\otimes \Av^*(\rv_2)\rangle
(\nabla_{r2}\times)^{T} \\ &\label{Hcor}\langle
\mathbf{H}(\mathbf{r}_1,\omega)\otimes
\mathbf{H}^*(\mathbf{r}_2,\omega) \rangle = \frac{1}{\mu_0^2} \langle
(\nabla_{r_1}\times\mathbf{A}(\mathbf{r}_1,\omega))
(\nabla_{r_2}\times\mathbf{A}^*(\mathbf{r}_2,\omega))^T \rangle
=\frac{1}{\mu_0^2} (\nabla_{r1}\times) \langle \Av(\rv_1)\otimes
\Av^*(\rv_2)\rangle (\nabla_{r2}\times)^{T}
\end{align}
The curl operator $\nabla\times$ acts only on the exponential phase
factor and not the $s,p$ polarization vectors
($\ev_{s\pm},\ev_{p\pm}$) and therefore leads to the above simplified
form. The matrix form of the curl operator therefore depends on the
exponential phase in each term. For instance, for a term in the
correlations $\langle \Av(\rv_1)\otimes\Av^*(\rv_2) \rangle$ that has
the phase factor $e^{ik_x(x_1-x_2)+ik_y(y_1-y_2)-ik_z^*(z_1-z_2)}$,
the curl operator $\nabla_{r2}\times$ is:
\begin{align*}
  \nabla_{r2}\times = \begin{bmatrix}0 & -\partial_z & \partial_y
    \\ \partial_z & 0 & -\partial_x \\ -\partial_y & \partial_x &
    0\end{bmatrix}e^{ik_x(x_1-x_2)+ik_y(y_1-y_2)-ik_z^*(z_1-z_2)} =
    i\begin{bmatrix} 0 & -k_z^* & -k_y \\ k_z^* & 0 & k_x \\ k_y &
    -k_x & 0\end{bmatrix}
\end{align*}
Here $k_x=k_{\parallel}\cos\phi$, $k_y=k_{\parallel}\sin\phi$ and
$k_z^2=k_0^2-k_{\parallel}^2$. By performing these matrix operations
one can obtain the above electromagnetic field correlations. The
electromagnetic field correlations are:
\begin{align}
  \langle \Ev(\rv_1)\otimes \Ev^*(\rv_2) \rangle =
  \mu_0\omega\Theta(\omega,T)\int
  &\frac{k_{\parallel}dk_{\parallel}d\phi}{2i(2\pi)^2}
  e^{i\kv_{\parallel}\cdot(\Rv_1-\Rv_2)}\bigg[
    \frac{i}{2k_z}e^{ik_z(z_1-z_2)}[\ev_{s+}\ev_{s+}^T+
      \ev_{p+}\ev_{p+}^T]\nonumber \\ &+
    \frac{i}{2k_z}e^{ik_z(z_1+z_2)}[(r_{ss}\ev_{s+}+
      r_{ps}\ev_{p+})\ev_{s-}^T +
      (r_{sp}\ev_{s+}+r_{pp}\ev_{p+})\ev_{p-}^T] \nonumber \\ &+
    \frac{i}{2k_z^*}e^{-ik_z^*(z_1-z_2)}[\ev_{s-}\ev_{s-}^T+
      \ev_{p-}\ev_{p-}^T]^{*^T} \nonumber \\ &+
    \frac{i}{2k_z^*}e^{-ik_z^*(z_1+z_2)}[(r_{ss}\ev_{s+}+
      r_{ps}\ev_{p+})\ev_{s-}^T +
      (r_{sp}\ev_{s+}+r_{pp}\ev_{p+})\ev_{p-}^T]^{*^T} \bigg]
  \\ \langle \Ev(\rv_1)\otimes \Hv^*(\rv_2) \rangle =
  -k_0\Theta(\omega,T) \int
  &\frac{k_{\parallel}dk_{\parallel}d\phi}{2i(2\pi)^2}
  e^{i\kv_{\parallel}\cdot(\Rv_1-\Rv_2)}\bigg[
    \frac{i}{2k_z}e^{ik_z(z_1-z_2)}[\ev_{s+}\ev_{p+}^T-
      \ev_{p+}\ev_{s+}^T]\nonumber \\ &+
    \frac{i}{2k_z}e^{ik_z(z_1+z_2)}[(r_{ss}\ev_{s+}+
      r_{ps}\ev_{p+})\ev_{p-}^T -
      (r_{sp}\ev_{s+}+r_{pp}\ev_{p+})\ev_{s-}^T] \nonumber \\ &+
    \frac{i}{2k_z^*}e^{-ik_z^*(z_1-z_2)}[\ev_{p-}\ev_{s-}^T-
      \ev_{s-}\ev_{p-}^T]^{*^T} \nonumber \\ &+
    \frac{i}{2k_z^*}e^{-ik_z^*(z_1+z_2)}[(r_{ss}\ev_{p+}-
      r_{ps}\ev_{s+})\ev_{s-}^T +
      (r_{sp}\ev_{p+}-r_{pp}\ev_{s+})\ev_{p-}^T]^{*^T} \bigg]
  \\ \langle \Hv(\rv_1)\otimes \Hv^*(\rv_2) \rangle =
  \frac{k_0^2\Theta(\omega,T)}{\mu_0\omega} \int
  &\frac{k_{\parallel}dk_{\parallel}d\phi}{2i(2\pi)^2}
  e^{i\kv_{\parallel}\cdot(\Rv_1-\Rv_2)}\bigg[
    \frac{i}{2k_z}e^{ik_z(z_1-z_2)}[\ev_{p+}\ev_{p+}^T+
      \ev_{s+}\ev_{s+}^T]\nonumber \\ &+
    \frac{i}{2k_z}e^{ik_z(z_1+z_2)}[(r_{ss}\ev_{p+}-
      r_{ps}\ev_{s+})\ev_{p-}^T -
      (r_{sp}\ev_{p+}-r_{pp}\ev_{s+})\ev_{s-}^T] \nonumber \\ &+
    \frac{i}{2k_z^*}e^{-ik_z^*(z_1-z_2)}[\ev_{p-}\ev_{p-}^T+
      \ev_{s-}\ev_{s-}^T]^{*^T} \nonumber \\ &+
    \frac{i}{2k_z^*}e^{-ik_z^*(z_1+z_2)}[(r_{ss}\ev_{p+}-
      r_{ps}\ev_{s+})\ev_{p-}^T -
      (r_{sp}\ev_{p+}-r_{pp}\ev_{s+})\ev_{s-}^T]^{*^T} \bigg]
\end{align}
In the following, we look at the spin angular momentum density and
Poynting flux along certain $\ev_z$ direction (perpendicular to
surface) and along $\ev_x$ direction (parallel to surface).  We
evaluate these quantities at spatial point $\rv_1=\rv_2=\rv=(0,0,d)$.

{\bf Calculation of spin density and Poynting flux perpendicular to
  surface}
\begin{align}
\epsilon_0\langle E_x(\rv)E_y^*(\rv)\rangle &=
\frac{\omega\Theta(\omega,T)}{c^2}\int \frac{k_{\parallel}
  dk_{\parallel}d\phi}{16\pi^2}
\bigg[\frac{1}{k_z}\big(-1+\frac{k_z^2}{k_0^2}\big)\sin\phi\cos\phi +
  \frac{1}{k_z^*}(-1+\frac{k_z^{*^2}}{k_0^2})\sin\phi\cos\phi
  \nonumber \\ &+ \frac{e^{2ik_z d}}{k_z}\bigg(
  -r_{ss}\sin\phi\cos\phi + r_{ps}\frac{k_z}{k_0}\cos^2\phi +
  r_{sp}\frac{k_z}{k_0}\sin^2\phi -
  r_{pp}\frac{k_z^2}{k_0^2}\sin\phi\cos\phi \bigg) \nonumber
  \\ &+\frac{e^{-2ik_z^*d}}{k_z^*}\bigg(-r_{ss}^*\sin\phi\cos\phi -
  r_{ps}^*\frac{k_z^*}{k_0}\sin^2\phi
  -r_{sp}^*\frac{k_z^*}{k_0}\cos^2\phi
  -r_{pp}^*\frac{k_z^{*^2}}{k_0^2}\sin\phi\cos\phi \bigg) \bigg]
\\ \epsilon_0\langle E_y(\rv)E_x^*(\rv)\rangle &=
\frac{\omega\Theta(\omega,T)}{c^2}\int \frac{k_{\parallel}
  dk_{\parallel}d\phi}{16\pi^2}
\bigg[\frac{1}{k_z}\big(-1+\frac{k_z^2}{k_0^2}\big)\sin\phi\cos\phi +
  \frac{1}{k_z^*}(-1+\frac{k_z^{*^2}}{k_0^2})\sin\phi\cos\phi
  \nonumber \\ &+ \frac{e^{2ik_z d}}{k_z}\bigg(
  -r_{ss}\sin\phi\cos\phi - r_{ps}\frac{k_z}{k_0}\sin^2\phi -
  r_{sp}\frac{k_z}{k_0}\cos^2\phi -
  r_{pp}\frac{k_z^2}{k_0^2}\sin\phi\cos\phi \bigg) \nonumber
  \\ &+\frac{e^{-2ik_z^*d}}{k_z^*}\bigg(-r_{ss}^*\sin\phi\cos\phi +
  r_{ps}^*\frac{k_z^*}{k_0}\cos^2\phi
  +r_{sp}^*\frac{k_z^*}{k_0}\sin^2\phi
  -r_{pp}^*\frac{k_z^{*^2}}{k_0^2}\sin\phi\cos\phi \bigg) \bigg]
\\ \mu_0\langle H_x(\rv)H_y^*(\rv)\rangle &=
\frac{\omega\Theta(\omega,T)}{c^2}\int \frac{k_{\parallel}
  dk_{\parallel}d\phi}{16\pi^2}
\bigg[\frac{1}{k_z}\big(-1+\frac{k_z^2}{k_0^2}\big)\sin\phi\cos\phi -
  \frac{1}{k_z^*}(-1+\frac{k_z^{*^2}}{k_0^2})\sin\phi\cos\phi
  \nonumber \\ &+ \frac{e^{2ik_z d}}{k_z}\bigg(
  -r_{ss}\frac{k_z^2}{k_0^2}\sin\phi\cos\phi -
  r_{ps}\frac{k_z}{k_0}\sin^2\phi - r_{sp}\frac{k_z}{k_0}\cos^2\phi -
  r_{pp}\sin\phi\cos\phi \bigg) \nonumber
  \\ &+\frac{e^{-2ik_z^*d}}{k_z^*}\bigg(-r_{ss}^*\frac{k_z^{*^2}}{k_0^2}
  \sin\phi\cos\phi + r_{ps}^*\frac{k_z^*}{k_0}\cos^2\phi
  +r_{sp}^*\frac{k_z^*}{k_0}\sin^2\phi -r_{pp}^*\sin\phi\cos\phi
  \bigg) \bigg] \\ \mu_0\langle H_y(\rv)H_x^*(\rv)\rangle &=
\frac{\omega\Theta(\omega,T)}{c^2}\int \frac{k_{\parallel}
  dk_{\parallel}d\phi}{16\pi^2}
\bigg[\frac{1}{k_z}\big(-1+\frac{k_z^2}{k_0^2}\big)\sin\phi\cos\phi -
  \frac{1}{k_z^*}(-1+\frac{k_z^{*^2}}{k_0^2})\sin\phi\cos\phi
  \nonumber \\ &+ \frac{e^{2ik_z d}}{k_z}\bigg(
  -r_{ss}\frac{k_z^2}{k_0^2}\sin\phi\cos\phi +
  r_{ps}\frac{k_z}{k_0}\cos^2\phi + r_{sp}\frac{k_z}{k_0}\sin^2\phi -
  r_{pp}\sin\phi\cos\phi \bigg) \nonumber
  \\ &+\frac{e^{-2ik_z^*d}}{k_z^*}\bigg(-r_{ss}^*\frac{k_z^{*^2}}{k_0^2}
  \sin\phi\cos\phi - r_{ps}^*\frac{k_z^*}{k_0}\sin^2\phi
  -r_{sp}^*\frac{k_z^*}{k_0}\cos^2\phi -r_{pp}^*\sin\phi\cos\phi
  \bigg) \bigg] \\ \langle E_x(\rv)H_y^*(\rv)\rangle &=
-k_0\Theta(\omega,T)\int \frac{k_{\parallel}
  dk_{\parallel}d\phi}{16\pi^2} \bigg[\frac{1}{k_z}(-k_z) +
  \frac{1}{k_z^*}(k_z^*) \nonumber \\ &+ \frac{e^{2ik_z
      d}}{k_z}\bigg(r_{ss}\frac{k_z}{k_0}\sin^2\phi -
  r_{ps}\frac{k_z^2}{k_0^2}\sin\phi\cos\phi + r_{sp}\sin\phi\cos\phi -
  r_{pp}\frac{k_z}{k_0}\cos^2\phi \bigg) \nonumber
  \\ &+\frac{e^{-2ik_z^*d}}{k_z^*}\bigg(-r_{ss}^*\frac{k_z^*}{k_0}\sin^2\phi
  + r_{ps}^*\sin\phi\cos\phi
  -r_{sp}^*\frac{k_z^{*^2}}{k_0^2}\sin\phi\cos\phi
  +r_{pp}^*\frac{k_z^*}{k_0}\cos^2\phi \bigg) \bigg] \\ \langle
E_y(\rv)H_x^*(\rv)\rangle &= -k_0\Theta(\omega,T)\int
\frac{k_{\parallel} dk_{\parallel}d\phi}{16\pi^2}
\bigg[\frac{1}{k_z}(k_z) + \frac{1}{k_z^*}(-k_z^*) \nonumber \\ &+
  \frac{e^{2ik_z d}}{k_z}\bigg(-r_{ss}\frac{k_z}{k_0}\cos^2\phi -
  r_{ps}\frac{k_z^2}{k_0^2}\sin\phi\cos\phi + r_{sp}\sin\phi\cos\phi +
  r_{pp}\frac{k_z}{k_0}\sin^2\phi \bigg) \nonumber
  \\ &+\frac{e^{-2ik_z^*d}}{k_z^*}\bigg(r_{ss}^*\frac{k_z^*}{k_0}\cos^2\phi
  + r_{ps}^*\sin\phi\cos\phi
  -r_{sp}^*\frac{k_z^{*^2}}{k_0^2}\sin\phi\cos\phi
  -r_{pp}^*\frac{k_z^*}{k_0}\sin^2\phi \bigg) \bigg]
\end{align}
The heat flux density along $\ev_z$ direction is given by the Poynting
flux $P_z = \langle\Re\{E_x^*H_y-E_y^*H_x\}\rangle$. From the above
expressions, we get:
\begin{align}
P_z = -k_0\Theta(\omega,T)\int\frac{k_{\parallel}dk_{\parallel}d\phi}{16
  \pi^2k_0}\Re\{\big[e^{-2ik_z^*d}(r_{ss}^*-r_{pp}^*)-
  e^{2ik_zd}(r_{ss}-r_{pp})\big]\} = 0 
\end{align}
The heat flux along $\ev_z$ direction is always zero irrespective of
the material type. Similarly, we obtain electric and magnetic
contributions to the spin angular momentum density along $\ev_z$
direction.
\begin{align}
S_{z}^{(\Ev)} &=\frac{\epsilon_0}{2\omega}\Im\langle E_x^*E_y -
E_y^*E_x \rangle =
\frac{\Theta(\omega,T)}{c^2}\int\frac{k_{\parallel}dk_{\parallel}d\phi}{16
  \pi^2 k_0}\Im\{-(r_{ps}+r_{sp})e^{2ik_z d}\} \\ S_{z}^{(\Hv)}
&=\frac{\mu_0}{2\omega}\Im\langle H_x^*H_y - H_y^*H_x \rangle =
\frac{\Theta(\omega,T)}{c^2}\int\frac{k_{\parallel}dk_{\parallel}d\phi}{16
  \pi^2 k_0}\Im\{(r_{ps}+r_{sp})e^{2ik_z d}\}
\end{align}
It follows that the total spin angular momentum density along $\ev_z$
is always zero. Note that the individual contributions above can be
nonzero in presence for $r_{sp}+r_{ps}\neq 0$ which is true for
nonreciprocal materials. The above results show that heat and total
angular momentum flux rates perpendicular to the surface are always
zero at thermal equilibrium. Since this is a thermodynamic
requirement, these results prove the consistency of fluctuational
electrodynamic theory with thermodynamics. \\

{\bf Calculation of spin density and Poynting flux parallel to the
  surface}
\begin{align}
\epsilon_0\langle E_y(\rv)E_z^*(\rv)\rangle &=
\frac{\omega\Theta(\omega,T)}{c^2}\int \frac{k_{\parallel}
  dk_{\parallel}d\phi}{16\pi^2} \bigg[\frac{1}{k_z}\bigg(\frac{-k_z
    k_{\parallel}}{k_0^2}\sin\phi \bigg) +
  \frac{1}{k_z^*}\bigg(\frac{k_z^*k_{\parallel}}{k_0^2}\sin\phi \bigg)
  \nonumber \\ &+ \frac{e^{2ik_z
      d}}{k_z}\bigg(-r_{sp}\frac{k_{\parallel}}{k_0}\cos\phi -
  r_{pp}\frac{k_z k_{\parallel}}{k_0^2}\sin\phi \bigg)
  +\frac{e^{-2ik_z^*d}}{k_z^*}\bigg(-r_{ps}^*\frac{k_{\parallel}}{k_0}
  \cos\phi + r_{pp}^*\frac{k_z^*k_{\parallel}}{k_0^2}\sin\phi \bigg)
  \bigg] \\ \epsilon_0\langle E_z(\rv)E_y^*(\rv)\rangle &=
\frac{\omega\Theta(\omega,T)}{c^2}\int \frac{k_{\parallel}
  dk_{\parallel}d\phi}{16\pi^2} \bigg[\frac{1}{k_z}\bigg(\frac{-k_z
    k_{\parallel}}{k_0^2}\sin\phi\bigg) +
  \frac{1}{k_z^*}\bigg(\frac{k_z^*k_{\parallel}}{k_0^2}\sin\phi\bigg)
  \nonumber \\ &+ \frac{e^{2ik_z
      d}}{k_z}\bigg(-r_{ps}\frac{k_{\parallel}}{k_0}\cos\phi +
  r_{pp}\frac{k_z k_{\parallel}}{k_0^2}\sin\phi \bigg)
  +\frac{e^{-2ik_z^*d}}{k_z^*}\bigg(-r_{sp}^*\frac{k_{\parallel}}{k_0}
  \cos\phi - r_{pp}^*\frac{k_z^*k_{\parallel}}{k_0^2}\sin\phi \bigg)
  \bigg] \\ \mu_0 \langle H_y(\rv)H_z^*(\rv)\rangle &=
\frac{\omega\Theta(\omega,T)}{c^2}\int \frac{k_{\parallel}
  dk_{\parallel}d\phi}{16\pi^2} \bigg[\frac{1}{k_z}\bigg(\frac{-k_z
    k_{\parallel}}{k_0^2}\sin\phi \bigg) +
  \frac{1}{k_z^*}\bigg(\frac{k_z^* k_{\parallel}}{k_0^2}\sin\phi
  \bigg) \nonumber \\ &+ \frac{e^{2ik_z d}}{k_z}\bigg(-r_{ss}\frac{k_z
    k_{\parallel}}{k_0}\sin\phi +
  r_{ps}\frac{k_{\parallel}}{k_0}\cos\phi \bigg)
  +\frac{e^{-2ik_z^*d}}{k_z^*}\bigg(r_{ss}^*\frac{k_z^*
    k_{\parallel}}{k_0^2}\sin\phi +
  r_{sp}^*\frac{k_{\parallel}}{k_0}\cos\phi \bigg) \bigg] \\ \mu_0
\langle H_z(\rv)H_y^*(\rv)\rangle &=
\frac{\omega\Theta(\omega,T)}{c^2}\int \frac{k_{\parallel}
  dk_{\parallel}d\phi}{16\pi^2} \bigg[\frac{1}{k_z}\bigg(\frac{-k_z
    k_{\parallel}}{k_0^2}\sin\phi \bigg) +
  \frac{1}{k_z^*}\bigg(\frac{k_z^* k_{\parallel}}{k_0^2}\sin\phi
  \bigg) \nonumber \\ &+ \frac{e^{2ik_z d}}{k_z}\bigg(r_{ss}\frac{k_z
    k_{\parallel}}{k_0}\sin\phi +
  r_{sp}\frac{k_{\parallel}}{k_0}\cos\phi \bigg)
  +\frac{e^{-2ik_z^*d}}{k_z^*}\bigg(-r_{ss}^*\frac{k_z^*
    k_{\parallel}}{k_0^2}\sin\phi +
  r_{ps}^*\frac{k_{\parallel}}{k_0}\cos\phi \bigg) \bigg] \\ \langle
E_y(\rv)H_z^*(\rv)\rangle &= -k_0\Theta(\omega,T)\int
\frac{k_{\parallel} dk_{\parallel}d\phi}{16\pi^2}
\bigg[\frac{1}{k_z}\frac{-k_{\parallel}}{k_0}\cos\phi +
  \frac{1}{k_z^*}\frac{-k_{\parallel}}{k_0}\cos\phi \nonumber \\ &+
  \frac{e^{2ik_z
      d}}{k_z}\bigg(-r_{ss}\frac{k_{\parallel}}{k_0}\cos\phi -
  r_{ps}\frac{k_z k_{\parallel}}{k_0^2}\sin\phi \bigg)+
  \frac{e^{-2ik_z^*d}}{k_z^*}\bigg(-r_{ss}^*\frac{\kp}{k_0}\cos\phi
  +r_{sp}^*\frac{k_z^* \kp}{k_0^2}\sin\phi \bigg) \bigg] \\ \langle
E_z(\rv)H_y^*(\rv)\rangle &= -k_0\Theta(\omega,T)\int
\frac{k_{\parallel} dk_{\parallel}d\phi}{16\pi^2} \bigg[
  \frac{1}{k_z}\frac{\kp}{k_0}\cos\phi +
  \frac{1}{k_z^*}\frac{\kp}{k_0}\cos\phi \nonumber \\ &+
  \frac{e^{2ik_z d}}{k_z}\bigg(r_{ps}\frac{k_z\kp}{k_0^2}\sin\phi +
  r_{pp}\frac{\kp}{k_0}\cos\phi \bigg)
  +\frac{e^{-2ik_z^*d}}{k_z^*}\bigg(-r_{sp}^*
  \frac{k_z^*\kp}{k_0^2}\sin\phi + r_{pp}^*\frac{\kp}{k_0}\cos\phi
  \bigg) \bigg]
\end{align}
Upon integration over angle $\phi$ for constant terms and cancellation
of various other terms, the simplified final expressions for spin
densities and flux rates are:
\begin{align}
S_x^{(\Ev)}&=\frac{1}{2\omega}\epsilon_0 \Im[\langle E_y^*E_z - E_z^*E_y
  \rangle] = \frac{\Theta(\omega,T)}{c^2}\int\frac{d\kp
  d\phi}{16\pi^2} \Im\bigg[(r_{sp}-r_{ps})e^{2ik_z d}\frac{\kp^2}{k_z
    k_0}\cos\phi + 2r_{pp}e^{2i k_z d}\frac{\kp}{k_0}\sin\phi\bigg]
\\ S_x^{(\Hv)}&=\frac{1}{2\omega}\epsilon_0 \Im[\langle H_y^*H_z -
  H_z^*H_y \rangle] = \frac{\Theta(\omega,T)}{c^2}\int\frac{d\kp
  d\phi}{16\pi^2} \Im\bigg[(r_{sp}-r_{ps})e^{2ik_z d}\frac{\kp^2}{k_z
    k_0}\cos\phi + 2r_{ss}e^{2i k_z d}\frac{\kp}{k_0}\sin\phi\bigg]
\\ P_x&=k_0\Theta(\omega,T)\int \frac{d\kp d\phi}{16\pi^2}
\Re\bigg[2(r_{ss}+r_{pp})e^{2ik_z d}\frac{\kp^2}{k_z k_0} +
  (r_{ps}-r_{sp}^*)\frac{\kp^2}{k_0^2}\sin\phi \bigg]
\end{align}
For isotropic media, $r_{sp}=r_{ps}=0$ and $r_{ss},r_{pp}$ do not
depend on the angle $\phi$. Upon integrating over $\phi$, all the
Poynting flux rates and spin densities in the vicinity of isotropic
media are zero. For generic biansiotropic media, one needs to compute
these quantities by integrating over $\phi$. We find that for
reciprocal bianisotropic media, the flux rates and spin-densities are
again zero. While this requires numerical validation by integrating
over angle $\phi$, it can also be proved based on time-reversal
symmetry arguments as discussed in the manuscript. For nonreciprocal
media (broken time-reversal symmetry), nonzero heat current and spin
angular momentum are expected at thermal equilibrium. We identify them
as persistent planar heat current (PPHC) and persistent thermal photon spin
(PTPS) in the manuscript. The presence of PTPS and PPHC parallel to the
surface does not lead to any thermodynamic contradiction.


\bibliography{photon}